\documentclass[trackchanges]{aastex701}
\usepackage{url}
\usepackage{multirow}
\usepackage{booktabs}
\usepackage{amsmath}
\usepackage{natbib}
\usepackage{enumitem}

\begin{document}

\title{COSINE (Cometary Object Study Investigating their Nature and Evolution) II: \\ Colors and Sizes of Comet Nuclei}
\shorttitle{COSINE II. Nucleus Properties of WISE/NEOWISE Comet Data}

\received{\today}
\revised{---} 
\accepted{---}

\author[orcid=0000-0002-8122-3606,sname=Kwon]{Yuna G. Kwon}
\affiliation{California Institute of Technology/IPAC, Pasadena, CA, USA}
\affiliation{Armagh Observatory \& Planetarium, College Hill, Armagh BT61 9DG, UK.}
\email[show]{ynkwontop@gmail.com}  

\author[orcid=0000-0003-1876-9988, sname=Dahlen]{Dar W. Dahlen} 
\affiliation{Technische Universit{\"a}t Braunschweig, Braunschweig, NI, Germany}
\email{dar.dahlen@tu-braunschweig.de}

\author[0000-0003-2638-720X, sname=Masiero]{Joseph R. Masiero}
\affiliation{California Institute of Technology/IPAC, Pasadena, CA, USA}
\email{jmasiero@ipac.caltech.edu}

\author[0000-0003-1156-9721, sname=Fern{\'a}ndez]{Yanga R. Fern{\'a}ndez}
\affiliation{University of Central Florida, Orlando, FL, USA}
\email{yanga.fernandez@ucf.edu}

\author[0000-0002-9347-8753, sname=Gicquel]{Adeline Gicquel}
\affiliation{University of Maryland, College Park, MD, USA}
\email{agicquel@umd.edu}

\author[0000-0001-9542-0953, sname=Bauer]{James M. Bauer}
\affiliation{University of Maryland, College Park, MD, USA}
\email{gerbsb@umd.edu}

\author[0000-0002-4676-2196, sname=Kim]{Yoonyoung Kim}
\affiliation{University of California, Los Angeles, CA, USA}
\email{yoonyoung@epss.ucla.edu}

\author[0000-0002-8532-9395, sname=Masci]{Frank Masci}
\affiliation{California Institute of Technology/IPAC, Pasadena, CA, USA}
\email{fmasci@ipac.caltech.edu}

\author[0000-0002-7578-3885, sname=Mainzer]{Amy K. Mainzer}
\affiliation{University of California, Los Angeles, CA, USA}
\email{mainzer@epss.ucla.edu}

\correspondingauthor{Yuna G. Kwon}

\begin{abstract}

Comet nuclei preserve a record of planetesimal formation, but their sizes are difficult to measure because the nucleus is usually blended with its coma. We present a catalog of nucleus sizes for 523 comets (235 long-period and 288 short-period comets; LPCs and SPCs) from the 14.5-year WISE/NEOWISE archive, supplemented by Spitzer photometry. A coma-nucleus decomposition of the radial profiles yields conservative upper limits for all 523 comets and diameters with uncertainties for 301 (107 LPCs and 194 SPCs), with median diameters of 5.3 and 2.5~km, respectively.  Nucleus colors across the four WISE bands follow the heliocentric trend expected from the transition between reflected and thermal emission, with no measurable difference between the two dynamical classes. 
The cumulative size-frequency distribution of the SPCs has a slope of $\gamma = 1.45_{-0.13}^{+0.15}$ (68\% confidence interval) over $D = 2$--15~km, steepening to $2.05_{-0.38}^{+0.47}$ for the perihelion $q < 2$~au subset. The LPC distribution is smoothly concave down: a broken power law places a break near 35~km, with slopes of $0.92_{-0.12}^{+0.13}$ below and $3.68_{-1.24}^{+1.96}$ above, while the $q < 2$~au subset gives $1.39_{-0.28}^{+0.35}$ over $D = 4$--20~km. All distributions flatten below the fitted ranges and level off near 1~km. We discuss the implications of our results and their connection with models of the destabilized trans-Neptunian population.

\end{abstract}

\keywords{\uat{Solar system}{1528} --- \uat{Comets}{280} --- \uat{Long period comets}{933} --- \uat{Short period comets}{1452} --- \uat{Comet nuclei}{2160} --- \uat{Comet surfaces}{2161}}

\section{Introduction \label{sec:intro}} 

Comets are among the least-processed planetesimals surviving from the formation of the Solar System and retain key records of its early history. Their constituents form a mixture of refractory material, including silicates and carbonaceous compounds (so-called CHON particles; \citealt{Levasseur-Regourd2018,Filacchione2024}), and volatile ices such as H$_2$O, CO$_2$, CO, and N$_2$ \citep{Bockelee-Morvan2017,Aikawa2024}. After formation, comets were gravitationally scattered by the giant planets into two primary reservoirs: the Kuiper Belt and scattered disk at $\sim 30$--$50$~au, and the Oort Cloud at $\sim 10^{3}$--$10^{5}$~au \citep[][and references therein]{Dones2015}. Residing beyond the condensation fronts of the major volatiles ($\lesssim 50$~K; e.g., \citealt{Bockelee-Morvan2004}), these reservoirs offer conditions favorable for long-term preservation, allowing comets to remain more pristine than inner Solar System bodies, which have experienced stronger solar processing.

This primitive state is progressively altered once comets are perturbed onto orbits that carry them into the inner Solar System. Solar irradiation depletes near-surface volatiles and produces vertically stratified ice layers \citep{Prialnik2004}. Cometary activity, driven by sublimation or, at large heliocentric distances, by the exothermic crystallization of amorphous water ice \citep[e.g.,][]{Meech2009,Meech2017}, further reworks the surface dust mantle and redistributes volatiles within the nucleus \citep{Vincent2019,Keller2020}. These processes complicate the reconstruction of intrinsic properties from observations. Nevertheless, surviving super-volatiles, isotopic ratios indicative of cryogenic origins, and fluffy dust aggregates all underscore the value of comets as accessible probes of early Solar System conditions \citep{Cochran2015,Blum2018,Bergin2024,Aikawa2024}. Their nuclei, as the source of all observable activity, exhibit a diversity that reflects this complex formation and evolutionary history.

This paper is the second installment of the COSINE (Cometary Object Study Investigating their Nature and Evolution) project. The first paper \citep[hereafter Paper~I]{Kwon2025} established a comet catalog from the full 14.5-year archive of the \emph{Wide-field Infrared Survey Explorer} (WISE; \citealt{Wright2010,Cutri2012}) and its planetary science extension, NEOWISE \citep{Mainzer2011,Cutri2013,Mainzer2014}. Building on that database, the present study separates the nucleus from the extended emission in each comet and isolates the point-source signal attributed to the nucleus, yielding colors, phase curves, sizes, and size distribution of comet nuclei. The resulting catalog provides the community with a large, internally consistent set of cometary nucleus sizes that extends earlier WISE/NEOWISE estimates from the cryogenic mission \citep{Bauer2015,Bauer2017}.

\section{Data Description \label{sec:data}}

\subsection{Observations \label{sec:obs}}

The WISE/NEOWISE spacecraft operated throughout all mission phases in a Sun-synchronous, low-Earth polar orbit, maintaining a fixed solar elongation of approximately 90\arcdeg. Its daily angular drift of $\sim$1\arcdeg\ relative to the Sun, combined with 10\% in-scan overlap, provided uniform and repeated sky coverage. The payload comprised a 40~cm telescope operating in continuous sky-scanning mode with a 47\arcmin\ $\times$ 47\arcmin\ field of view (FoV). Observations were obtained in four infrared bands centered at 3.4, 4.6, 12, and 22~$\mu$m (W1--W4; \citealt{Wright2010}). The pixel scale was 2.75\arcsec\ for W1--W3 and 5.5\arcsec\ for the 2$\times$2-binned W4 channel; motion blur of $\sim$0.05\arcsec\ remained well below the pixel sampling in all bands. The azimuthally averaged point-spread function (PSF) had full width at half maximum (FWHM) values of 6.1\arcsec, 6.4\arcsec, 6.5\arcsec, and 12.0\arcsec\ for W1 through W4, respectively. Exposure times were 7.7~s for W1 and W2 and 8.8~s for W3 and W4.

The mission began on 2010 January 14 and concluded on 2024 July 31, progressing through several operational phases defined by depletion of its dual-stage solid hydrogen cryostat \citep{Mainzer2011,Mainzer2014}. All four bands were active during the initial Cryo phase. As cryogen levels declined, W4 and subsequently W3 were deactivated, defining the 3-Band and Post-Cryo phases. After a two-year hibernation beginning in 2011 February 1, observations resumed in 2013 December 13 with only W1 and W2 active, initiating the decade-long Reactivation phase. Full descriptions of the optical system and mission operations are given by \citet{Wright2010}, \citet{Mainzer2011}, and \citet{Mainzer2014}.

\subsection{Baseline Dataset \label{sec:final_data}}

This study builds on the comet catalog compiled in Paper~I \citep{Kwon2025} from the complete 14.5-year WISE/NEOWISE dataset. After applying selection thresholds of heliocentric distance $<11.5$~au and coadded-image signal-to-noise ratio (SNR) $>4$, the WISE/NEOWISE sample comprises 1,633 coadded images produced from 22,197 frames across 966 epochs of 484 unique comets. Following the dynamical classification defined in Section~3.1.1 of Paper~I, the sample divides into 234 long-period comets (LPCs) and 250 short-period comets (SPCs). The LPCs comprise hyperbolic comets (HCs), near-parabolic comets (NPCs), and Halley-type comets (HTCs); the SPCs comprise Jupiter-family comets (JFCs) and Encke-type comets (ETCs).

We supplemented this sample with the cometary-nucleus dataset of \citet{Fernandez2013}, the Survey of the Ensemble Physical Properties of Cometary Nuclei (SEPPCoN), conducted with the \textit{Spitzer Space Telescope} \citep{Werner2004}. We adopted the published thermal-infrared fluxes, measured with two instruments: the Multiband Imaging Photometer for Spitzer (MIPS; \citealt{Rieke2004}) and the Infrared Spectrograph (IRS; \citealt{Houck2004}). Each MIPS target was imaged twice, the second visit timed so that the comet had shifted on the sky; differencing the two frames removes the static background and isolates the comet by its motion. MIPS provides imaging at 23.68~$\mu$m, and IRS contributes blue and red channels at 15.77 and 22.33~$\mu$m, respectively (color-corrected monochromatic wavelengths; \citealt{Fernandez2013}). This dataset adds flux measurements from 39 new comets across 98 epochs (43 pre-perihelion and 55 post-perihelion) and 196 frames.

\begin{figure}[!thb]
\centering
\includegraphics[width=0.73\textwidth]{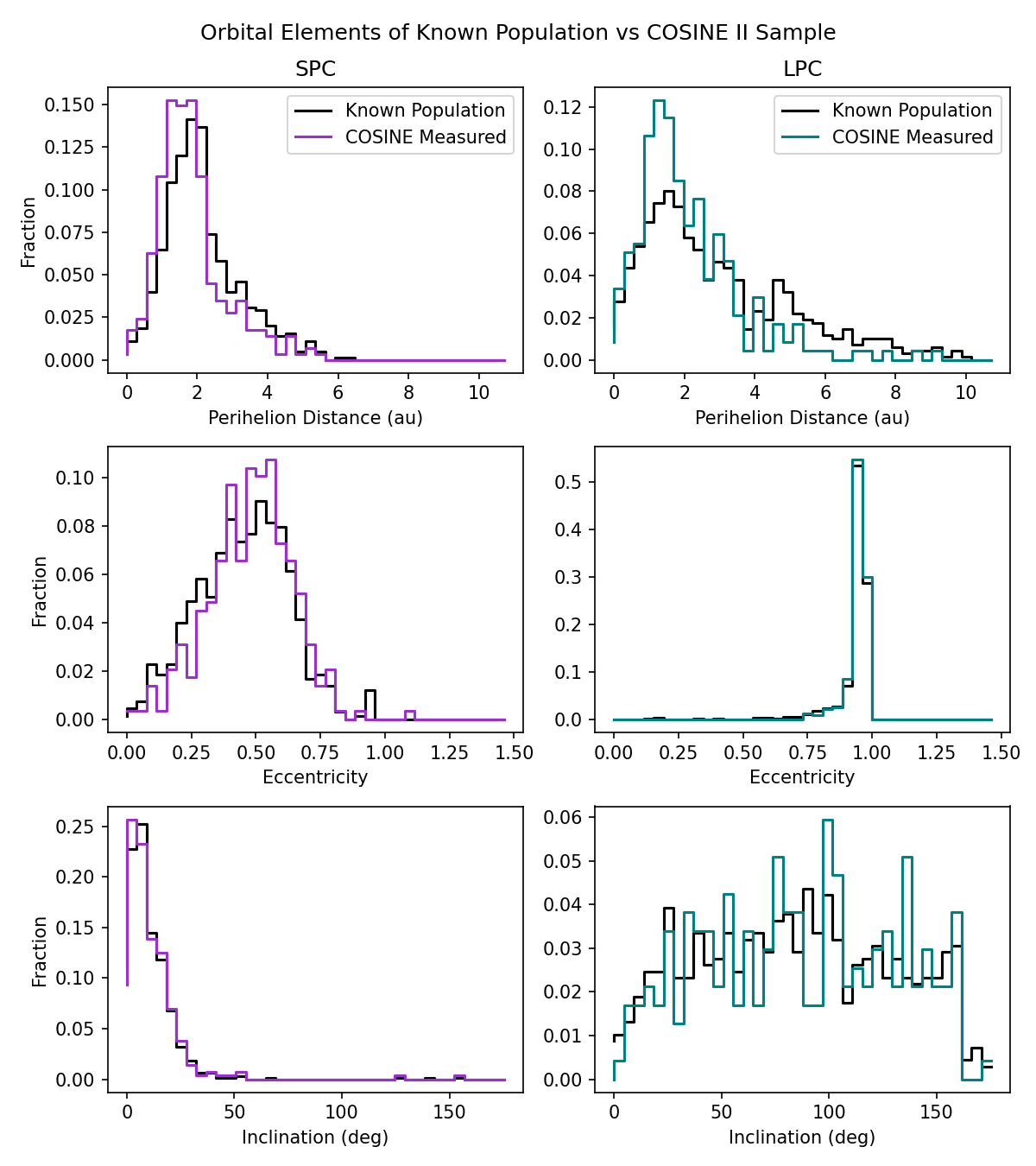}
\caption{Orbital-element distributions of the currently known comet population and of our baseline dataset, the latter combining the Paper~I and \citet{Fernandez2013} Spitzer samples. For each dynamical group (LPCs and SPCs), the histograms give the fraction of comets per bin in perihelion distance, eccentricity, and inclination, normalized to the total number in that group. The known population is taken from Paper~I (Section~2.1 of \citealt{Kwon2025}).
\label{Fig01}}
\end{figure}

Combining the two infrared surveys yields 523 unique comets (235 LPCs and 288 SPCs) observed at 1,064 epochs (465 pre-perihelion and 599 post-perihelion) from 1,829 coadded images. Figure~\ref{Fig01} compares the orbital-element distributions of the baseline dataset with those of the currently known comet population, shown separately for LPCs and SPCs. The known population comprises the 3,768 comet designations compiled in Paper~I before the selection cuts, as of October 2024\footnote{This sample does not include sungrazer comets nor dormant comet candidates.}. In perihelion distance, the dataset preferentially samples comets at small values and underrepresents those at larger perihelia in both dynamical groups, as expected for a flux-limited survey. Eccentricity and inclination, by contrast, are sampled in fair proportion to the known population.

This combined sample is the dataset for the present study and constitutes the homogeneous thermal-infrared compilation of comets, with a well-balanced representation of LPCs and SPCs. Because it samples a large comet population at consistent thermal-infrared wavelengths from comparable space platforms, it provides a robust basis for comparative studies within the cometary population and across small-body populations \citep{Bauer2024}. Table~\ref{tab:db_summary} summarizes its key properties.

\begin{deluxetable*}{c|cccc|ccc}[thb]
\tablewidth{0pt} 
\tablecaption{Key Properties of the Comet Database for This Study \label{tab:db_summary}}
\tablehead{
\multicolumn{8}{c}{\bf Census Summary}
}
\startdata 
\multicolumn{8}{l}{In total:} \\
\multicolumn{8}{l}{\qquad$\checkmark$~523 comets (235 LPCs $+$ 288 SPCs)} \\
\multicolumn{8}{l}{\qquad$\checkmark$~1,064 epochs (465 pre-perihelion $+$ 599 post-perihelion)} \\
\multicolumn{8}{l}{\qquad$\checkmark$~1,829 coadded images} \\ [.1cm]
\hline
\hline
\textbf{Numbers} & \multicolumn{4}{c}{\textbf{WISE Band}} & \multicolumn{3}{c}{\textbf{Spitzer Band}$^{(a)}$} \\ [.05cm]
\cline{2-8}
(LPC/SPC)$^{(b)}$ & \textbf{W1} & \textbf{W2} & \textbf{W3} & \textbf{W4} & \textbf{IRS-blue} & \textbf{IRS-red} & \textbf{MIPS} \\ [.05cm]
\hline
Comets & 140/143 & 189/204 & 41/88 & 52/85 & 62 & 62 & 36 \\
Epochs & 294/258 & 373/394 & 55/100 & 66/93 & 62 & 62 & 36  \\
Frames & 3,707/3,990 & 4,643/5,908 & 689/1,270 & 857/1,133 & 62 & 62 & 72  \\ [.1cm]
\enddata
\tablenotetext{(a)}{\quad Color-corrected monochromatic wavelengths of the IRS and MIPS instruments aboard the \textit{Spitzer Space Telescope}: `IRS-blue' and `IRS-red' correspond to 15.77~$\mu$m and 22.33~$\mu$m, and `MIPS' to 23.68~$\mu$m \citep{Fernandez2013}.}
\tablenotetext{(b)}{\quad The dynamical classification into long-period comets (LPCs) and short-period comets (SPCs) follows Paper~I. All targets from \citet{Fernandez2013} are Jupiter-family comets and therefore SPCs, except C/2005~W2 (Christensen), which we assign to the LPC class for consistency with the Paper~I criteria.}
\end{deluxetable*}

\section{Models of Nucleus Extraction \label{sec:models}}

A comet's observed flux is the superposition of an unresolved point source (the nucleus) and spatially extended emission from the coma and, when present, an asymmetric dust tail or trail. Recovering the nuclear signal is a prerequisite for any size or albedo estimate and requires separating the compact and extended components. We performed this separation on each WISE coadded cutout from Paper~I \citep{Kwon2025} with two complementary forward models. This section presents the two models and their rationale; the radial-profile construction and the fitting algebra are detailed in Appendix~\ref{sec:app1}.

Both models build on the standard combination of point-spread function (PSF) fitting and azimuthal (wedge) profile analysis used in cometary nucleus photometry \citep[e.g.,][]{Lamy1998a,Lamy2004,Fernandez2013,Bauer2015}. We divided each cutout into $N = 24$ angular wedges of width $\Delta\theta = 2\pi/N = 15\arcdeg$ about the centroid and, within radial annuli of width $\Delta\rho = 1$~pix, computed the area-weighted mean intensity in each (wedge, annulus) cell. Averaging these cells over a selected set of wedges yields a one-dimensional radial profile, which we modeled at projected distance $\rho$ from the centroid as
\begin{equation}
  M(\rho) \;=\; A\,P(\rho) \;+\; B\,C(\rho;s) \;+\; b~,
  \label{eq:phot_obs_signal}
\end{equation}
\noindent where $P(\rho)$ is the instrumental PSF radial profile; $C(\rho;s)$ is the coma profile, a surface-brightness law $\propto \rho^{-s}$ of unknown power-law index $s$ convolved with the same PSF; $A$ and $B$ are the nucleus and coma amplitudes in data number (DN), respectively; and $b$ is the background level adopted from Paper~I. Both the $P(\rho)$ and $C(\rho;s)$ templates were precomputed and normalized to unit sum, so that the scalar amplitudes $A$ and $B$ carry all of the flux information and the PSF convolution enters only through the fixed template shapes.

Fitting $A$ and $B$ freely against Eq.~\ref{eq:phot_obs_signal} is ill-conditioned: a sufficiently steep convolved coma is nearly collinear with the PSF, so a broad range of $(A, B)$ pairs can reproduce the observed profile to within the noise. This near-degeneracy is most severe near $s \simeq 2$, where the convolved coma template becomes nearly indistinguishable from the PSF (Fig.~\ref{Figap_01}). We regularized the decomposition by anchoring the model at the most nucleus-dominated point of the source, its intensity peak, requiring the model peak to equal the background-subtracted data peak,
\begin{equation}
  A\,\max P \;+\; B\,\max[\,C(\cdot;s)\,] \;=\; d_{\rm peak} - b~,
  \label{eq:peak}
\end{equation}
\noindent where $d_{\rm peak}$ is the peak data value of the cutout image (Appendix~\ref{sec:app1}). This constraint removes one degree of freedom and thereby fixes how the observed peak intensity is partitioned between nucleus and coma. The two models correspond to two treatments of this partition: assigning the peak entirely to the nucleus (Model~1; Section~\ref{sec:model1}), or admitting a coma contribution (Model~2; Section~\ref{sec:model2}).

\subsection{Model 1: Upper Limit Case \label{sec:model1}}

Model~1 represents the source as a single point source on a uniform background. It is the case when the radial profile is statistically indistinguishable from the PSF, either because the comet is inactive or because its coma is too compact to resolve, leaving the profile too PSF-like (the $s \simeq 2$ degeneracy) for $B$ to be constrained (Fig.~\ref{Figap_01}). Setting $B \rightarrow 0$ in Eq.~\ref{eq:peak} assigns the entire central excess to the nucleus, and the peak constraint fixes the nuclear amplitude at $A_{\rm M1} = (d_{\rm peak} - b)/\max P$, so that Eq.~\ref{eq:phot_obs_signal} reduces to
\begin{equation}
  M_1(\rho) \;=\; A_{\rm M1}\,P(\rho) \;+\; b~.
  \label{eq:model1}
\end{equation}
\noindent Because any unresolved coma would also contribute to the central intensity, attributing all of it to the nucleus can only overstate the nuclear flux; Eq.~\ref{eq:model1} therefore provides a conservative \emph{upper limit}. The size implied by $M_1$ served as a ceiling in the thermal modeling (Section~\ref{sec:phot_thermal}), yielding strict upper bounds on the fitted nuclei diameters.

\begin{figure}[!bht]
\centering
\includegraphics[width=0.87\textwidth]{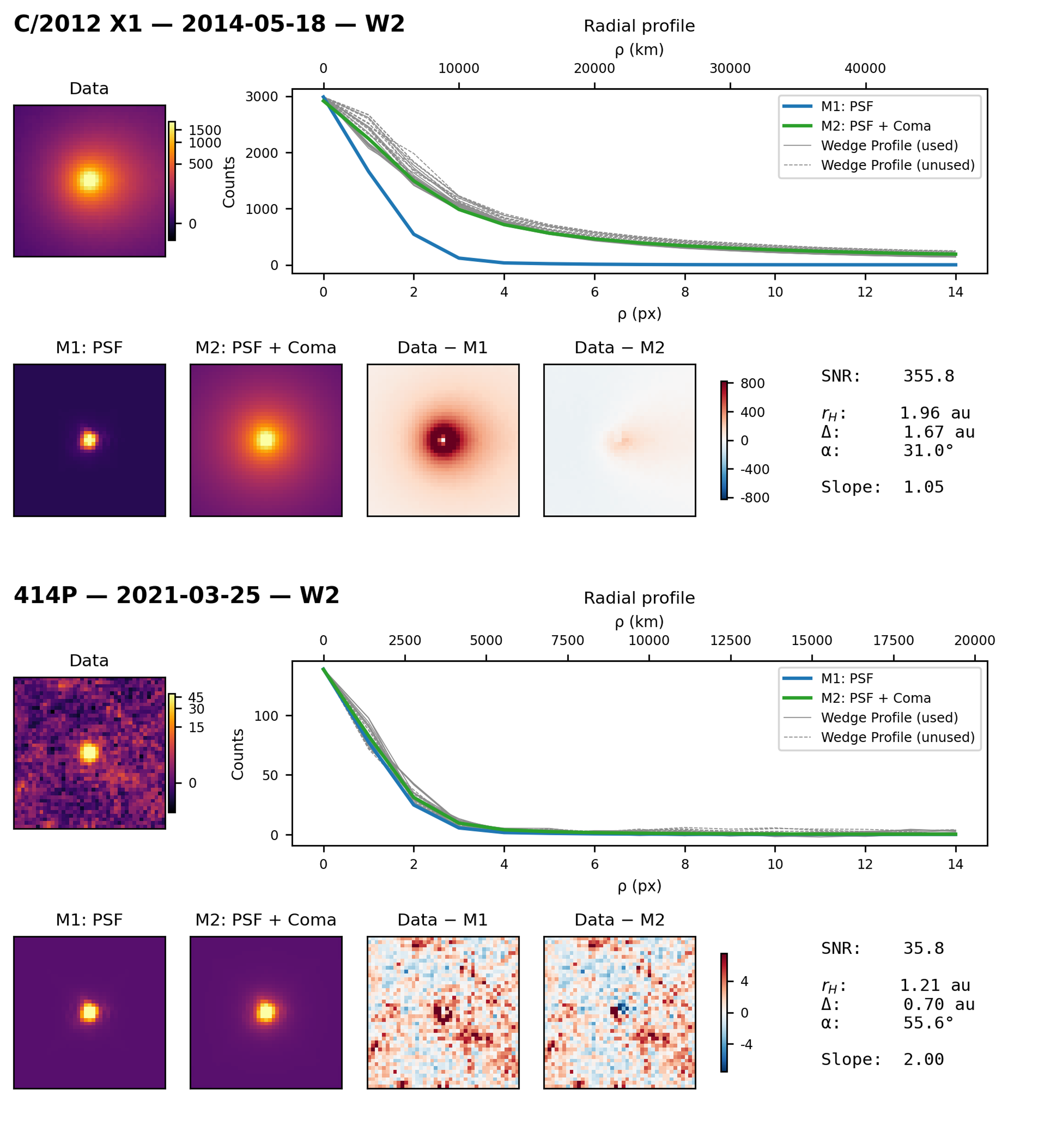}
\caption{Decomposition examples for high-SNR comets. \emph{Upper rows:} C/2012 X1 (LINEAR) (UT 2014 May 18, W2 band), a Model~2 (resolved-coma, non-degenerate) case whose profile is shallower than the PSF. From left to right, the color maps show the observed image (\texttt{Data}), the Model~1 (\texttt{PSF}) image, the Model~2 (\texttt{PSF $+$ Coma}) image, and the residual of each model after subtraction from the data; the radial-profile panel then compares the mean observed profile (green), the PSF (blue), and the per-wedge profiles (gray). Solid gray lines show the symmetric profiles used for the slope fit; dashed gray lines show the asymmetric profiles excluded from it, namely those contaminated by, for instance, dust tails (Appendix~\ref{sec:app1}). In the rightmost panel, $r_{\rm H}$, $\Delta$, and $\alpha$ denote the heliocentric distance, geocentric distance, and phase angle, respectively; \texttt{Slope} is the best-fit slope $s$ from the $\chi^2$ minimization. \emph{Lower rows:} 414P/STEREO (UT 2021 March 25, W2 band), a Model~1 (upper-limit, degenerate) case whose profile is indistinguishable from the PSF. The layout and symbols are the same as in the upper rows.\label{Fig02_1}}
\end{figure}
\begin{figure}[!thb]
\centering
\includegraphics[width=0.87\textwidth]{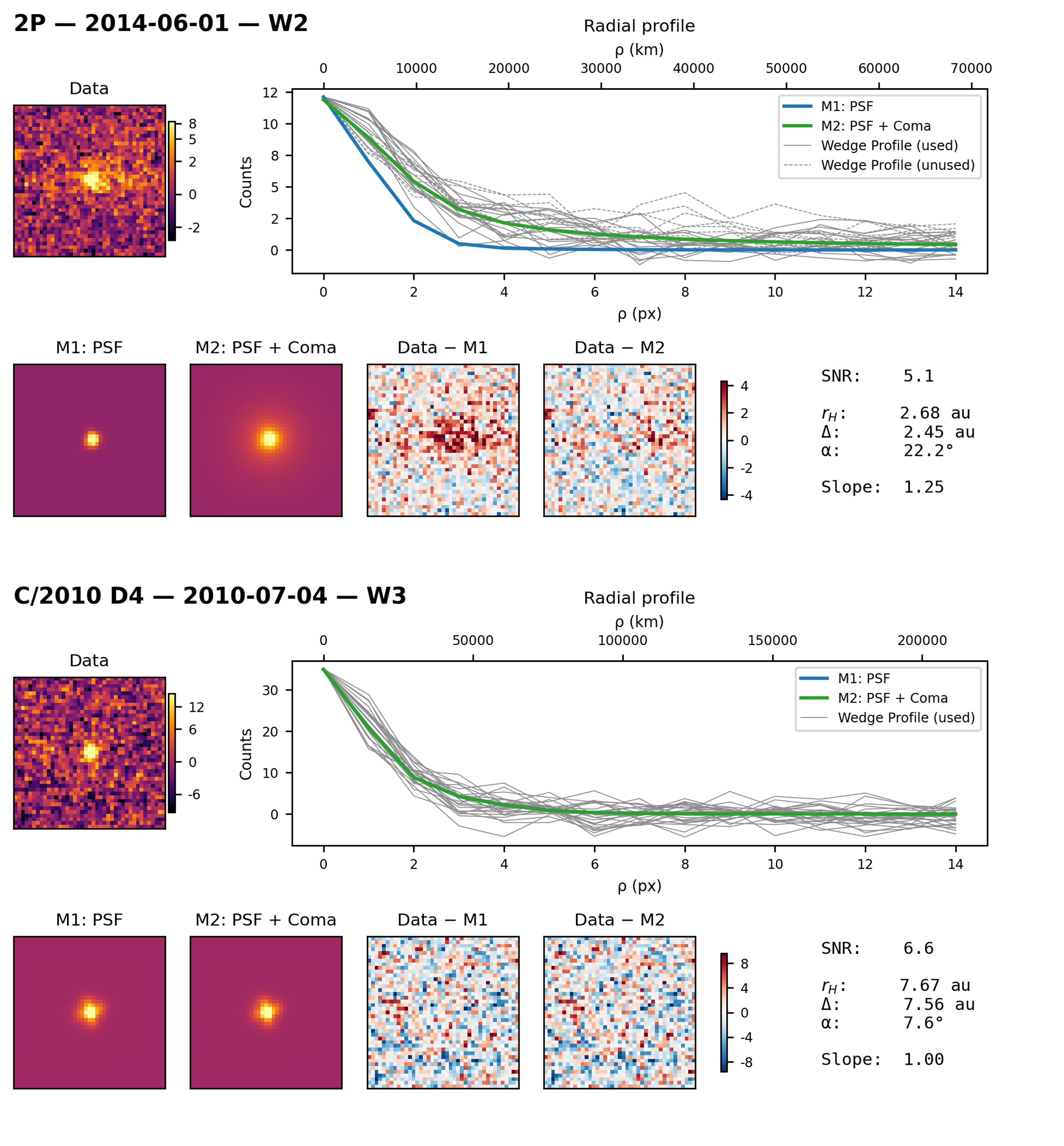}
\caption{Same as Figure~\ref{Fig02_1}, but for 2P/Encke and C/2010 D4 (WISE) at lower ($<10$) SNR.
\label{Fig02_2}}
\end{figure}

The lower panels of Figures~\ref{Fig02_1} and \ref{Fig02_2} show Model~1 (degenerate) cases at high and low signal-to-noise ratio (SNR), respectively. The observed mean-intensity profiles (green) are statistically indistinguishable from the PSF (blue), so the Model~1 (PSF-only) and Model~2 (PSF plus coma) images are nearly identical.

\subsection{Model 2: PSF + Coma Fitting Case \label{sec:model2}}

Model~2 treats the coma as resolvable, that is, when the radial profile is measurably shallower than the PSF (Fig.~\ref{Figap_01}); this is the case for sufficiently active comets. Retaining the coma term in Eq.~\ref{eq:phot_obs_signal} and imposing the peak constraint of Eq.~\ref{eq:peak} makes $A$ a function of $B$, reducing the decomposition to a single free amplitude $B$ at each slope $s$. The resulting $\chi^2$ is quadratic in $B$ and is minimized in closed form by weighted least squares on the radial profile, with $B$ restricted to the physical range $0 \le B \le B_{\rm max}(s)$ that keeps both amplitudes nonnegative (Appendix~\ref{sec:app1}).

The coma slope is not known or assumed a priori. For an optically thin coma in steady radial outflow, mass continuity predicts a column density, and hence a projected surface brightness, $\propto \rho^{-1}$ (i.e., $s \simeq 1$); solar radiation pressure on the dust steepens the profile toward $s \simeq 1.5$, and observed profiles span roughly $1 \le s \le 2$ \citep{Jewitt1987}. We therefore evaluated the closed-form fit on a fixed grid spanning and slightly exceeding this range, $s \in \{1.00, 1.05, \ldots, 2.50\}$ (31 values), and adopted the $(A, B, s)$ with the smallest $\chi^2(s)$. As the slope becomes steeper, the coma becomes increasingly PSF-like and therefore more degenerate with the nucleus signal. Each per-slope fit is analytic, so exhaustive grid evaluation is both robust and inexpensive.
 
A symmetric source (PSF plus symmetric coma) contributes equal mean intensity to every wedge, whereas an asymmetric feature such as a dust tail appears in only a subset of wedges. We identified and excluded these tail wedges before averaging the wedge profiles, so that the profile fed to the slope fit represents the symmetric (coma) component; the detailed procedure is given in Appendix~\ref{sec:app1}. Subtracting the Model~2 image from the data then leaves the residual asymmetric structure, predominantly dust tails and trails.

The upper panels of Figures~\ref{Fig02_1} and \ref{Fig02_2} show Model~2 (non-degenerate) cases at high and low SNR, respectively. The mean-intensity profiles (green) are statistically shallower than the PSF (smaller $s$), so the comets are spatially extended and the Model~2 image differs markedly from the Model~1 image. Table~\ref{tab:m1m2_io} in Appendix~\ref{sec:app1} summarizes the inputs and outputs of each step of the procedure.

Both models are fit independently to every source, each anchored to the same observed peak, so Model~1 and Model~2 yield two estimates of the nuclear amplitude at the same central pixel. Where the coma is resolved, part of the peak is assigned to the coma ($B > 0$), so the Model~2 amplitude $A_{\rm M2}$ falls below the Model~1 value and carries its own uncertainty, turning the Model~1 ceiling into a measurement. Where the coma is unresolved, the two model images are nearly identical, $B$ is essentially unconstrained, and the Model~2 nuclear amplitude for such comets spans from zero to the full peak; the Model~1 upper limit then remains the only meaningful statement about the nucleus. In both regimes the thermal modeling of Section~\ref{sec:phot_thermal} propagates the fitted amplitude with its uncertainty, so a poorly constrained decomposition naturally yields a weak constraint on the diameter.

\subsection{Nucleus Flux Uncertainty of Model 2 \label{sec:model2-unc}}

Fitting Model~2 to the data returns a slope and a best-fit amplitude $A_{\rm M2}$ along with uncertainties. These uncertainties were derived from an analytic covariance matrix based upon how the idealized model matches the actual data. We found that in general the uncertainties computed this way are optimistic, as the model is idealized and cannot fit the data perfectly. To combat this issue, we performed an additional processing step to numerically estimate the true uncertainty. We inject additional flux in the shape of the PSF at the location of the nucleus signal in the stacked image, then refit Model~2 to this altered image. If Model~2 successfully extracts the nucleus signal, the recovered amplitude should equal the original amplitude plus the injected flux. Methods similar to this have been used previously to characterize photometric error and completeness of astronomical survey pipelines \citep{stetson1987daophot,stetson1988ccd,Frohmaier2017,Huang2018}, and also to check reliability of extracting galactic centers from the parent galaxy \citep{Simmons2008}.

We performed a Monte-Carlo numerical draw 100 times per coadded image, with additional Gaussian noise with the same standard deviation as our background, while sweeping the amplitude of the injected flux (details can be found in Appendix~\ref{sec:fluxinj}). From this, we computed the uncertainty as a function of nucleus amplitude, which is typically higher than the idealized analytic fitted uncertainty. We used the maximum of these two uncertainties, $\sigma_{\rm eff} = \max(\sigma_A, \sigma_{\rm inj})$, along with the fitted nucleus flux in the diameter fitting step below.

\subsection{Methodological Differences from Previous Surveys  \label{sec:uniqueness}}

Our approach refines the standard radial-profile analysis of cometary nuclei \citep[e.g.,][]{Lamy2004, Fernandez2013, Bauer2015, Bauer2017} in four respects. First, the nucleus-coma decomposition is regularized against peak flux rather than fit freely: anchoring the model to the observed peak intensity reduces the problem to a single amplitude that is solved in closed form by weighted least squares over an exhaustive slope grid, and the PSF-coma degeneracy near $s \simeq 2$ is identified analytically. This removes the dependence on discrete fitting-window choices whose scatter previously entered the error budget \citep{Bauer2017}. Second, non-detections are treated statistically rather than annotated or discarded: a decomposition is demoted to an upper limit by explicit criteria, and such limits enter the thermal fit preserving the information they carry. Third, the systematic error of each extraction is measured rather than assumed, using per-image injection-recovery tests that quantify the scatter of the recovered nucleus flux under matched noise realizations, in order for better management of the limitations of conventional nucleus extraction techniques pointed out by \cite{Hui2018}. Fourth, diameters derive from a joint posterior over all bands and epochs of each comet, sampled by MCMC with priors on the albedo, beaming, and slope parameters, replacing two-point fits with discrete beaming values and yielding credible intervals with per-object convergence diagnostics. Further details for our methodology can be found in Appendices~\ref{sec:app1} and \ref{sec:fluxinj}. Together these elements convert the dominant systematics of nucleus photometry into quantified, reproducible terms of the error budget.

\subsection{Potential Impact of Gas Emission on Size Estimation \label{sec:gas_impact}}

Nuclei size estimates calculated from W1 and W2 measurements may be impacted by the presence of gas emission. Both bandpasses contain fluorescent emission bands of cometary volatiles: H$_2$O ($\sim$2.94~$\mu$m), HCN ($\sim$3.03~$\mu$m), CH$_4$ ($\sim$3.32~$\mu$m), C$_2$H$_6$ ($\nu_7$ at $\sim$3.35~$\mu$m and $\nu_5$ at $\sim$3.45~$\mu$m), and CH$_3$OH ($\sim$3.48~$\mu$m) in W1, and CO$_2$ ($\sim$4.3~$\mu$m) and CO ($\sim$4.7~$\mu$m) in W2 \citep{Dello2011,Reach2013,Paganini2015}. Gas emission can contribute up to several tens of percent of the in-band flux, most notably the CO$_2$ band in W2 \citep{Ootsubo2012}, with the relative contribution varying with heliocentric distance \citep{Womack2017,Harrington2022}. Broadband fluxes in W1 and W2 must therefore be interpreted as a possible blend of dust and gas components.

The impact of gas emission on our radial profile fitting differs between the two models. For Model~2 (PSF plus coma fitting), the effect is minor: a gas coma follows a shallow column density profile, with a logarithmic slope near $-1$ interior to the photo-dissociation scale length \citep{Haser1957,Combi2004}, and such an extended component is readily separated from the PSF (Fig.~\ref{Figap_01}). It is the dust coma, whose profile can steepen toward the PSF-like $s \simeq 2$ degeneracy, that dominates the separation problem. For Model~1, by contrast, any gas emission concentrated within the PSF disk would add to the central amplitude that this model attributes entirely to the nucleus, inflating the derived upper limit. We therefore caution that the Model~1 radii of gas-rich comets, particularly CO$_2$- or CO-rich objects observed in W2 (e.g., C/2017 K2; \citealt{Ejecta2025,Hmiddouch2025}), may be overestimated.

\subsection{Nucleus Colors \label{sec:phot_color}}

For each epoch in which two bands both yield an accepted Model~2 fit, we derived a color directly from the ratio of the nucleus amplitudes,
\begin{equation}
  W_a - W_b = (\mathrm{ZP}_a - \mathrm{ZP}_b)
              - 2.5 \log_{10}\!\left( A_a / A_b \right),
  \label{eq:color}
\end{equation}
where subscripts a and b denote each of WISE band, $A$ is the Model~2 nucleus amplitude in DN and $\mathrm{ZP}$ the zero point of the corresponding coadd. Since every band of an epoch is built from the same set of frames, the two measurements share one observing geometry and no differential correction is required. Because a symmetric flux uncertainty becomes strongly asymmetric in magnitude at low SNR, we propagated by Monte Carlo rather than linearly. We drew each amplitude per band from $\mathcal{N}(A, \sigma_{\rm eff})$, kept those within the physical bounds $0 < A \le A_{\rm M1}$ set by the peak constraint (Eq.~\ref{eq:peak}), and evaluate Eq.~\ref{eq:color} for each retained pair. We report the color of the nominal amplitudes with the 16th and 84th percentiles of this distribution.

\section{Result 1: Colors and Phase Curves \label{sec:phot_results}}

\begin{figure}[!thb]
\centering
\includegraphics[width=\textwidth]{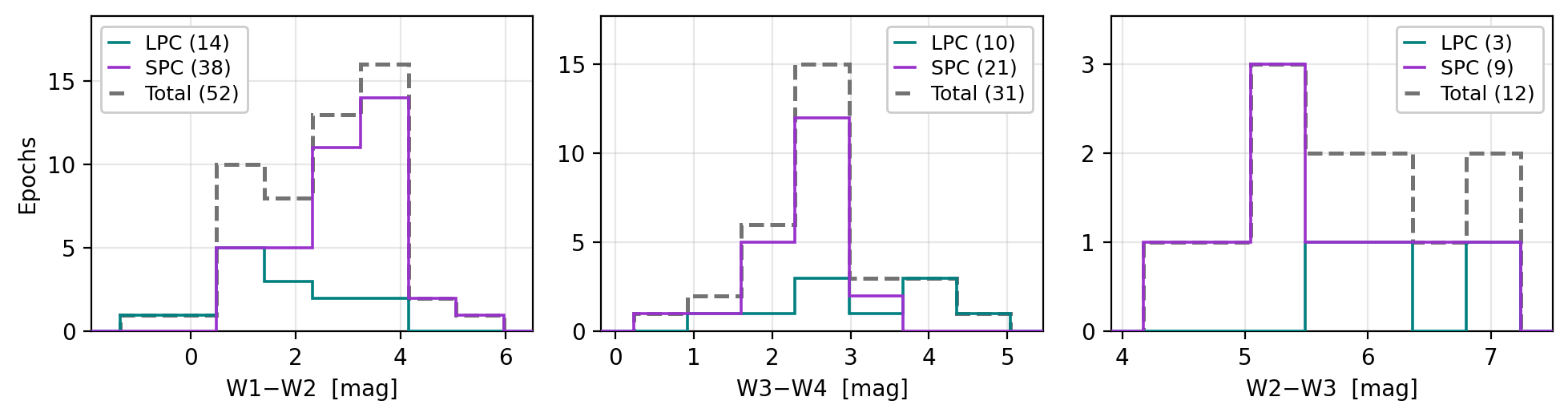}
\caption{Distributions of the epoch-matched WISE colors of the comet nuclei extracted with Model~2 (Section~\ref{sec:phot_color}). From left to right: $W1-W2$, $W3-W4$, and $W2-W3$. Each color is formed from the two nucleus amplitudes of a single epoch, so an object observed at several epochs contributes several entries, and the legend counts are numbers of epochs rather than of comets. Teal and purple histograms show LPCs and SPCs, and the dashed gray histogram their sum.}
\label{Fig_nuchist}
\end{figure}

Figure~\ref{Fig_nuchist} shows the epoch-matched $W1-W2$, $W3-W4$, and $W2-W3$ colors of the nuclei extracted with Model~2 (Section~\ref{sec:phot_color}). We do not cover other color combinations ($W1-W3$, $W2-W4$, $W1-W4$) here since there are fewer than 10 data points available. Each color is formed from the two nucleus amplitudes of a single epoch, so a comet observed at several epochs contributes several entries. The available statistics differ between panels: $W1-W2$ is available at 52 epochs across the full mission, whereas $W3-W4$ requires the four-band cryogenic phase and is limited to 31 epochs, and $W2-W3$ to 12.

All three colors are positive, indicating that the flux density rises toward longer wavelengths. To set a reference, we compute the color of a spectrally neutral reflector from the Vega-based definition of the WISE magnitudes, $m_X = -2.5 \log_{10}(F_X / F_X^{0})$, where $F_X^{0}$ is the zero-magnitude flux density of band $X$ \citep[309.540, 171.787, 31.674, and 8.363~Jy for W1 to W4;][]{Wright2010}. For a shorter band $a$ and a longer band $b$,
\begin{equation}
  m_a - m_b = -2.5 \log_{10}\!\left[
    \frac{F_a}{F_b}\,\frac{F_b^{0}}{F_a^{0}} \right] ,
  \label{eq:neutral_color}
\end{equation}
\noindent where $F_a/F_b$ for a neutral surface equals the ratio of the solar flux densities, evaluated as $B_\nu(5772~\mathrm{K})$ at the isophotal wavelengths 3.3526, 4.6028, 11.5608, and 22.0883~$\mu$m \citep{Wright2010}. The albedo cancels. This gives $W1-W2 = +0.07$, $W2-W3 = +0.02$, and $W3-W4 = +0.10$~mag, all within 0.1~mag of zero as expected for a Vega-based system.

The observed colors are far larger than this reference, with $W1-W2$ peaking near $+4$~mag, so calibration effects at the 0.1~mag level cannot account for them. One of the plausible origins is thermal emission: the observed flux ratios in all three colors correspond to blackbody color temperatures of roughly 200 to 300~K, the range expected for nucleus surfaces at 1 to $\sim$4~au \citep{Harris1998, Delbo2002}. Given the range of our observing geometry, thermal emission can dominate all bands longward of W1 for most of these nuclei, where the indices trace the thermal continuum, and hence the surface temperature at each epoch rather than surface composition. We will discuss this point in more detail in Section~\ref{sec:general_phot}. Within this limitation, the LPC and SPC distributions overlap in all three colors, with no evidence for a separation between the two dynamical classes, although the samples remain small.

The Model~2 nucleus fluxes also allow the dependence on phase angle (the Sun-comet-observer angle) to be examined. We found that the fluxes of LPCs and SPCs form a single diffuse cloud in all four bands that brightens toward small phase angles. The distribution is qualitatively identical to that of absolute magnitude in Figure~G1 of Paper I, including a sharp truncation at the faint end. We attribute the apparent trend to selection rather than to intrinsic surface property: at the fixed 90~degree solar elongation of the WISE survey, small phase angles correspond to large heliocentric distances (Fig.~F1 of Paper I), where the brightest nuclei preferentially remain above the detection limit. An ensemble of single-epoch snapshots of different comets is therefore shaped by observing geometry, and phase curves that constrain surface properties must instead come from monitoring individual nuclei over multiple epochs \citep{Muinonen2015}.

\section{Result 2: Nucleus Size Estimation \label{sec:size}}

The radial-profile decomposition of Section~\ref{sec:models} yields, for each WISE band $b\in\{W1,W2,W3,W4\}$ and each epoch, the amplitude of the unresolved nucleus after separation from the surrounding coma. Each amplitude provides a nucleus flux, which we propagate with its uncertainty into a thermal model to recover the nucleus diameter $D$.

The diameter and intrinsic surface properties of a nucleus do not change between epochs, so every band-epoch measurement of a given object constrains the same $D$. We therefore fit all band-epoch fluxes of an object jointly in a single inference rather than epoch by epoch. However, since our band-epochs sample different rotational phases, the recovered $D$ is an effective, rotationally averaged diameter. Previous comet surveys find that such an average recovers $\gtrsim$90\% of the true mean diameter, which is sufficient for the ensemble-level characterization of large comet samples pursued here (e.g., \citealt{Lamy2004,Snodgrass2011}).

For every object, we run two parallel analyses. In Model~1 (Section~\ref{sec:model1}), each band-epoch flux enters as a $3\sigma$ upper limit from the pure-PSF, nucleus-only fit. Because that model assigns the entire central excess to the nucleus, it overstates the nucleus flux and returns a conservative upper bound on $D$ that is valid regardless of the coma. In Model~2 (Section~\ref{sec:model2}), the nucleus flux from the PSF-plus-coma decomposition enters as a detection where the decomposition is reliable, giving a tighter estimate of $D$, or as an upper limit where the data do not require a nucleus. A detection is demoted to an upper limit when the two model images become nearly degenerate, either $A_{\rm M2} < \sigma_{\rm eff} $ so the nucleus is indistinguishable from zero, or $A_{\rm M2} > 0.95\, A_{\rm M1}$ so the nucleus is indistinguishable from a bare PSF. In either regime, the decomposition adds no usable information, and the conservative Model~1 limit is retained.

For each object we report the maximum a posteriori (MAP) diameter and the 16th and 84th percentile confidence interval of $D$, together with the number of band-epochs and distinct epochs entering the fit (detailed in Section~\ref{sec:size_result}). Fits that fail standard convergence diagnostics are flagged.

\subsection{Thermal Modeling \label{sec:phot_thermal}}

We model the band fluxes with the Near-Earth Asteroid Thermal Model (NEATM; \citealt{Harris1998}), augmented by a reflected-light component and by WISE color corrections applied facet by facet \citep{Wright2010}. The nucleus is treated as a spherical body whose surface facets are in instantaneous equilibrium with insolation. The beaming parameter $\eta$ rescales the subsolar temperature and absorbs, in a single empirical factor, the combined effects of surface roughness, thermal inertia, and viewing geometry \citep{Delbo2002}. Reflected sunlight contributes appreciably in the shorter-wavelength bands (W1 and W2) and is carried by the reflected-light component, while thermal emission dominates W3 and W4. The exact reflected-to-thermal flux ratio depends on heliocentric distance. The thermal emission depends on the Bond albedo $A_{\rm Bond}$, which we obtain from the geometric $V$-band albedo $p_V$ and the photometric slope parameter $G$ through the standard $H-G$ relation \citep{Bowell1989}. The fixing $p_V$ ties the reflected-light scaling to $D$, while the thermal channel constrains $D$ directly. 

Most objects are observed in only a few bands, often in a single band-epoch, so a fully free multi-parameter fit is unconstrained. Since the diameter predominantly sets the thermal flux while the remaining quantities act as second-order corrections, we constrain all parameters except $D$ with priors (Table~\ref{tab:neatm_priors}), each consistent with published values. We impose a Gaussian prior on $p_V$ centered at $0.04$ with a width small enough to hold it effectively fixed \citep{Fernandez2013}, together with $\eta\sim\mathcal{N}(1.03,0.11)$ \citep{Fernandez2013} and $G\sim\mathcal{N}(0.15,0.02)$. The reflected-light component uses an infrared albedo $p_{\rm IR}=1.6\,p_V$, consistent with the ratios found for WISE and NEOWISE asteroid samples (e.g., \citealt{Masiero2011}). We found that varying this ratio over the published range of roughly $1.5$ to $2.0$ shifts the inferred diameter by far less than the flux uncertainty, so the result is insensitive to its precise value. We hold the emissivity fixed at $\varepsilon=0.9$ \citep{Guilbert2011}.

\begin{deluxetable}{llcc}
\tablecaption{Treatment of the NEATM and reflected-light parameters in the joint diameter fit. All parameters except $D$ are constrained by Gaussian priors or held fixed.\label{tab:neatm_priors}}
\tablehead{
\colhead{Parameter} & \colhead{Symbol} &
\colhead{Prior / value} & \colhead{Reference}
}
\startdata
Diameter              & $D$              & free                          & \nodata \\
Geometric $V$ albedo  & $p_V$            & $0.04$\tablenotemark{*}       & \citet{Fernandez2013} \\
Slope parameter       & $G$              & $\mathcal{N}(0.15,\,0.02)$    & \citet{Bowell1989,Pravec2012} \\
Beaming parameter     & $\eta$           & $\mathcal{N}(1.03,\,0.11)$    & \citet{Fernandez2013} \\
IR-to-$V$ albedo ratio& $p_{\rm IR}/p_V$ & $1.6$                         & \citet{Masiero2011} \\
Emissivity            & $\varepsilon$    & $0.9$                         & \citet{Guilbert2011} \\
\enddata
\tablenotetext{*}{Imposed as a Gaussian prior centered on $0.04$ with negligible width, holding $p_V$ effectively fixed.}
\end{deluxetable}

We sample the joint posterior $\text{Prob}(D,\eta,p_V,H,G\mid\{F_b\})$ with Hamiltonian Monte Carlo, which efficiently handles the correlated parameters, using the NEATM implementation in the \texttt{kete} package \citep{Dahlen2025}. Each band-epoch contributes a flux $F_b$ with uncertainty $\sigma_{\rm b}$, and the per-band log-likelihood distinguishes detections from upper limits,
\begin{equation}
  \ln\mathcal{L}(F_b\mid\theta) \;=\;
  \begin{cases}
    -\dfrac{1}{2}\left(\dfrac{F_b - F_{\rm mod}(b)}{\sigma_{\rm b}}\right)^{2}, & \text{detection,}\\[3mm]
    \ln\Phi\!\left(\dfrac{F_b - F_{\rm mod}(b)}{\sigma_{\rm b}}\right), & \text{upper limit,}
  \end{cases}
  \label{eq:like}
\end{equation}
\noindent where $F_{\rm mod}(b)$ is the model flux in band $b$, parameter $\theta=(D,\eta,p_V,H,G)$, and $\Phi$ is the standard-normal cumulative distribution function. For an upper limit, a model flux at or below the limit contributes a nearly flat log-likelihood, while a model that exceeds the limit is penalized quadratically. Detections retain the Gaussian form. The resulting joint log-likelihood is the sum of Equation~(\ref{eq:like}) over all band-epochs of the object. Model~1 and Model~2 thus share a single likelihood and differ only in whether a given band-epoch enters as a detection or as an upper limit.

\subsection{Sizes of Comet Nuclei \label{sec:size_result}}

The joint fit of Section~\ref{sec:phot_thermal} returns, for each comet, 10,000 Markov Chain Monte Carlo (MCMC) draws of the parameter vector $(D,\eta,p_V,H,G)$. All available bands and epochs enter a single fit, so the derived diameter is \emph{effectively averaged over all observations of that object}. We adopt as the nominal diameter the maximum a posteriori (MAP) value, that is, the peak of the marginal posterior on $D$, which is equivalent to the best residual fit once the priors are included. We prefer the posterior peak to the median or the mean because, for objects with few thermal detections, $D$ and the beaming parameter $\eta$ are strongly correlated along a curved degeneracy. Projecting this locus onto $D$ yields a skewed marginal whose median and mean are pulled toward the extended tail, whereas the peak remains at the most probable diameter and reproduces the point estimate returned by classical minimum-$\chi^{2}$ thermal fits \citep{Fernandez2013}. For well-constrained objects with multiple non-degenerate thermal band-epochs, the marginal is nearly symmetric and the peak, median, and mean converge, so the choice of estimator matters only in the least-constrained regime. We characterize the uncertainty on $D$ by the 16th and 84th percentiles of the same marginal posterior, a $1\sigma$-equivalent 68\% credible interval (Section~\ref{sec:model2-unc}).

Table~\ref{tab:size_result} summarizes the diameter estimates and the census of band-epoch data entering the size fits for the 235 LPCs and 288 SPCs in our sample (Table~\ref{tab:db_summary}). For each comet, the table lists the number of epochs per WISE band, distinguishing epochs that yield only a Model~1 upper limit from those with a valid Model~2 nucleus extraction (values in parentheses). Nucleus fluxes adopted from Spitzer observations \citep{Fernandez2013} and included in the Model~2 fits are listed in a separate column. Because that sample consists entirely of JFCs, no LPC has an entry there. Our sample also contains most, but not all, of the WISE cryogenic-phase comets analyzed by \citet{Bauer2017}. The differences trace back to the target-selection criteria and the coadded-image analysis adopted here (Section~\ref{sec:uniqueness}).

\begin{figure}[hbt]
\centering
\includegraphics[width=0.87\textwidth]{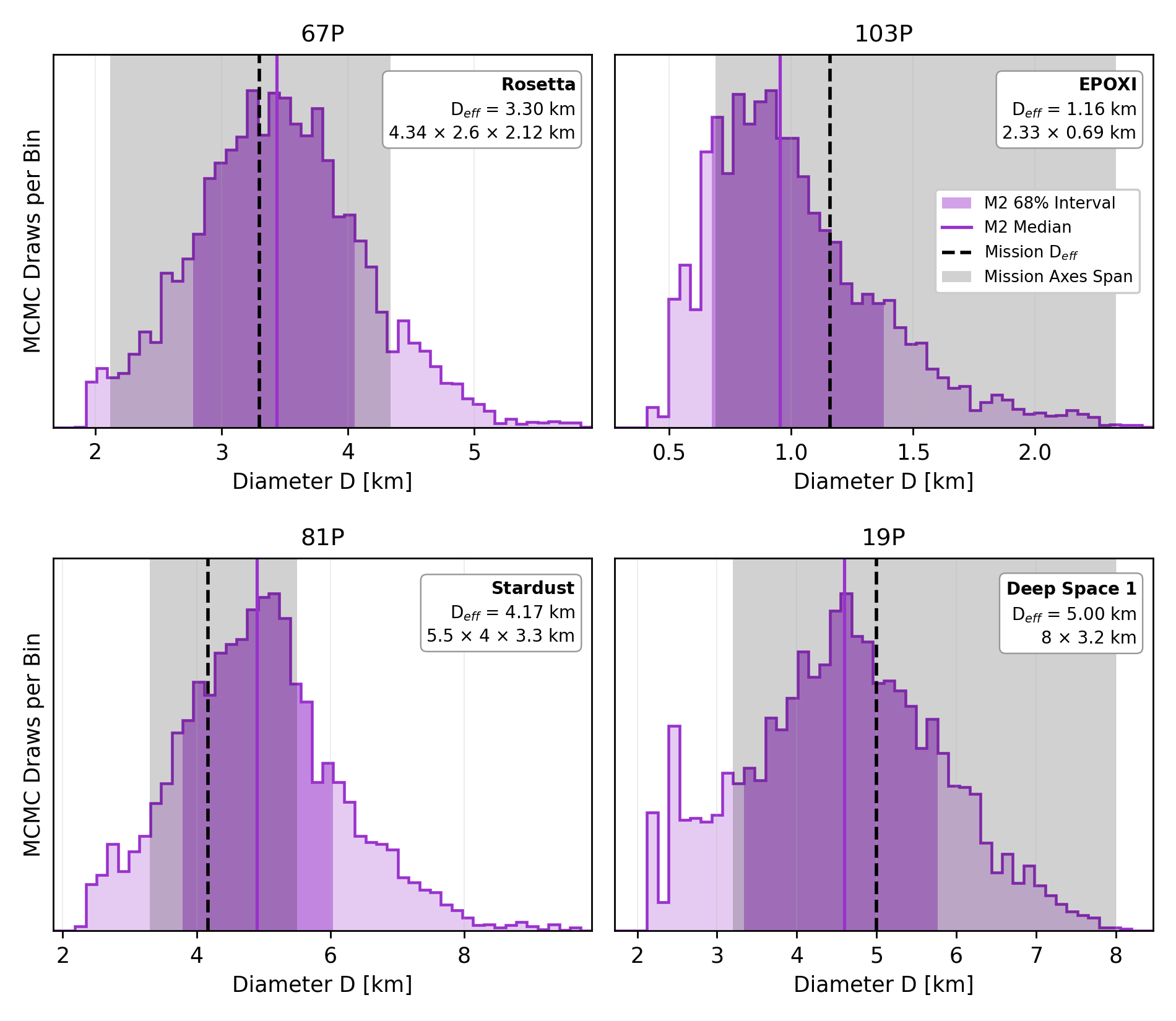}
\caption{Comparison of our Model~2 diameters with the nucleus dimensions measured in situ for four comets in our sample visited by spacecraft. In each panel, the purple histogram shows the marginal posterior on the diameter, sampled by MCMC, and the solid purple line marks its MAP (\texttt{Diameter} in Table~\ref{tab:size_result}), with the shaded band giving the 16th to 84th percentile interval. The dashed black line is the effective diameter $D_{\rm eff}$ reported by the mission, and the gray band spans the range of the individual axis lengths, which are quoted in each legend as three semi-axes or two projected dimensions depending on the source. References: 67P/Churyumov-Gerasimenko (Rosetta; \citealt{Jorda2016}), 103P/Hartley~2 (EPOXI; \citealt{Thomas2013}), 81P/Wild~2 (Stardust; \citealt{Duxbury2004}), and 19P/Borrelly (Deep Space~1; \citealt{Buratti2004, Kokotanekova2017}).
\label{Fig_comparison}}
\end{figure}

A Model~1 upper-limit diameter (95\% confidence level) is available for every comet, since Model~1 requires no coma decomposition and is always valid (as a ceiling). Model~2 point estimates with 1$\sigma$-equivalent uncertainties (Section~\ref{sec:model2-unc}) are available for 107 LPCs and 194 SPCs. Throughout this section, we quote diameters as $D_{\mathrm{unc\_lo}}^{\mathrm{unc\_hi}}$, where the subscript and superscript denote the 16th and 84th uncertainties (68\% CI), respectively. We caution that for comets observed only in W1 and/or W2, with no thermal-infrared data (W3, W4, or Spitzer), the posterior is weakly constrained and the resulting uncertainties are correspondingly large. Such estimates should be treated as indicative only. By contrast, for comets with non-degenerate thermal band-epochs, our effective diameters agree well with in situ spacecraft determinations (Fig.~\ref{Fig_comparison}). 

\begin{figure}[!thb]
\centering
\includegraphics[width=0.7\textwidth]{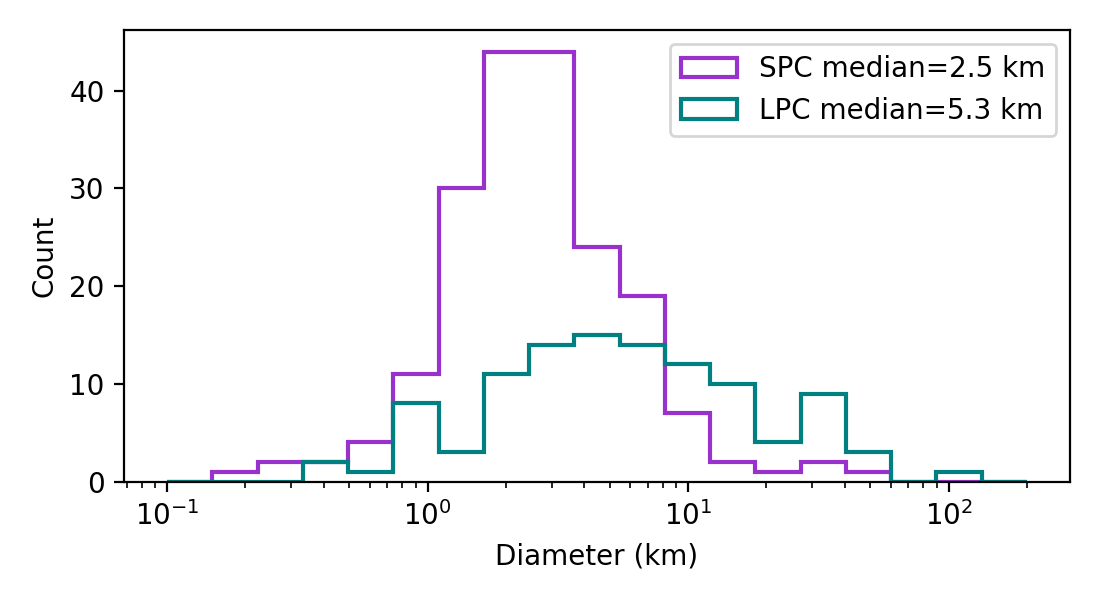}
\caption{Distribution of the nucleus diameters for the 301 comets (107 LPCs and 194 SPCs) with valid Model~2 estimates and uncertainties. The 222 comets (128 LPCs and 94 SPCs) with only Model~1 upper limits are not shown. Median values for each dynamical group are given in the legend.
\label{Fig03}}
\end{figure}

The Model~2 diameters of the SPCs range from $0.21_{-0.05}^{+0.06}$~km (320P/McNaught) to $54.74_{-8.85}^{+10.62}$~km (29P/Schwassmann-Wachmann~1)\footnote{Although its transitional, Centaur-like orbit sets it apart dynamically from JFCs, it satisfies our superset SPC criteria (and is listed as a JFC in JPL Horizons), and we therefore include it in statements about the SPC size distribution. Including or excluding this object does not change the results reported in this study.}, while the LPCs span $0.45_{-0.10}^{+0.11}$~km (C/2010 J4 (WISE)) to $76.00_{-24.35}^{+17.51}$~km (C/2014 B1 (Schwartz)). Figure~\ref{Fig03} shows the corresponding size distributions, built from the Model~2 point estimates only (upper limits are excluded). Both dynamical groups show broadly unimodal, Gaussian-like distributions, but the SPC distribution is narrower, clustering near $\sim$3~km. The median diameters are 2.5~km for the SPCs and 5.3~km for the LPCs. At face value, LPC nuclei are roughly twice as large as SPC nuclei, in qualitative agreement with previous studies \citep[e.g.,][]{Meech2004,Bauer2015,Bauer2017}. We defer the interpretation of this difference to Section~\ref{sec:discuss}.

\startlongtable
\begin{deluxetable*}{c|ccccc|c|c}
\tabletypesize{\scriptsize}
\tablewidth{0pt}
\tablecaption{Summary of nucleus size estimates for the comet sample. Model~2 \texttt{Diameter} values are reported as medians with 16th and 84th percentile uncertainties (68\% CI). Model~1 diameters are 95\% CI upper limits. The stacked images, the size estimates, and the journal of data used for the size estimation will all be shared via the NASA Planetary Data System after publication. \label{tab:size_result}}
\tablehead{
Comet &
\multicolumn{5}{c}{Number of Epochs\tablenotemark{a}} \vline&
\multirow{2}{*}{Diameter\tablenotemark{c}} &
\multirow{2}{*}{Upper Limit\tablenotemark{d}} \\
\cline{2-6}
Designation & \colhead{W1} & \colhead{W2} & \colhead{W3} & \colhead{W4} &
   \colhead{Spitzer\tablenotemark{b}} \vline& (km) & (km)
}
\startdata
\multicolumn{8}{c}{{\bf Long-Period Comets (LPCs)}} \\
\cline{1-8}
12P & 1(1) & 2 & \nodata & \nodata & \nodata & $17.00_{-5.02}^{+7.12}$ & 59.16  \\
13P & 1 & 1 & \nodata & \nodata & \nodata &  \nodata & 23.31  \\
38P & 2(1) & 2(2) & \nodata & \nodata & \nodata & $6.87_{-1.07}^{+1.08}$ & 13.28  \\
C/2005 L3 & \nodata & \nodata & 1 & 1 & \nodata &  \nodata & 61.04  \\
C/2005 W2 & \nodata & \nodata & \nodata & \nodata & 2(2) & $1.02_{-0.19}^{+0.18}$ & 1.48  \\
C/2006 OF2 & \nodata & \nodata & 1 & 1 & \nodata &  \nodata & 27.89  \\
C/2006 Q1 & \nodata & \nodata & 1 & 1 & \nodata &  \nodata & 30.02  \\
C/2006 S3 & 1 & 1 & 1 & 1 & \nodata &  \nodata & 60.58  \\
C/2006 W3 & 1(1) & 1(1) & 2(1) & 1 & \nodata & $41.31_{-7.31}^{+5.91}$ & 68.99  \\
C/2007 B2 & \nodata & \nodata & \nodata & 1 & \nodata &  \nodata & 29.83  \\
C/2007 D1 & \nodata & \nodata & \nodata & 1 & \nodata &  \nodata & 81.12  \\
C/2007 G1 & \nodata & \nodata & 1 & 1 & \nodata &  \nodata & 16.80  \\
C/2007 M1 & \nodata & \nodata & \nodata & 2 & \nodata &  \nodata & 41.12  \\
C/2007 M2 & \nodata & \nodata & 1 & 1 & \nodata &  \nodata & 21.34  \\
C/2007 N3 & \nodata & \nodata & \nodata & 1 & \nodata &  \nodata & 26.80  \\
C/2007 Q3 & 2 & 2(1) & 2 & 2 & \nodata & $10.14_{-2.42}^{+3.57}$ & 23.42  \\
C/2007 VO53 & \nodata & \nodata & 1 & 1(1) & \nodata & $8.07_{-2.65}^{+3.43}$ & 15.37  \\
C/2008 A1 & \nodata & \nodata & \nodata & 1 & \nodata &  \nodata & 22.38  \\
C/2008 E3 & \nodata & \nodata & \nodata & 2 & \nodata &  \nodata & 22.83  \\
C/2008 FK75 & 2 & 1 & 2 & 1 & \nodata &  \nodata & 22.74  \\
C/2008 N1 & \nodata & 1 & 2 & 2 & \nodata &  \nodata & 8.61  \\
C/2008 Q1 & \nodata & \nodata & 1 & 2(1) & \nodata & $8.30_{-2.81}^{+2.33}$ & 14.24  \\
C/2008 Q3 & \nodata & \nodata & 1 & 1 & \nodata &  \nodata & 11.19  \\
C/2008 S3 & \nodata & \nodata & \nodata & 1(1) & \nodata & $23.41_{-7.42}^{+6.89}$ & 52.18  \\
C/2008 T2 & \nodata & \nodata & \nodata & 1(1) & \nodata & $3.20_{-1.17}^{+1.37}$ & 11.59  \\
C/2009 F2 & \nodata & \nodata & 2 & 2(1) & \nodata & $12.01_{-4.61}^{+5.22}$ & 25.06  \\
C/2009 F4 & \nodata & \nodata & 2(1) & 1(1) & \nodata & $32.23_{-9.69}^{+7.74}$ & 53.01  \\
C/2009 F5 & \nodata & \nodata & \nodata & 1 & \nodata &  \nodata & 16.71  \\
C/2009 F6 & \nodata & \nodata & 1(1) & 1 & \nodata & $3.59_{-1.25}^{+1.12}$ & 8.80  \\
C/2009 G1 & \nodata & \nodata & 1 & 1 & \nodata &  \nodata & 7.21  \\
C/2009 K2 & \nodata & \nodata & 1 & 1 & \nodata &  \nodata & 6.97  \\
C/2009 K5 & 1 & 1 & \nodata & \nodata & \nodata &  \nodata & 27.76  \\
C/2009 P1 & 1 & 1 & 1 & 1 & \nodata &  \nodata & 53.88  \\
C/2009 T1 & \nodata & \nodata & 1(1) & 1(1) & \nodata & $11.66_{-2.95}^{+2.17}$ & 22.91  \\
C/2009 T3 & \nodata & \nodata & 1(1) & 1 & \nodata & $1.76_{-0.63}^{+0.82}$ & 4.55  \\
C/2009 U3 & 1 & 1(1) & 1(1) & 1 & \nodata & $1.76_{-0.32}^{+0.32}$ & 3.38  \\
C/2009 U5 & \nodata & \nodata & \nodata & 1(1) & \nodata & $11.40_{-6.80}^{+4.23}$ & 24.98  \\
C/2009 UG89 & \nodata & \nodata & 1(1) & 1 & \nodata & $5.62_{-2.20}^{+1.77}$ & 11.70  \\
C/2009 Y1 & \nodata & \nodata & 1(1) & 1(1) & \nodata & $6.59_{-1.49}^{+1.14}$ & 11.13  \\
C/2010 A4 & \nodata & \nodata & 1 & 1(1) & \nodata & $5.46_{-2.30}^{+1.93}$ & 8.96  \\
C/2010 B1 & 1 & 1 & 1(1) & 1(1) & \nodata & $8.67_{-2.36}^{+2.14}$ & 14.85  \\
C/2010 D3 & \nodata & \nodata & 2 & 2 & \nodata &  \nodata & 7.94  \\
C/2010 D4 & \nodata & \nodata & 2 & 2(1) & \nodata & $22.47_{-4.84}^{+3.79}$ & 36.36  \\
C/2010 DG56 & \nodata & 1 & 2(1) & 2(1) & \nodata & $0.99_{-0.39}^{+0.17}$ & 1.67  \\
C/2010 E1 & \nodata & \nodata & 1(1) & 1 & \nodata & $4.29_{-1.21}^{+0.86}$ & 4.85  \\
C/2010 F1 & \nodata & \nodata & 1 & 1 & \nodata &  \nodata & 9.15  \\
C/2010 FB87 & 1 & 1 & 1(1) & 1(1) & \nodata & $5.33_{-1.52}^{+0.82}$ & 8.19  \\
C/2010 G2 & \nodata & \nodata & 1(1) & 2(1) & \nodata & $6.74_{-2.48}^{+2.08}$ & 13.80  \\
C/2010 G3 & \nodata & \nodata & 2(1) & 2(1) & \nodata & $5.79_{-2.20}^{+1.93}$ & 12.29  \\
C/2010 H1 & \nodata & \nodata & \nodata & 1(1) & \nodata & $2.52_{-0.95}^{+0.82}$ & 5.33  \\
C/2010 J1 & \nodata & 1(1) & 2(1) & 2(1) & \nodata & $4.97_{-1.13}^{+0.78}$ & 6.38  \\
C/2010 J2 & \nodata & \nodata & 1 & 1(1) & \nodata & $7.27_{-2.16}^{+2.36}$ & 11.41  \\
C/2010 J4 & 1(1) & 1(1) & 2(1) & 2(1) & \nodata & $0.45_{-0.10}^{+0.11}$ & 0.71  \\
C/2010 KW7 & \nodata & \nodata & 2 & 1 & \nodata &  \nodata & 4.12  \\
C/2010 L4 & \nodata & \nodata & 1 & 1 & \nodata &  \nodata & 4.22  \\
C/2010 L5 & \nodata & 2(2) & 2 & 2 & \nodata & $1.20_{-0.16}^{+0.15}$ & 1.82  \\
C/2011 J2 & 3 & 3 & \nodata & \nodata & \nodata &  \nodata & 69.89  \\
C/2011 KP36 & 3(2) & 8 & \nodata & \nodata & \nodata & $39.66_{-9.48}^{+12.95}$ & 92.50  \\
C/2011 L4 & 1 & \nodata & \nodata & \nodata & \nodata &  \nodata & 91.03  \\
C/2012 F3 & 1 & 1 & \nodata & \nodata & \nodata &  \nodata & 62.72  \\
C/2012 K1 & 5(2) & 5(1) & \nodata & \nodata & \nodata & $5.95_{-1.73}^{+1.91}$ & 20.21  \\
C/2012 K6 & 1 & 1 & \nodata & \nodata & \nodata &  \nodata & 55.37  \\
C/2012 X1 & 2(1) & 2 & \nodata & \nodata & \nodata & $18.37_{-5.86}^{+8.14}$ & 45.59  \\
C/2013 A1 & 5 & 5(1) & \nodata & \nodata & \nodata & $4.75_{-1.47}^{+2.33}$ & 10.76  \\
C/2013 N4 & \nodata & 1(1) & \nodata & \nodata & \nodata & $5.28_{-1.88}^{+1.91}$ & 18.65  \\
C/2013 PE67 & \nodata & 1 & \nodata & \nodata & \nodata &  \nodata & 6.50  \\
C/2013 R1 & \nodata & 1 & \nodata & \nodata & \nodata &  \nodata & 56.68  \\
C/2013 UQ4 & 2(1) & 4 & \nodata & \nodata & \nodata & $6.17_{-2.03}^{+1.30}$ & 7.64  \\
C/2013 US10 & 7(1) & 7(3) & \nodata & \nodata & \nodata & $6.16_{-1.40}^{+1.84}$ & 25.52  \\
C/2013 V1 & 1 & 1 & \nodata & \nodata & \nodata &  \nodata & 9.39  \\
C/2013 V2 & 2(1) & 2(1) & \nodata & \nodata & \nodata & $17.66_{-6.00}^{+6.33}$ & 41.25  \\
C/2013 V4 & 2 & 2 & \nodata & \nodata & \nodata &  \nodata & 81.47  \\
C/2013 V5 & 1 & 1(1) & \nodata & \nodata & \nodata & $16.58_{-8.51}^{+6.15}$ & 39.02  \\
C/2013 X1 & 4(1) & 5 & \nodata & \nodata & \nodata & $7.88_{-2.46}^{+4.52}$ & 23.61  \\
C/2013 Y2 & 1 & 1 & \nodata & \nodata & \nodata &  \nodata & 9.31  \\
C/2014 A4 & 3(1) & \nodata & \nodata & \nodata & \nodata & $31.95_{-9.95}^{+9.64}$ & 62.46  \\
C/2014 B1 & \nodata & 2(1) & \nodata & \nodata & \nodata & $76.00_{-24.35}^{+17.51}$ & 95.76  \\
C/2014 C3 & 1 & 1 & \nodata & \nodata & \nodata &  \nodata & 5.08  \\
C/2014 E2 & 1 & 3(1) & \nodata & \nodata & \nodata & $8.95_{-3.02}^{+3.20}$ & 12.54  \\
C/2014 G1 & 1 & \nodata & \nodata & \nodata & \nodata &  \nodata & 95.76  \\
C/2014 N3 & 3 & 3(1) & \nodata & \nodata & \nodata & $31.76_{-9.54}^{+10.79}$ & 59.29  \\
C/2014 Q1 & 1 & 1 & \nodata & \nodata & \nodata &  \nodata & 53.28  \\
C/2014 Q2 & 3 & 2 & \nodata & \nodata & \nodata &  \nodata & 29.79  \\
C/2014 Q3 & 2 & 2(1) & \nodata & \nodata & \nodata & $1.91_{-0.62}^{+1.06}$ & 8.10  \\
C/2014 R1 & 1 & 1 & \nodata & \nodata & \nodata &  \nodata & 12.05  \\
C/2014 R4 & 1(1) & 1 & \nodata & \nodata & \nodata & $2.94_{-0.84}^{+1.07}$ & 5.07  \\
C/2014 S2 & 4(1) & 4(1) & \nodata & \nodata & \nodata & $18.20_{-5.93}^{+6.52}$ & 45.81  \\
C/2014 W2 & 2 & 2 & \nodata & \nodata & \nodata &  \nodata & 42.76  \\
C/2014 W5 & \nodata & 2 & \nodata & \nodata & \nodata &  \nodata & 11.93  \\
C/2014 W9 & 1(1) & 1 & \nodata & \nodata & \nodata & $2.51_{-0.79}^{+0.76}$ & 3.91  \\
C/2014 Y1 & \nodata & 1 & \nodata & \nodata & \nodata &  \nodata & 11.22  \\
C/2015 A1 & \nodata & 1(1) & \nodata & \nodata & \nodata & $1.95_{-0.61}^{+0.69}$ & 4.37  \\
C/2015 B2 & \nodata & 1(1) & \nodata & \nodata & \nodata & $16.81_{-5.60}^{+6.39}$ & 46.01  \\
C/2015 ER61 & 3 & 1 & \nodata & \nodata & \nodata &  \nodata & 11.60  \\
C/2015 F2 & 1 & 1 & \nodata & \nodata & \nodata &  \nodata & 2.60  \\
C/2015 F4 & 2(1) & 2(1) & \nodata & \nodata & \nodata & $1.96_{-0.56}^{+0.92}$ & 8.78  \\
C/2015 G2 & 1 & 1 & \nodata & \nodata & \nodata &  \nodata & 15.73  \\
C/2015 GX & 2 & 3(2) & \nodata & \nodata & \nodata & $4.90_{-1.30}^{+1.05}$ & 8.15  \\
C/2015 H1 & 1 & 2(1) & \nodata & \nodata & \nodata & $4.05_{-1.21}^{+1.17}$ & 6.17  \\
C/2015 K1 & 1 & 2 & \nodata & \nodata & \nodata &  \nodata & 21.97  \\
C/2015 O1 & 5 & 4 & \nodata & \nodata & \nodata &  \nodata & 62.74  \\
C/2015 T4 & 1 & 1(1) & \nodata & \nodata & \nodata & $7.56_{-2.67}^{+2.92}$ & 15.82  \\
C/2015 V1 & 3 & 3(2) & \nodata & \nodata & \nodata & $34.56_{-11.37}^{+9.88}$ & 59.85  \\
C/2015 V2 & 6 & 5 & \nodata & \nodata & \nodata &  \nodata & 17.85  \\
C/2015 V3 & \nodata & 1 & \nodata & \nodata & \nodata &  \nodata & 86.21  \\
C/2015 VL62 & 2 & 2 & \nodata & \nodata & \nodata &  \nodata & 30.12  \\
C/2015 W1 & \nodata & 2 & \nodata & \nodata & \nodata &  \nodata & 13.40  \\
C/2015 WZ & 1 & 2 & \nodata & \nodata & \nodata &  \nodata & 5.35  \\
C/2015 X8 & 1(1) & 1 & \nodata & \nodata & \nodata & $1.98_{-0.67}^{+0.80}$ & 3.49  \\
C/2015 Y1 & 1 & 1 & \nodata & \nodata & \nodata &  \nodata & 11.96  \\
C/2015 YG1 & 1 & 1 & \nodata & \nodata & \nodata &  \nodata & 9.36  \\
C/2016 A8 & 1 & 1(1) & \nodata & \nodata & \nodata & $5.17_{-1.56}^{+2.04}$ & 13.57  \\
C/2016 B1 & \nodata & 4 & \nodata & \nodata & \nodata &  \nodata & 87.86  \\
C/2016 C2 & \nodata & 1 & \nodata & \nodata & \nodata &  \nodata & 3.51  \\
C/2016 J2 & 1 & 2(2) & \nodata & \nodata & \nodata & $0.86_{-0.23}^{+0.27}$ & 2.57  \\
C/2016 K1 & \nodata & 1 & \nodata & \nodata & \nodata &  \nodata & 9.34  \\
C/2016 M1 & 6 & 6(1) & \nodata & \nodata & \nodata & $13.58_{-4.03}^{+6.29}$ & 54.86  \\
C/2016 N4 & 2 & 2 & \nodata & \nodata & \nodata &  \nodata & 35.33  \\
C/2016 N6 & 6 & 5(1) & \nodata & \nodata & \nodata & $40.34_{-14.24}^{+11.29}$ & 50.15  \\
C/2016 R2 & 2 & 6(1) & \nodata & \nodata & \nodata & $26.12_{-9.58}^{+8.40}$ & 40.84  \\
C/2016 T2 & \nodata & 3(1) & \nodata & \nodata & \nodata & $2.66_{-0.89}^{+1.14}$ & 5.04  \\
C/2016 T3 & \nodata & 1 & \nodata & \nodata & \nodata &  \nodata & 17.06  \\
C/2016 U1 & 1 & 2 & \nodata & \nodata & \nodata &  \nodata & 2.05  \\
C/2016 VZ18 & 1(1) & 2(1) & \nodata & \nodata & \nodata & $0.94_{-0.16}^{+0.15}$ & 1.19  \\
C/2017 B3 & 3 & 2 & \nodata & \nodata & \nodata &  \nodata & 70.33  \\
C/2017 C1 & \nodata & 1(1) & \nodata & \nodata & \nodata & $0.73_{-0.24}^{+0.30}$ & 1.84  \\
C/2017 D2 & 1 & 1 & \nodata & \nodata & \nodata &  \nodata & 13.82  \\
C/2017 E5 & \nodata & 1 & \nodata & \nodata & \nodata &  \nodata & 3.57  \\
C/2017 K2 & 8(1) & 8 & \nodata & \nodata & \nodata & $24.43_{-8.33}^{+12.55}$ & 71.88  \\
C/2017 K6 & 1 & 1 & \nodata & \nodata & \nodata &  \nodata & 13.96  \\
C/2017 M4 & 2 & 2 & \nodata & \nodata & \nodata &  \nodata & 57.44  \\
C/2017 O1 & 1 & 2(1) & \nodata & \nodata & \nodata & $2.60_{-0.90}^{+1.48}$ & 9.87  \\
C/2017 S6 & \nodata & 1 & \nodata & \nodata & \nodata &  \nodata & 3.50  \\
C/2017 T1 & 1 & 1 & \nodata & \nodata & \nodata &  \nodata & 6.44  \\
C/2017 T2 & 2(1) & 2 & \nodata & \nodata & \nodata & $15.05_{-4.23}^{+6.75}$ & 35.08  \\
C/2018 A3 & 1 & 1 & \nodata & \nodata & \nodata &  \nodata & 38.17  \\
C/2018 A6 & 3 & 2 & \nodata & \nodata & \nodata &  \nodata & 49.37  \\
C/2018 C2 & 1 & 1 & \nodata & \nodata & \nodata &  \nodata & 4.31  \\
C/2018 DO4 & \nodata & 2(1) & \nodata & \nodata & \nodata & $8.78_{-3.72}^{+2.72}$ & 17.05  \\
C/2018 EN4 & \nodata & 1 & \nodata & \nodata & \nodata &  \nodata & 3.42  \\
C/2018 F4 & 1 & 3(1) & \nodata & \nodata & \nodata & $39.71_{-12.54}^{+13.18}$ & 61.32  \\
C/2018 K1 & 1 & 2(1) & \nodata & \nodata & \nodata & $3.33_{-0.98}^{+1.16}$ & 6.62  \\
C/2018 M1 & \nodata & 1(1) & \nodata & \nodata & \nodata & $2.74_{-0.67}^{+0.65}$ & 3.66  \\
C/2018 N1 & 2(1) & 2(1) & \nodata & \nodata & \nodata & $0.54_{-0.16}^{+0.20}$ & 3.36  \\
C/2018 N2 & 6 & 5(1) & \nodata & \nodata & \nodata & $12.70_{-4.46}^{+6.08}$ & 63.83  \\
C/2018 W1 & \nodata & 1 & \nodata & \nodata & \nodata &  \nodata & 3.38  \\
C/2018 W2 & 2 & 2(1) & \nodata & \nodata & \nodata & $3.90_{-0.60}^{+0.65}$ & 8.35  \\
C/2018 Y1 & 2 & 2(2) & \nodata & \nodata & \nodata & $1.31_{-0.33}^{+0.47}$ & 6.52  \\
C/2019 B1 & \nodata & 2 & \nodata & \nodata & \nodata &  \nodata & 2.40  \\
C/2019 D1 & 1 & 1 & \nodata & \nodata & \nodata &  \nodata & 3.92  \\
C/2019 F1 & 2 & 2 & \nodata & \nodata & \nodata &  \nodata & 47.70  \\
C/2019 H1 & \nodata & 2 & \nodata & \nodata & \nodata &  \nodata & 4.65  \\
C/2019 J3 & 1 & 2(1) & \nodata & \nodata & \nodata & $2.97_{-0.82}^{+1.25}$ & 11.23  \\
C/2019 JU6 & \nodata & 1 & \nodata & \nodata & \nodata &  \nodata & 4.61  \\
C/2019 K4 & 1 & 1 & \nodata & \nodata & \nodata &  \nodata & 9.50  \\
C/2019 K5 & 1 & 1 & \nodata & \nodata & \nodata &  \nodata & 10.92  \\
C/2019 K7 & 2 & \nodata & \nodata & \nodata & \nodata &  \nodata & 76.57  \\
C/2019 L3 & 6(2) & 5(3) & \nodata & \nodata & \nodata & $35.93_{-5.38}^{+6.43}$ & 95.76  \\
C/2019 N1 & 2 & 1 & \nodata & \nodata & \nodata &  \nodata & 44.24  \\
C/2019 Q4 & \nodata & 1 & \nodata & \nodata & \nodata &  \nodata & 9.09  \\
C/2019 T2 & 1 & 1(1) & \nodata & \nodata & \nodata & $4.38_{-1.33}^{+2.02}$ & 12.37  \\
C/2019 T4 & 4 & 4 & \nodata & \nodata & \nodata &  \nodata & 97.27  \\
C/2019 U5 & 7 & 6 & \nodata & \nodata & \nodata &  \nodata & 77.80  \\
C/2019 U6 & \nodata & 1(1) & \nodata & \nodata & \nodata & $4.92_{-1.52}^{+2.02}$ & 13.48  \\
C/2019 Y1 & \nodata & 1 & \nodata & \nodata & \nodata &  \nodata & 5.73  \\
C/2019 Y4 & 1 & 1 & \nodata & \nodata & \nodata &  \nodata & 18.64  \\
C/2020 F3 & \nodata & 1(1) & \nodata & \nodata & \nodata & $6.42_{-1.31}^{+1.38}$ & 11.60  \\
C/2020 F5 & 3 & 3 & \nodata & \nodata & \nodata &  \nodata & 73.72  \\
C/2020 H6 & 2 & 2 & \nodata & \nodata & \nodata &  \nodata & 68.58  \\
C/2020 J1 & 5 & 5 & \nodata & \nodata & \nodata &  \nodata & 67.36  \\
C/2020 K1 & 3 & 3 & \nodata & \nodata & \nodata &  \nodata & 55.88  \\
C/2020 K3 & 2 & 2(1) & \nodata & \nodata & \nodata & $0.96_{-0.31}^{+0.45}$ & 3.49  \\
C/2020 M3 & \nodata & 1(1) & \nodata & \nodata & \nodata & $2.27_{-0.73}^{+1.14}$ & 8.28  \\
C/2020 M5 & 2 & 2(1) & \nodata & \nodata & \nodata & $25.20_{-7.73}^{+7.44}$ & 40.74  \\
C/2020 N1 & 1 & 1 & \nodata & \nodata & \nodata &  \nodata & 4.58  \\
C/2020 P1 & \nodata & 1 & \nodata & \nodata & \nodata &  \nodata & 3.28  \\
C/2020 PV6 & 3(1) & 3 & \nodata & \nodata & \nodata & $10.37_{-3.31}^{+2.74}$ & 15.54  \\
C/2020 Q1 & 1 & 1 & \nodata & \nodata & \nodata &  \nodata & 2.61  \\
C/2020 R4 & 1 & 2 & \nodata & \nodata & \nodata &  \nodata & 4.02  \\
C/2020 R7 & 3 & 1 & \nodata & \nodata & \nodata &  \nodata & 46.72  \\
C/2020 S2 & \nodata & 1 & \nodata & \nodata & \nodata &  \nodata & 2.34  \\
C/2020 S3 & 1 & 1(1) & \nodata & \nodata & \nodata & $17.69_{-5.76}^{+9.49}$ & 43.66  \\
C/2020 S4 & 1 & 2(1) & \nodata & \nodata & \nodata & $16.67_{-4.93}^{+6.86}$ & 30.23  \\
C/2020 S8 & 1 & 2 & \nodata & \nodata & \nodata &  \nodata & 8.81  \\
C/2020 T2 & 2 & 3 & \nodata & \nodata & \nodata &  \nodata & 26.05  \\
C/2020 V2 & 7 & 7(1) & \nodata & \nodata & \nodata & $36.20_{-13.85}^{+13.06}$ & 63.70  \\
C/2020 Y2 & 4 & 3(1) & \nodata & \nodata & \nodata & $16.30_{-5.97}^{+7.66}$ & 41.68  \\
C/2020 Y3 & \nodata & 1 & \nodata & \nodata & \nodata &  \nodata & 6.49  \\
C/2021 A1 & \nodata & 1 & \nodata & \nodata & \nodata &  \nodata & 58.82  \\
C/2021 A2 & 1 & 2(1) & \nodata & \nodata & \nodata & $5.00_{-0.90}^{+0.87}$ & 6.43  \\
C/2021 A4 & \nodata & 1 & \nodata & \nodata & \nodata &  \nodata & 2.33  \\
C/2021 A7 & 2 & 2 & \nodata & \nodata & \nodata &  \nodata & 13.45  \\
C/2021 A10 & \nodata & 1 & \nodata & \nodata & \nodata &  \nodata & 2.55  \\
C/2021 B3 & \nodata & 2(1) & \nodata & \nodata & \nodata & $2.08_{-0.73}^{+1.11}$ & 6.63  \\
C/2021 E3 & 4(2) & 5(2) & \nodata & \nodata & \nodata & $4.49_{-1.26}^{+1.89}$ & 17.63  \\
C/2021 F1 & 1 & 1 & \nodata & \nodata & \nodata &  \nodata & 12.71  \\
C/2021 G2 & 1 & 1 & \nodata & \nodata & \nodata &  \nodata & 95.37  \\
C/2021 O3 & \nodata & 1 & \nodata & \nodata & \nodata &  \nodata & 23.09  \\
C/2021 S3 & 3 & 2 & \nodata & \nodata & \nodata &  \nodata & 18.75  \\
C/2021 T4 & 2(1) & 2 & \nodata & \nodata & \nodata & $3.75_{-0.87}^{+1.15}$ & 10.54  \\
C/2021 U5 & \nodata & 1 & \nodata & \nodata & \nodata &  \nodata & 10.58  \\
C/2021 X1 & 4 & 4 & \nodata & \nodata & \nodata &  \nodata & 38.60  \\
C/2021 Y1 & 1 & 1 & \nodata & \nodata & \nodata &  \nodata & 17.89  \\
C/2022 A2 & 2(1) & 2(1) & \nodata & \nodata & \nodata & $10.24_{-2.91}^{+3.97}$ & 33.50  \\
C/2022 A3 & 1 & \nodata & \nodata & \nodata & \nodata &  \nodata & 48.06  \\
C/2022 E2 & 3 & 3 & \nodata & \nodata & \nodata &  \nodata & 88.56  \\
C/2022 E3 & 5(2) & 4(2) & \nodata & \nodata & \nodata & $3.73_{-0.94}^{+1.11}$ & 13.31  \\
C/2022 F2 & 1(1) & 1(1) & \nodata & \nodata & \nodata & $3.03_{-0.79}^{+0.88}$ & 4.78  \\
C/2022 J1 & 1(1) & 1 & \nodata & \nodata & \nodata & $3.37_{-1.05}^{+1.17}$ & 7.75  \\
C/2022 J2 & \nodata & 1 & \nodata & \nodata & \nodata &  \nodata & 4.43  \\
C/2022 JK5 & 2 & 2 & \nodata & \nodata & \nodata &  \nodata & 23.18  \\
C/2022 L1 & \nodata & 2 & \nodata & \nodata & \nodata &  \nodata & 4.35  \\
C/2022 L2 & 3 & 3(1) & \nodata & \nodata & \nodata & $13.67_{-5.75}^{+6.11}$ & 29.84  \\
C/2022 N1 & \nodata & 1 & \nodata & \nodata & \nodata &  \nodata & 2.17  \\
C/2022 P1 & 2 & 3(2) & \nodata & \nodata & \nodata & $3.05_{-0.66}^{+0.83}$ & 9.91  \\
C/2022 P3 & \nodata & 1 & \nodata & \nodata & \nodata &  \nodata & 15.04  \\
C/2022 S3 & \nodata & 1 & \nodata & \nodata & \nodata &  \nodata & 6.94  \\
C/2022 S4 & 1 & \nodata & \nodata & \nodata & \nodata &  \nodata & 39.33  \\
C/2022 U2 & 1 & 1 & \nodata & \nodata & \nodata &  \nodata & 3.90  \\
C/2022 V2 & 1(1) & 2 & \nodata & \nodata & \nodata & $8.58_{-3.05}^{+2.52}$ & 13.63  \\
C/2022 W3 & \nodata & 2 & \nodata & \nodata & \nodata &  \nodata & 3.06  \\
C/2023 A1 & 1 & 2 & \nodata & \nodata & \nodata &  \nodata & 5.24  \\
C/2023 A3 & 2 & 2 & \nodata & \nodata & \nodata &  \nodata & 41.32  \\
C/2023 B2 & 1 & 1 & \nodata & \nodata & \nodata &  \nodata & 5.64  \\
C/2023 C2 & \nodata & 1 & \nodata & \nodata & \nodata &  \nodata & 54.90  \\
C/2023 E1 & 3(1) & 3(3) & \nodata & \nodata & \nodata & $0.85_{-0.19}^{+0.21}$ & 2.72  \\
C/2023 F1 & \nodata & 1(1) & \nodata & \nodata & \nodata & $1.25_{-0.46}^{+0.35}$ & 2.94  \\
C/2023 K1 & \nodata & 1 & \nodata & \nodata & \nodata &  \nodata & 10.69  \\
C/2023 P1 & \nodata & 1(1) & \nodata & \nodata & \nodata & $6.03_{-1.97}^{+3.23}$ & 23.53  \\
C/2023 S3 & 1 & 1(1) & \nodata & \nodata & \nodata & $2.14_{-0.64}^{+0.56}$ & 3.47  \\
C/2023 V4 & 1 & \nodata & \nodata & \nodata & \nodata &  \nodata & 19.30  \\
C/2023 X1 & 1 & 1(1) & \nodata & \nodata & \nodata & $1.35_{-0.45}^{+0.56}$ & 2.76  \\
C/2024 J2 & \nodata & 1 & \nodata & \nodata & \nodata &  \nodata & 40.43  \\
P/2010 JC81 & \nodata & 1 & 1(1) & 1(1) & \nodata & $11.33_{-1.25}^{+1.19}$ & 14.65  \\
P/2016 WM48 & \nodata & 1(1) & \nodata & \nodata & \nodata & $2.14_{-0.74}^{+0.55}$ & 3.91  \\
\cline{1-8}
\multicolumn{8}{c}{{\bf Short-Period Comets (SPCs)}} \\
\cline{1-8}
2P & 2(2) & 6(3) & \nodata & \nodata & \nodata & $4.06_{-0.71}^{+0.81}$ & 5.33  \\
4P & 4 & 3 & \nodata & \nodata & \nodata &  \nodata & 11.51  \\
6P & 2(1) & 2(1) & \nodata & \nodata & 2(2) & $2.63_{-0.55}^{+0.63}$ & 3.28  \\
7P & 2(2) & 3(2) & 1 & 1(1) & 2(2) & $5.44_{-0.32}^{+0.34}$ & 6.68  \\
8P & \nodata & 1(1) & \nodata & \nodata & \nodata & $8.96_{-2.80}^{+4.50}$ & 25.87  \\
9P & 2 & 4(4) & 1(1) & 1(1) & \nodata & $2.83_{-0.68}^{+0.87}$ & 6.60  \\
10P & 4(3) & 8(6) & 1(1) & 1(1) & \nodata & $9.62_{-1.20}^{+1.03}$ & 10.20  \\
11P & 1 & 2(1) & \nodata & \nodata & 2(2) & $1.17_{-0.08}^{+0.07}$ & 1.81  \\
14P & \nodata & \nodata & 1(1) & 1(1) & 2(2) & $5.48_{-0.32}^{+0.28}$ & 6.57  \\
15P & 1 & 1 & \nodata & \nodata & 2(2) & $1.84_{-0.14}^{+0.12}$ & 3.79  \\
16P & \nodata & \nodata & \nodata & \nodata & 2(2) & $1.36_{-0.20}^{+0.18}$ & 1.64  \\
17P & 1 & 3(1) & \nodata & 1(1) & \nodata & $6.44_{-1.84}^{+1.39}$ & 15.07  \\
19P & 2 & 4(3) & 2(2) & 2(1) & \nodata & $4.59_{-1.30}^{+1.21}$ & 10.18  \\
21P & 3 & 3 & \nodata & \nodata & \nodata &  \nodata & 6.06  \\
22P & 2 & 4(2) & \nodata & \nodata & 2(2) & $4.32_{-0.41}^{+0.38}$ & 10.77  \\
24P & 1 & 1 & \nodata & \nodata & \nodata &  \nodata & 6.66  \\
26P & \nodata & 1(1) & \nodata & \nodata & \nodata & $3.26_{-1.40}^{+1.23}$ & 9.37  \\
28P & 1 & 2(2) & \nodata & \nodata & \nodata & $21.44_{-4.61}^{+3.94}$ & 29.04  \\
29P & 15(5) & 18(4) & 1 & 1 & \nodata & $54.74_{-8.85}^{+10.62}$ & 96.97  \\
30P & 1 & 1 & 1(1) & 1 & \nodata & $3.40_{-0.88}^{+0.87}$ & 6.70  \\
31P & \nodata & \nodata & 1(1) & 1 & 2(2) & $3.29_{-0.28}^{+0.29}$ & 8.62  \\
32P & 3 & 3(1) & \nodata & \nodata & 2(2) & $4.75_{-0.49}^{+0.49}$ & 11.33  \\
33P & \nodata & 2 & \nodata & \nodata & 2(2) & $2.36_{-0.17}^{+0.17}$ & 8.46  \\
37P & 2 & 2 & \nodata & \nodata & 2(2) & $2.47_{-0.25}^{+0.24}$ & 6.09  \\
42P & \nodata & 1 & \nodata & \nodata & \nodata &  \nodata & 5.99  \\
43P & 2 & 2(1) & \nodata & \nodata & 2 & $2.13_{-0.70}^{+0.75}$ & 9.48  \\
44P & \nodata & 2(1) & 1 & 1 & \nodata & $5.02_{-1.59}^{+1.55}$ & 8.59  \\
45P & \nodata & 1(1) & \nodata & \nodata & \nodata & $1.41_{-0.41}^{+0.66}$ & 3.53  \\
46P & 1 & 1 & \nodata & \nodata & \nodata &  \nodata & 4.37  \\
47P & 1 & 1 & 1(1) & 1 & 2(2) & $6.08_{-0.52}^{+0.50}$ & 7.68  \\
48P & 2(1) & 2 & 1(1) & 1(1) & 2(2) & $4.08_{-0.44}^{+0.31}$ & 7.46  \\
49P & \nodata & 2(1) & 1(1) & 1(1) & \nodata & $7.38_{-1.67}^{+1.12}$ & 12.20  \\
50P & 1 & 1 & \nodata & \nodata & 2(2) & $3.03_{-0.28}^{+0.25}$ & 5.21  \\
51P & 1 & 1 & \nodata & \nodata & 2(2) & $0.83_{-0.07}^{+0.07}$ & 4.07  \\
52P & 1 & 2 & \nodata & \nodata & \nodata &  \nodata & 5.01  \\
53P & 3 & 2 & \nodata & \nodata & \nodata &  \nodata & 16.92  \\
54P & \nodata & \nodata & \nodata & \nodata & 2 &  \nodata & 1.89  \\
56P & 1 & 1 & \nodata & \nodata & 2(2) & $3.85_{-0.44}^{+0.38}$ & 16.64  \\
57P & 1 & 2(1) & \nodata & \nodata & 2(2) & $1.91_{-0.15}^{+0.15}$ & 3.97  \\
58P & \nodata & 1 & \nodata & \nodata & \nodata &  \nodata & 6.61  \\
59P & \nodata & 1 & \nodata & \nodata & \nodata &  \nodata & 6.44  \\
60P & 1 & 1 & \nodata & \nodata & \nodata &  \nodata & 5.30  \\
61P & 2 & 3 & \nodata & \nodata & \nodata &  \nodata & 5.71  \\
62P & 2 & 2(1) & \nodata & \nodata & 2(2) & $1.19_{-0.16}^{+0.14}$ & 4.47  \\
64P & 1 & 1 & 1 & 1 & \nodata &  \nodata & 6.50  \\
65P & 5 & 7(1) & 1 & 1 & \nodata & $10.92_{-4.26}^{+3.56}$ & 16.09  \\
66P & 2 & 2(1) & \nodata & \nodata & \nodata & $1.38_{-0.49}^{+0.72}$ & 4.72  \\
67P & 4(1) & 3(2) & 2(1) & 2 & \nodata & $3.44_{-0.63}^{+0.65}$ & 9.02  \\
68P & 1 & 1 & 1 & 1 & 2(2) & $5.51_{-0.43}^{+0.45}$ & 8.23  \\
69P & \nodata & \nodata & \nodata & \nodata & 2(2) & $1.74_{-0.14}^{+0.14}$ & 1.87  \\
70P & 2 & 3(1) & \nodata & \nodata & \nodata & $3.23_{-1.08}^{+1.11}$ & 6.73  \\
71P & 2 & 3(1) & \nodata & \nodata & \nodata & $2.15_{-0.59}^{+1.10}$ & 7.47  \\
73P & 2 & 2(1) & \nodata & \nodata & \nodata & $1.65_{-0.46}^{+0.73}$ & 4.56  \\
74P & 4 & 1(1) & 2 & 2(1) & 2(2) & $6.68_{-1.11}^{+0.99}$ & 12.18  \\
76P & \nodata & 1 & \nodata & \nodata & \nodata &  \nodata & 2.11  \\
77P & 4(1) & 4(1) & 1 & 1 & 2(2) & $3.40_{-0.25}^{+0.24}$ & 6.60  \\
78P & 2 & 2(1) & \nodata & \nodata & 2(2) & $2.68_{-0.34}^{+0.34}$ & 12.81  \\
79P & \nodata & \nodata & \nodata & \nodata & 2(2) & $1.38_{-0.15}^{+0.14}$ & 1.56  \\
80P & \nodata & 1 & \nodata & \nodata & \nodata &  \nodata & 10.44  \\
81P & 3 & 5(2) & 1(1) & 1 & \nodata & $4.90_{-1.14}^{+1.20}$ & 13.19  \\
84P & 1 & 1 & \nodata & \nodata & \nodata &  \nodata & 6.73  \\
87P & \nodata & 1 & \nodata & \nodata & \nodata &  \nodata & 6.71  \\
88P & 3 & 5 & \nodata & \nodata & \nodata &  \nodata & 7.66  \\
89P & 1 & 1 & 1 & 1 & 2(2) & $2.81_{-0.23}^{+0.26}$ & 5.13  \\
93P & 1 & 1 & \nodata & \nodata & 2(2) & $4.94_{-0.27}^{+0.40}$ & 8.31  \\
94P & 2 & 3(1) & 1 & 1(1) & 2(2) & $4.57_{-0.28}^{+0.27}$ & 5.85  \\
96P & \nodata & 2(2) & \nodata & \nodata & \nodata & $3.22_{-0.53}^{+0.53}$ & 4.67  \\
97P & \nodata & 1(1) & \nodata & \nodata & \nodata & $4.37_{-2.49}^{+1.95}$ & 14.37  \\
100P & \nodata & 1 & 1(1) & 1 & \nodata & $1.80_{-0.61}^{+0.57}$ & 3.03  \\
101P & 1 & 1 & \nodata & \nodata & 2(2) & $1.97_{-0.36}^{+0.27}$ & 11.08  \\
103P & 1 & 4(2) & 1 & 1 & \nodata & $0.95_{-0.26}^{+0.41}$ & 2.94  \\
104P & 1 & 1 & \nodata & \nodata & \nodata &  \nodata & 1.48  \\
105P & 1 & 1 & \nodata & \nodata & \nodata &  \nodata & 6.23  \\
106P & 1 & 2(2) & \nodata & \nodata & \nodata & $2.06_{-0.59}^{+0.65}$ & 3.96  \\
107P & \nodata & \nodata & \nodata & \nodata & 2(2) & $2.96_{-0.22}^{+0.19}$ & 2.98  \\
108P & 2 & 4(2) & \nodata & \nodata & \nodata & $1.72_{-0.48}^{+0.70}$ & 6.58  \\
110P & 3 & 4(1) & \nodata & \nodata & \nodata & $6.58_{-2.49}^{+2.16}$ & 11.40  \\
112P & \nodata & 1(1) & \nodata & \nodata & \nodata & $1.76_{-0.54}^{+0.59}$ & 3.57  \\
113P & \nodata & \nodata & \nodata & \nodata & 2(2) & $3.42_{-0.23}^{+0.23}$ & 3.54  \\
114P & 1 & 1 & \nodata & \nodata & \nodata &  \nodata & 3.75  \\
115P & 1 & 1 & \nodata & \nodata & \nodata &  \nodata & 7.53  \\
116P & 4 & 4 & 1 & 1 & \nodata &  \nodata & 8.67  \\
117P & 8(1) & 5(1) & 1 & 1 & \nodata & $9.38_{-4.21}^{+3.21}$ & 14.40  \\
118P & 4 & 5(1) & 1 & 1 & 2(2) & $2.62_{-0.44}^{+0.48}$ & 6.47  \\
119P & 2 & 3(2) & \nodata & \nodata & 2(2) & $1.99_{-0.28}^{+0.30}$ & 13.46  \\
120P & \nodata & \nodata & \nodata & \nodata & 2 &  \nodata & 1.51  \\
121P & \nodata & \nodata & \nodata & \nodata & 2(2) & $7.88_{-0.60}^{+0.56}$ & 8.28  \\
123P & 3(1) & 3(1) & \nodata & \nodata & 2(2) & $4.48_{-0.39}^{+0.42}$ & 9.38  \\
124P & 2(1) & 3(3) & \nodata & \nodata & 2(2) & $5.09_{-0.59}^{+0.53}$ & 5.42  \\
125P & 1 & 2(1) & \nodata & \nodata & \nodata & $1.59_{-0.51}^{+0.55}$ & 3.53  \\
126P & 1 & 2 & \nodata & \nodata & \nodata &  \nodata & 8.36  \\
127P & \nodata & \nodata & 1 & 1 & 2(2) & $1.78_{-0.19}^{+0.19}$ & 3.61  \\
129P & \nodata & \nodata & \nodata & \nodata & 2(2) & $2.46_{-0.28}^{+0.27}$ & 2.76  \\
130P & 2 & 2 & 2 & 1 & 2(2) & $4.49_{-0.39}^{+0.36}$ & 5.05  \\
131P & \nodata & \nodata & \nodata & \nodata & 2(2) & $2.20_{-0.19}^{+0.21}$ & 2.35  \\
132P & 1 & 1 & \nodata & \nodata & 2(2) & $1.63_{-0.13}^{+0.12}$ & 4.46  \\
134P & 2 & 2 & \nodata & \nodata & \nodata &  \nodata & 17.15  \\
137P & 1 & 1(1) & 1 & 1 & 2(2) & $7.91_{-0.49}^{+0.49}$ & 8.14  \\
138P & \nodata & \nodata & \nodata & \nodata & 2(2) & $1.51_{-0.15}^{+0.15}$ & 1.49  \\
139P & \nodata & \nodata & \nodata & \nodata & 2(2) & $2.72_{-0.21}^{+0.21}$ & 2.80  \\
141P & \nodata & 1 & \nodata & \nodata & \nodata &  \nodata & 1.96  \\
142P & \nodata & \nodata & 1 & 1 & \nodata &  \nodata & 2.71  \\
143P & \nodata & 2(1) & 1(1) & 1(1) & 2(2) & $9.69_{-0.65}^{+0.62}$ & 11.22  \\
144P & 2(1) & 2 & \nodata & 1 & 2(2) & $1.68_{-0.10}^{+0.12}$ & 5.61  \\
145P & 1 & 2 & 1(1) & 1(1) & \nodata & $3.78_{-0.68}^{+0.48}$ & 6.39  \\
146P & \nodata & 1 & \nodata & \nodata & 2(2) & $1.91_{-0.19}^{+0.18}$ & 3.62  \\
148P & \nodata & \nodata & \nodata & \nodata & 2(2) & $2.30_{-0.12}^{+0.12}$ & 2.35  \\
149P & \nodata & \nodata & 2 & 1 & 2(2) & $2.80_{-0.21}^{+0.24}$ & 4.55  \\
150P & 1(1) & 3(3) & 1 & 1 & \nodata & $5.57_{-0.82}^{+0.94}$ & 7.05  \\
152P & \nodata & \nodata & \nodata & \nodata & 2(2) & $2.03_{-0.46}^{+0.33}$ & 2.87  \\
155P & 1 & 1 & \nodata & \nodata & \nodata &  \nodata & 3.02  \\
156P & 1 & 1 & \nodata & \nodata & \nodata &  \nodata & 11.67  \\
157P & \nodata & 1(1) & \nodata & \nodata & \nodata & $1.44_{-0.44}^{+0.43}$ & 3.44  \\
158P & \nodata & \nodata & \nodata & 1(1) & \nodata & $6.49_{-2.02}^{+1.47}$ & 15.20  \\
159P & \nodata & \nodata & \nodata & \nodata & 2(2) & $2.91_{-0.68}^{+0.58}$ & 3.76  \\
160P & \nodata & 1 & \nodata & \nodata & 2(2) & $2.02_{-0.20}^{+0.19}$ & 6.70  \\
162P & 2(2) & 4(3) & \nodata & \nodata & 2(2) & $15.34_{-1.06}^{+0.60}$ & 12.51  \\
163P & \nodata & \nodata & \nodata & \nodata & 2(2) & $2.67_{-0.18}^{+0.16}$ & 2.75  \\
164P & 1 & 1(1) & 1 & \nodata & \nodata & $3.89_{-1.19}^{+1.42}$ & 6.95  \\
168P & \nodata & \nodata & \nodata & \nodata & 2(2) & $0.96_{-0.11}^{+0.09}$ & 1.14  \\
169P & \nodata & 1 & 1(1) & 1(1) & 2(2) & $4.50_{-0.22}^{+0.24}$ & 4.69  \\
170P & \nodata & 2 & \nodata & \nodata & \nodata &  \nodata & 23.77  \\
171P & 1 & 1 & \nodata & \nodata & 2(2) & $2.48_{-0.23}^{+0.23}$ & 3.61  \\
172P & \nodata & \nodata & 2(1) & 2(1) & 2(2) & $8.93_{-0.74}^{+0.58}$ & 11.19  \\
173P & \nodata & \nodata & \nodata & \nodata & 2(2) & $8.37_{-1.93}^{+2.07}$ & 7.97  \\
178P & 1(1) & 1(1) & \nodata & \nodata & \nodata & $3.54_{-0.94}^{+1.09}$ & 6.67  \\
183P & \nodata & \nodata & \nodata & 1(1) & \nodata & $6.87_{-1.96}^{+1.16}$ & 12.65  \\
189P & \nodata & 2(2) & \nodata & \nodata & \nodata & $0.93_{-0.20}^{+0.17}$ & 1.35  \\
195P & \nodata & \nodata & 2 & 2 & \nodata &  \nodata & 14.55  \\
197P & \nodata & \nodata & \nodata & \nodata & 2(2) & $1.85_{-0.15}^{+0.15}$ & 1.99  \\
199P & \nodata & \nodata & 1 & 1 & \nodata &  \nodata & 10.37  \\
201P & 1 & 1 & \nodata & \nodata & \nodata &  \nodata & 3.45  \\
203P & 1 & \nodata & 1 & 1 & 2 &  \nodata & 12.14  \\
204P & 1 & 1 & \nodata & \nodata & \nodata &  \nodata & 3.91  \\
207P & 2(2) & 3(2) & \nodata & \nodata & \nodata & $0.73_{-0.18}^{+0.17}$ & 1.46  \\
209P & 3(3) & 3(3) & \nodata & \nodata & \nodata & $1.77_{-0.28}^{+0.29}$ & 2.01  \\
211P & \nodata & \nodata & 2 & 1 & \nodata &  \nodata & 3.59  \\
213P & \nodata & 1 & 1 & 2 & 2(2) & $2.97_{-0.57}^{+0.47}$ & 4.45  \\
215P & \nodata & \nodata & 1 & 1(1) & 2(2) & $2.58_{-0.22}^{+0.21}$ & 7.74  \\
216P & \nodata & \nodata & \nodata & \nodata & 2(2) & $1.18_{-0.16}^{+0.13}$ & 1.38  \\
217P & 1 & 1 & \nodata & \nodata & \nodata &  \nodata & 6.31  \\
219P & \nodata & \nodata & 1 & \nodata & 2(2) & $2.01_{-0.15}^{+0.16}$ & 8.74  \\
220P & 1 & 1 & \nodata & \nodata & \nodata &  \nodata & 2.63  \\
221P & \nodata & 1 & \nodata & \nodata & 2(2) & $2.07_{-0.11}^{+0.11}$ & 3.94  \\
222P & \nodata & \nodata & 1(1) & \nodata & \nodata & $1.22_{-0.53}^{+0.40}$ & 3.01  \\
223P & \nodata & 2(1) & 1(1) & \nodata & 2(2) & $5.93_{-0.34}^{+0.31}$ & 5.76  \\
225P & \nodata & 2(2) & 1 & \nodata & \nodata & $1.60_{-0.26}^{+0.26}$ & 2.28  \\
226P & 2 & 2(1) & \nodata & \nodata & \nodata & $1.84_{-0.56}^{+0.93}$ & 4.80  \\
227P & 1 & 1(1) & 1(1) & 1 & \nodata & $2.17_{-0.55}^{+0.34}$ & 3.59  \\
228P & \nodata & \nodata & \nodata & \nodata & 2(2) & $2.45_{-0.36}^{+0.32}$ & 2.57  \\
229P & \nodata & \nodata & 1(1) & 1 & \nodata & $2.03_{-0.64}^{+0.81}$ & 4.26  \\
230P & 1 & 3(3) & 1(1) & 1(1) & \nodata & $4.60_{-0.38}^{+0.33}$ & 5.95  \\
232P & \nodata & \nodata & 1 & 1 & \nodata &  \nodata & 5.41  \\
233P & \nodata & \nodata & 1(1) & 1 & \nodata & $1.32_{-0.24}^{+0.15}$ & 1.71  \\
234P & \nodata & \nodata & 2(2) & 2 & \nodata & $2.40_{-0.69}^{+0.49}$ & 4.11  \\
235P & \nodata & \nodata & 1(1) & 1 & \nodata & $2.11_{-0.85}^{+0.61}$ & 4.03  \\
236P & \nodata & 1 & 1(1) & 1(1) & \nodata & $1.73_{-0.31}^{+0.24}$ & 2.38  \\
237P & 2 & 3(1) & 1(1) & 1(1) & \nodata & $2.59_{-0.95}^{+0.59}$ & 4.39  \\
239P & \nodata & 1(1) & 1 & 1 & \nodata & $2.07_{-0.52}^{+0.45}$ & 3.22  \\
240P & 4 & 6(1) & 1 & \nodata & \nodata & $5.39_{-1.49}^{+1.80}$ & 10.12  \\
241P & \nodata & 2 & \nodata & \nodata & \nodata &  \nodata & 6.12  \\
243P & 1 & 1(1) & 1 & \nodata & 2 & $1.06_{-0.33}^{+0.39}$ & 5.09  \\
244P & \nodata & \nodata & \nodata & \nodata & 2 &  \nodata & 3.60  \\
245P & \nodata & \nodata & 1 & 1(1) & \nodata & $1.45_{-0.47}^{+0.69}$ & 2.86  \\
246P & 8(1) & 5 & \nodata & \nodata & 2(2) & $8.48_{-1.15}^{+1.09}$ & 30.61  \\
247P & \nodata & 1 & \nodata & \nodata & \nodata &  \nodata & 3.19  \\
248P & \nodata & \nodata & 1 & \nodata & \nodata &  \nodata & 4.49  \\
249P & 1 & 1(1) & \nodata & \nodata & \nodata & $1.30_{-0.29}^{+0.22}$ & 1.70  \\
251P & \nodata & 1 & \nodata & \nodata & \nodata &  \nodata & 1.94  \\
252P & 2(2) & 2(2) & \nodata & \nodata & \nodata & $0.29_{-0.06}^{+0.06}$ & 0.65  \\
254P & 1 & 1 & 1(1) & \nodata & \nodata & $2.06_{-0.93}^{+0.61}$ & 5.15  \\
256P & \nodata & \nodata & \nodata & \nodata & 2(2) & $1.47_{-0.23}^{+0.23}$ & 1.89  \\
257P & 2 & 1(1) & \nodata & \nodata & \nodata & $2.61_{-0.97}^{+0.77}$ & 5.02  \\
260P & 2 & 2 & \nodata & \nodata & 2(2) & $3.11_{-0.16}^{+0.16}$ & 4.67  \\
261P & 1 & 1 & \nodata & \nodata & \nodata &  \nodata & 6.69  \\
263P & \nodata & 1 & \nodata & \nodata & \nodata &  \nodata & 2.51  \\
266P & 1 & 1 & \nodata & \nodata & \nodata &  \nodata & 9.29  \\
269P & 1 & \nodata & \nodata & \nodata & \nodata &  \nodata & 58.57  \\
277P & 1 & 1(1) & \nodata & \nodata & \nodata & $1.85_{-0.62}^{+0.82}$ & 4.76  \\
278P & \nodata & 1 & \nodata & \nodata & \nodata &  \nodata & 6.31  \\
280P & \nodata & \nodata & \nodata & \nodata & 2(2) & $2.42_{-0.30}^{+0.29}$ & 2.93  \\
284P & 3 & 3(2) & \nodata & \nodata & \nodata & $4.66_{-1.58}^{+1.62}$ & 9.21  \\
286P & \nodata & \nodata & \nodata & \nodata & 2(2) & $3.73_{-0.29}^{+0.28}$ & 3.87  \\
289P & 1 & 1(1) & \nodata & \nodata & \nodata & $0.34_{-0.10}^{+0.10}$ & 0.60  \\
290P & 1 & 1 & \nodata & \nodata & \nodata &  \nodata & 14.44  \\
291P & \nodata & \nodata & \nodata & \nodata & 2 &  \nodata & 3.08  \\
292P & 1 & 1 & \nodata & \nodata & \nodata &  \nodata & 12.34  \\
296P & \nodata & 2(2) & \nodata & \nodata & \nodata & $2.21_{-0.51}^{+0.50}$ & 3.24  \\
299P & 1 & \nodata & \nodata & \nodata & \nodata &  \nodata & 34.38  \\
300P & \nodata & 1 & \nodata & \nodata & 2(2) & $1.36_{-0.14}^{+0.14}$ & 4.39  \\
304P & 1 & 2(1) & \nodata & \nodata & \nodata & $1.07_{-0.32}^{+0.25}$ & 1.70  \\
305P & 1 & 1 & \nodata & \nodata & \nodata &  \nodata & 2.02  \\
306P & 1 & 1(1) & \nodata & \nodata & 2(2) & $1.12_{-0.15}^{+0.10}$ & 1.26  \\
307P & \nodata & 1 & \nodata & \nodata & \nodata &  \nodata & 4.51  \\
309P & \nodata & 1(1) & \nodata & \nodata & 2(2) & $3.05_{-0.13}^{+0.12}$ & 4.18  \\
312P & \nodata & 1 & \nodata & \nodata & \nodata &  \nodata & 3.07  \\
315P & 2 & 2(1) & \nodata & \nodata & 2(2) & $11.10_{-0.53}^{+0.47}$ & 14.96  \\
317P & 1 & 2 & 1(1) & 1(1) & \nodata & $1.40_{-0.11}^{+0.09}$ & 1.43  \\
318P & 2 & 1 & \nodata & \nodata & \nodata &  \nodata & 10.72  \\
319P & 3(2) & 4(4) & \nodata & \nodata & \nodata & $1.98_{-0.33}^{+0.33}$ & 2.45  \\
320P & \nodata & 2(1) & \nodata & \nodata & \nodata & $0.21_{-0.05}^{+0.06}$ & 0.43  \\
324P & \nodata & \nodata & 1 & \nodata & \nodata &  \nodata & 2.77  \\
325P & \nodata & 1 & \nodata & \nodata & \nodata &  \nodata & 1.41  \\
327P & 1(1) & 2(1) & \nodata & 1 & \nodata & $1.65_{-0.48}^{+0.44}$ & 2.72  \\
328P & \nodata & 1(1) & \nodata & \nodata & \nodata & $1.59_{-0.65}^{+0.64}$ & 4.08  \\
329P & \nodata & 1 & \nodata & \nodata & \nodata &  \nodata & 4.19  \\
333P & 1(1) & 3(3) & \nodata & \nodata & \nodata & $5.20_{-0.74}^{+0.73}$ & 6.46  \\
337P & \nodata & 1(1) & 1(1) & 1(1) & \nodata & $1.39_{-0.17}^{+0.15}$ & 1.65  \\
338P & \nodata & 1 & \nodata & 1 & \nodata &  \nodata & 6.19  \\
343P & \nodata & 1(1) & \nodata & \nodata & \nodata & $3.49_{-1.02}^{+1.07}$ & 5.82  \\
344P & \nodata & \nodata & \nodata & \nodata & 2(2) & $2.91_{-0.27}^{+0.23}$ & 3.05  \\
346P & \nodata & 1(1) & \nodata & \nodata & \nodata & $3.29_{-1.36}^{+1.12}$ & 7.36  \\
349P & \nodata & \nodata & 1 & 1 & \nodata &  \nodata & 3.52  \\
353P & \nodata & \nodata & 1 & 1(1) & \nodata & $2.67_{-0.78}^{+0.57}$ & 4.19  \\
354P & \nodata & \nodata & 1 & 1 & \nodata &  \nodata & 1.75  \\
355P & \nodata & \nodata & \nodata & \nodata & 2 &  \nodata & 2.29  \\
356P & \nodata & \nodata & 1 & 1 & \nodata &  \nodata & 4.10  \\
360P & \nodata & \nodata & 1 & \nodata & \nodata &  \nodata & 1.44  \\
361P & \nodata & 1 & \nodata & \nodata & \nodata &  \nodata & 21.06  \\
364P & 3 & 4(2) & \nodata & \nodata & \nodata & $1.37_{-0.28}^{+0.26}$ & 1.99  \\
368P & \nodata & 1 & \nodata & \nodata & 2(2) & $3.31_{-0.23}^{+0.22}$ & 6.92  \\
369P & \nodata & 1(1) & 1 & 1 & \nodata & $2.19_{-0.69}^{+0.71}$ & 3.50  \\
371P & \nodata & \nodata & \nodata & \nodata & 2(2) & $1.46_{-0.20}^{+0.18}$ & 1.73  \\
376P & \nodata & \nodata & \nodata & \nodata & 2(2) & $5.53_{-0.46}^{+0.44}$ & 5.64  \\
378P & \nodata & \nodata & \nodata & \nodata & 2(2) & $10.20_{-0.88}^{+0.92}$ & 11.59  \\
382P & \nodata & 4(1) & \nodata & \nodata & \nodata & $37.15_{-11.90}^{+15.68}$ & 84.61  \\
387P & \nodata & 1 & \nodata & \nodata & \nodata &  \nodata & 2.88  \\
389P & \nodata & 1 & \nodata & \nodata & \nodata &  \nodata & 2.63  \\
395P & 1 & \nodata & \nodata & \nodata & 2(2) & $5.65_{-0.59}^{+0.49}$ & 55.52  \\
398P & 1 & 1(1) & 1 & 1 & \nodata & $1.15_{-0.34}^{+0.38}$ & 2.17  \\
400P & \nodata & 1 & \nodata & \nodata & \nodata &  \nodata & 6.19  \\
405P & 2 & 3(3) & \nodata & \nodata & \nodata & $0.63_{-0.18}^{+0.24}$ & 1.31  \\
407P & \nodata & 1(1) & \nodata & \nodata & \nodata & $1.92_{-0.77}^{+0.89}$ & 5.78  \\
408P & 1 & \nodata & \nodata & \nodata & \nodata &  \nodata & 38.55  \\
409P & 1 & 2(2) & \nodata & \nodata & 2(2) & $5.52_{-0.36}^{+0.35}$ & 6.54  \\
412P & \nodata & \nodata & 2(1) & 1(1) & \nodata & $0.68_{-0.16}^{+0.25}$ & 1.54  \\
413P & 1 & 3 & \nodata & \nodata & \nodata &  \nodata & 5.81  \\
414P & \nodata & 1(1) & \nodata & \nodata & \nodata & $1.02_{-0.24}^{+0.22}$ & 1.75  \\
417P & \nodata & 1(1) & \nodata & \nodata & \nodata & $1.34_{-0.42}^{+0.34}$ & 2.47  \\
418P & 1(1) & 1(1) & 2(1) & 1(1) & \nodata & $3.38_{-0.79}^{+0.70}$ & 5.25  \\
424P & 1 & 1(1) & \nodata & \nodata & \nodata & $2.99_{-0.50}^{+0.42}$ & 3.45  \\
425P & \nodata & \nodata & \nodata & \nodata & 2(2) & $2.30_{-0.48}^{+0.46}$ & 3.62  \\
430P & \nodata & 1 & \nodata & \nodata & \nodata &  \nodata & 1.75  \\
431P & \nodata & 1(1) & \nodata & \nodata & \nodata & $1.70_{-0.68}^{+0.60}$ & 3.32  \\
440P & 1(1) & 1(1) & \nodata & \nodata & \nodata & $1.82_{-0.51}^{+0.53}$ & 3.97  \\
444P & 1(1) & 3(2) & 1(1) & 1(1) & \nodata & $1.57_{-0.49}^{+0.26}$ & 2.19  \\
446P & \nodata & 1(1) & \nodata & \nodata & \nodata & $1.68_{-0.43}^{+0.39}$ & 2.78  \\
450P & \nodata & \nodata & \nodata & \nodata & 2(2) & $7.02_{-0.65}^{+0.66}$ & 7.89  \\
453P & \nodata & \nodata & 1 & 1(1) & \nodata & $1.98_{-0.63}^{+0.51}$ & 3.04  \\
459P & 2(2) & 3(3) & \nodata & \nodata & \nodata & $3.05_{-0.43}^{+0.48}$ & 3.93  \\
460P & 2(2) & 2(2) & \nodata & \nodata & \nodata & $0.82_{-0.11}^{+0.10}$ & 0.90  \\
461P & 1(1) & 2(2) & 1(1) & 1(1) & \nodata & $1.65_{-0.16}^{+0.15}$ & 1.84  \\
463P & 1(1) & 2(1) & \nodata & \nodata & \nodata & $0.32_{-0.09}^{+0.13}$ & 0.96  \\
467P & \nodata & \nodata & \nodata & 1(1) & \nodata & $6.71_{-2.34}^{+2.55}$ & 20.30  \\
468P & \nodata & \nodata & \nodata & \nodata & 2(2) & $1.82_{-0.17}^{+0.17}$ & 2.01  \\
471P & 1 & 1 & 1 & \nodata & \nodata &  \nodata & 6.69  \\
475P & \nodata & \nodata & \nodata & \nodata & 2(2) & $2.12_{-0.20}^{+0.19}$ & 2.30  \\
477P & \nodata & 1 & \nodata & \nodata & \nodata &  \nodata & 2.46  \\
479P & 1 & 1(1) & \nodata & \nodata & \nodata & $0.97_{-0.19}^{+0.19}$ & 3.08  \\
496P & 1 & \nodata & 1(1) & \nodata & \nodata & $3.03_{-0.60}^{+0.40}$ & 3.73  \\
2005 T5 & \nodata & \nodata & \nodata & \nodata & 2(2) & $2.63_{-0.23}^{+0.22}$ & 2.94  \\
C/2007 S2 & \nodata & \nodata & 1 & 1 & \nodata &  \nodata & 49.02  \\
C/2008 E1 & \nodata & \nodata & \nodata & 1 & \nodata &  \nodata & 25.66  \\
C/2014 W11 & 2 & 1 & \nodata & \nodata & \nodata &  \nodata & 53.82  \\
C/2021 K1 & 1 & 1 & \nodata & \nodata & \nodata &  \nodata & 15.18  \\
P/1998 VS24 & \nodata & \nodata & \nodata & \nodata & 2 &  \nodata & 3.23  \\
P/2004 V5-A & \nodata & \nodata & \nodata & \nodata & 2(2) & $3.28_{-0.76}^{+0.64}$ & 3.37  \\
P/2008 Y3 & \nodata & \nodata & \nodata & 2 & \nodata &  \nodata & 16.32  \\
P/2009 T2 & 1 & 1(1) & 1(1) & 1 & \nodata & $3.11_{-0.79}^{+0.74}$ & 4.50  \\
P/2009 WX51 & \nodata & \nodata & 1 & 1 & \nodata &  \nodata & 0.78  \\
P/2009 Y2 & \nodata & \nodata & 1(1) & 1(1) & \nodata & $3.12_{-0.84}^{+0.56}$ & 4.57  \\
P/2010 C1 & \nodata & \nodata & 1 & 1 & \nodata &  \nodata & 14.95  \\
P/2010 D2 & \nodata & \nodata & 1 & 1 & \nodata &  \nodata & 6.29  \\
P/2010 E2 & \nodata & \nodata & 1(1) & 1 & \nodata & $2.17_{-0.67}^{+0.42}$ & 3.74  \\
P/2010 H2 & \nodata & \nodata & 1 & 1 & \nodata &  \nodata & 7.95  \\
P/2010 J3 & 1(1) & 1(1) & 2(2) & 1(1) & \nodata & $4.86_{-0.43}^{+0.41}$ & 6.28  \\
P/2010 U1 & \nodata & \nodata & \nodata & 1 & \nodata &  \nodata & 15.35  \\
P/2011 P1 & 1 & 1 & \nodata & \nodata & \nodata &  \nodata & 92.82  \\
P/2012 B1 & \nodata & 1 & \nodata & \nodata & \nodata &  \nodata & 73.75  \\
P/2013 W1 & \nodata & 1 & \nodata & \nodata & \nodata &  \nodata & 0.99  \\
P/2014 L2 & 1 & 2 & \nodata & \nodata & \nodata &  \nodata & 14.66  \\
P/2014 X1 & 1(1) & 1(1) & \nodata & \nodata & \nodata & $2.82_{-0.92}^{+1.11}$ & 6.13  \\
P/2020 T3 & \nodata & 1(1) & \nodata & \nodata & \nodata & $0.40_{-0.15}^{+0.14}$ & 1.13  \\
P/2020 U2 & 1 & 1 & \nodata & \nodata & \nodata &  \nodata & 4.89  \\
P/2021 HS & \nodata & 2(2) & \nodata & \nodata & \nodata & $0.59_{-0.11}^{+0.11}$ & 0.81  \\
P/2021 N2 & 1(1) & 1 & \nodata & \nodata & \nodata & $28.27_{-10.33}^{+8.86}$ & 49.12  \\
P/2021 PE20 & \nodata & 1 & \nodata & \nodata & \nodata &  \nodata & 1.37  \\
P/2022 L3 & 2 & 2 & \nodata & \nodata & \nodata &  \nodata & 9.31  \\
\enddata
\tablenotetext{a}{Total number of band-epochs entering the size-estimation fit; the number in parentheses gives those with a non-degenerate Model~2 (PSF $+$ coma) fit (Section~\ref{sec:models}). Bands with no parenthetical value have no Model~2 viable epoch.}
\tablenotetext{b}{Number of Spitzer epochs contributing independent nucleus flux measurements \citep{Fernandez2013}, listed for comparison.}
\tablenotetext{c}{Maximum a posteriori (MAP) diameter from the joint Model~2 fit. The subscript and superscript give the uncertainties covering the 16th and 84th posterior
percentiles (1$\sigma$-equivalent; Section~\ref{sec:phot_thermal}).
Reported only for comets with at least one Model~2 viable epoch.}
\tablenotetext{d}{Upper-limit diameter from Model~1 at the 95\% confidence level. Available for every comet, as Model~1 is always valid as a ceiling (Section~\ref{sec:model1}).}
\tablecomments{Comets are grouped by dynamical class. Model~1 provides a conservative diameter ceiling for all objects, while the Model~2 measurement is available only where the coma is resolved (Section~\ref{sec:models}).}
\end{deluxetable*}

\subsection{Comparison with Other Survey Data  \label{sec:verify}}

We compare our diameter estimates with those of five previous comet size surveys for the comets in common: two thermal-infrared surveys, \citet{Fernandez2013} and \citet{Bauer2017} (the latter including \citealt{Bauer2015}), and three visible-wavelength surveys, \citet{Lamy2004}, \citet{Meech2004}, and \citet{Snodgrass2011}. All literature diameters were taken directly from the published tables of each study.

The two classes of survey probe our results in complementary ways. The infrared surveys observed the (nearly) same thermal portion of the spectral energy distribution as this work, applied comparable nucleus-extraction techniques, and adopted similar albedo and beaming assumptions; agreement with them therefore tests the internal consistency of our methodology. The visible surveys instead derive diameters from the absolute magnitude under an assumed geometric albedo \citep{Bowell1989,Fowler1992}, in some
cases selecting apparently inactive comets at large heliocentric distances, where unresolved dust within the PSF may still contaminate the nucleus flux \citep[e.g.,][]{Meech2004}. The systematic effects in those studies are thus largely independent of ours, and agreement with them constitutes a more demanding cross-check of the underlying assumptions of both approaches.

\begin{figure}[htb]
\centering
\includegraphics[width=0.53\textwidth]{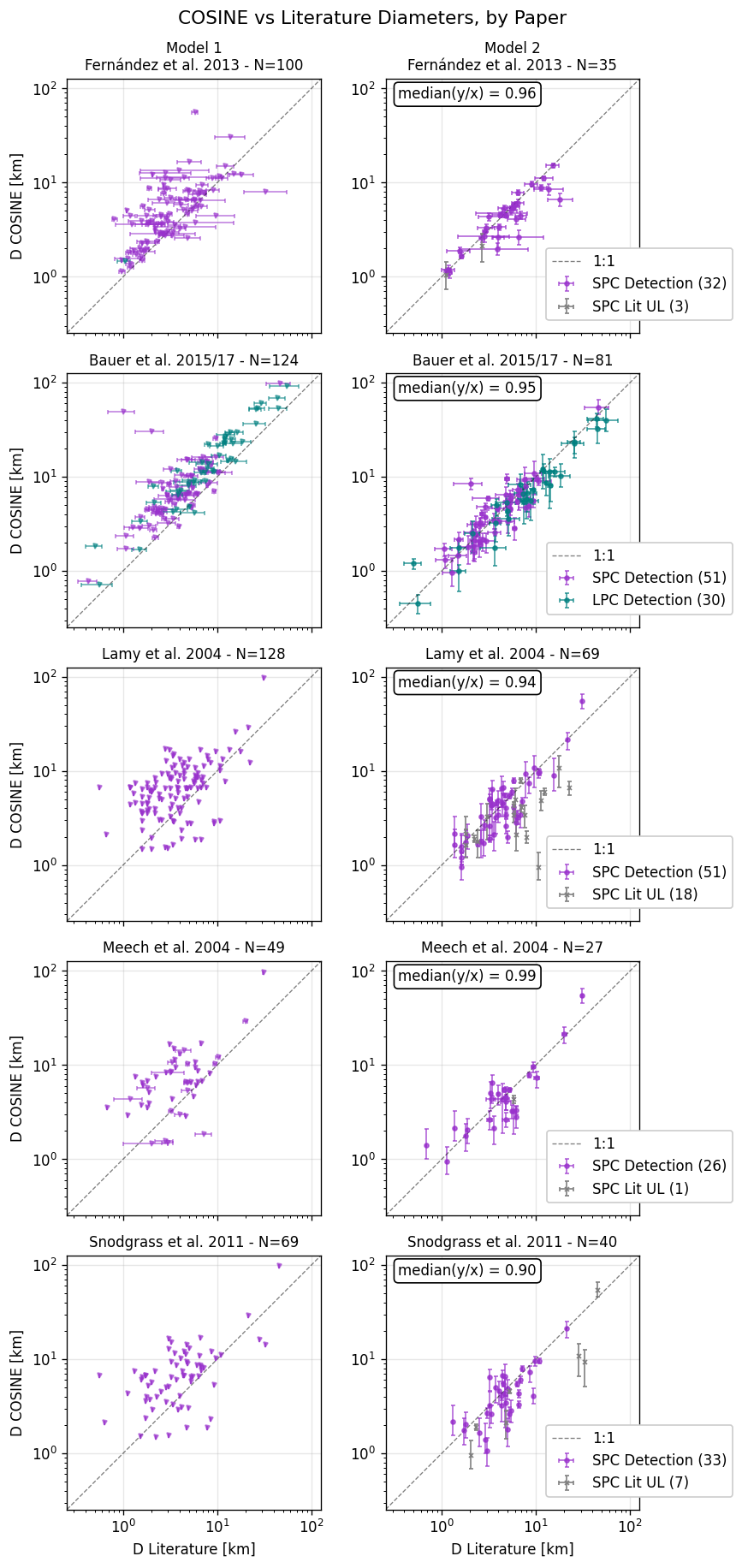}
\caption{Comparison of the COSINE nucleus diameters (Table~\ref{tab:size_result};
$y$-axis) with published values ($x$-axis) for the comets common to both samples. All diameters are in km, and each row corresponds to one literature study. Left column: our Model~1 upper limits, which lie above the one-to-one line by construction. Right column: our Model~2 diameters, plotted against literature detections (\texttt{Detection}, purple and teal circles) and literature upper limits (\texttt{Lit UL}, gray asterisks); the latter lie below the one-to-one line, likewise by construction. The median ratio of our diameters to the literature values is quoted in each right-column panel. Error bars are shown where the source study reports uncertainties. Colors follow Figure~\ref{Fig03}.}
\label{Fig04}
\end{figure}

Figure~\ref{Fig04} presents the comparison of sizes of comets covered in both this and previous studies, with the number of common objects $N$ indicated in each panel. Each row corresponds to one literature study, plotted on the x-axis against the COSINE diameters on the y-axis. Colors follow Figure~\ref{Fig03}, where LPCs are teal and SPCs are purple. The left column compares our Model~1 upper limits (\texttt{Upper Limit} in Table~\ref{tab:size_result}), which by construction fall above the one-to-one line. The right column compares our Model~2 diameters (\texttt{Diameter} in Table~\ref{tab:size_result}) and lists the median of the ratio of our diameters to the literature values. Within each panel, literature detections (the counterpart of our Model~2 estimates) and literature upper limits are plotted with different symbols, with the number of comets in each category given in parentheses. The literature upper limits by construction fall below the one-to-one line. Error bars are shown wherever the source study provides uncertainties, which is the case for the detections in the two infrared surveys but for few of the visible-survey diameters.

Despite the distinct, largely independent systematic effects of the individual surveys, the median diameter ratios between this work and every compared study are consistent with unity to within 10\%. This agreement supports two conclusions. First, our nucleus-extraction methodology yields stable diameters that are compatible with both thermal-infrared and visible-wavelength determinations, rather than introducing method-specific biases. Second, the comet population is consistent with a narrow range of the surface properties relevant to these methods, in particular geometric albedo and beaming parameter \citep{Fernandez2013}, in contrast to the broad diversity observed among asteroids \citep{DeMeo2014}. On this basis, we proceed to a more detailed examination of the results in the following sections.

\subsection{Limitations \label{sec:limits}}

Although our nucleus-extraction and size-derivation procedure is applied uniformly across the sample and yields diameters consistent with previous surveys (Section~\ref{sec:verify}), it is not free of limitations. The caveats we consider most important fall into three categories: the instrument, the observing geometry, and the nature of the comets themselves.

The first limitation is instrumental. The PSF-plus-coma fitting technique we employ was developed for, and has mainly been applied on, Hubble Space Telescope observations \citep{Lamy1996,Lamy1998a,Lamy1998b}, whose fine pixel scale makes it possible to pin down the photocenter and cleanly separate the point-source and extended components \citep{Hui2018}. In adopting the technique for WISE, we are deliberately pushing it beyond its native regime: the WISE detectors have a comparatively coarse pixel scale and a correspondingly broad PSF. Even though our analysis oversamples the templates by a factor of 40 (0.025~pix), the coarse native sampling weakens the contrast between the PSF and coma templates, leaving the decomposition more sensitive to the choice of fitting parameters and the extracted nucleus flux more uncertain than it would be at finer resolution.

The second limitation comes from the observing geometry. The detected comets span heliocentric distances of 0.996 to 10.804~au (median 2.548~au); by dynamical group, the LPCs were observed at 1.081 to 10.804~au (median 3.203~au) and the SPCs at 0.996 to 6.544~au (median 2.279~au) \citep{Kwon2025}. At these distances, a single pixel projects to up to $\sim$$10^{3}$~km at the comet, so any residual near-nucleus dust (or gas) can remain spatially unresolved and is absorbed into the point-source component of the fit. With finite pixels, this contamination can never be fully excluded, and it systematically overestimates the nucleus size \citep{Fernandez2013}.

The third limitation arises from the comets themselves. When a bright nucleus is superposed on a weak coma, the decomposition is well conditioned and the fit is straightforward. When the nucleus is faint or the coma dominates the total flux, however, the nucleus signal is buried in the coma, and the extracted flux becomes prone to over- or under-subtraction, with a correspondingly larger fractional uncertainty. This ambiguity is more severe for the LPCs than for the SPCs, as expected from their generally higher activity and lower nucleus-to-coma contrast, consistent with \citet{Hui2018}. The uncertainties we report are therefore systematically much larger for the LPCs in our sample.

Beyond these three, subtler systematics may well be hiding in the results. As one test, we compared the diameters of comets with cryogenic-phase (W3/W4) data against those constrained only by Reactivation-phase (W1/W2) data, using the overlapping literature comets of Section~\ref{sec:verify} as an external reference. The Reactivation-only diameters carry larger uncertainties, as expected, but we found no systematic offset between the two subsets. Our results are therefore not bias-free, but what they offer instead is the homogeneously analyzed database of comet nucleus sizes, a foundation we consider well worth having for studying population-level trends. A more detailed investigation tailored to the dynamical subgroups of comets is the subject of our future work.

\section{Discussion \label{sec:discuss}}

\subsection{General Trends in Photometric Properties of Comet Nuclei \label{sec:general_phot}}

The infrared color of a nucleus combines reflected sunlight with thermal emission. The thermal component depends on surface temperature, which is set primarily by heliocentric distance ($r_{\rm H}$) and modulated by rotation, spin-axis orientation, shape, and thermal inertia. A nucleus therefore changes color as it heats and cools along its orbit, and a sample spanning a range of $r_{\rm H}$ can show a spread in color even if all surfaces were identical.

\begin{figure}[htb]
\centering
\includegraphics[width=0.7\textwidth]{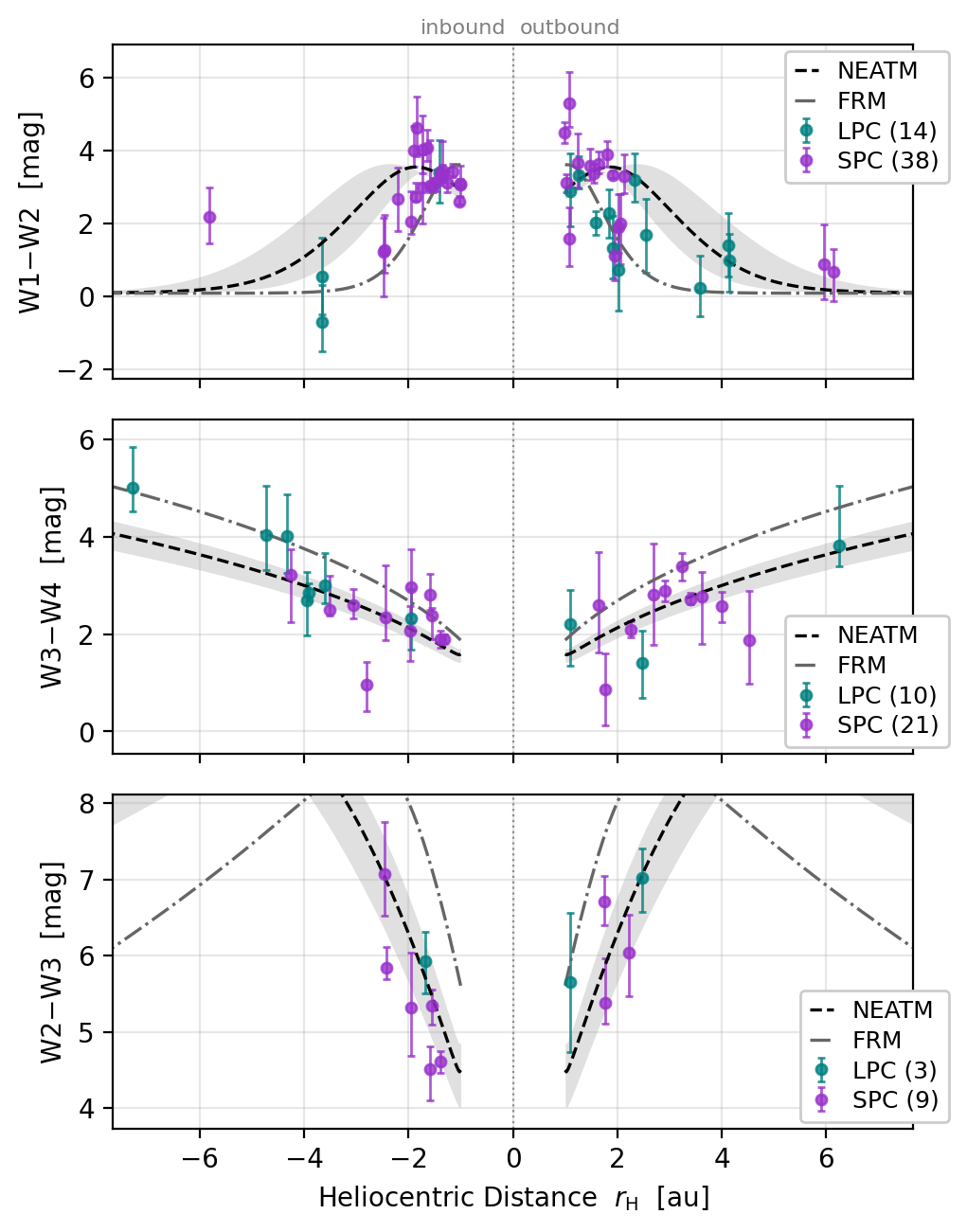}
\caption{Nucleus colors $W1-W2$, $W3-W4$, and $W2-W3$ from Model~2 (Fig.~\ref{Fig_nuchist}) as a function of signed heliocentric distance, negative for inbound and positive for outbound observations. The dashed and dash-dot curves are the colors predicted by NEATM \citep{Harris1998} and by the fast-rotating model \citep[FRM;][]{Lebofsky1989} for the albedo and beaming priors at the 90~degree solar elongation of the WISE survey (the FRM has no beaming parameter). The gray band spans NEATM beaming values from 0.70 to 1.36 (3$\sigma$ of the beaming priors in Table~\ref{tab:neatm_priors}). Modeling details are given in Appendix~\ref{sec:app2}.}
\label{Fig_nuchelio}
\end{figure}

To separate this geometric effect from intrinsic differences, Figure~\ref{Fig_nuchelio} compares the measured colors with the NEATM \citep{Harris1998} and the fast-rotating model \citep[FRM;][]{Lebofsky1989} predictions for a nucleus with the median albedo, slope parameter, and beaming of our sample. The two models bracket the plausible surface temperature distribution: NEATM concentrates emission near the subsolar point, whereas the FRM assumes isothermal latitude bands and therefore predicts cooler surfaces at a given distance, shifting each color transition inward. Modeling details are given in Appendix~\ref{sec:app2}.

Both LPCs and SPCs broadly follow the model curves over the full range of $r_{\rm H}$, in all three colors and on both sides of perihelion. The shape of each curve reflects which spectral component dominates the two bands involved. Over the distances sampled here, W1 is dominated by reflected sunlight except inside $\sim$2~au, and W3 and W4 are always thermal, whereas W2 is the band in which the two components trade dominance with $r_{\rm H}$ (Appendix~\ref{sec:app2}). That the heliocentric trend is set by the thermal component illustrates why thermal infrared photometry is an efficient size diagnostic: to first order the flux depends on temperature and size of the object, with only a weak dependence on albedo and hence on surface composition \citep{Harris2002, Delbo2015}.

The residual scatter about the curves is not yet interpretable, given the sparse sampling in $r_{\rm H}$ and the size of the individual uncertainties. One of the outliers is 29P/Schwassmann-Wachmann~1 at $r_{\rm H} \simeq 6$~au inbound, whose $W1-W2 \simeq 2$~mag exceeds both models by more than 1~mag (top row in Fig.~\ref{Fig_nuchelio}).  We confirm that its profile fit is well behaved and yields a diameter consistent with published values, so the excess more likely reflects its intrinsic color and/or the unresolved gas emission in W2 (Section~\ref{sec:gas_impact}) in the PSF. Elsewhere the residuals are distributed roughly symmetrically about the curves, with no measurable inbound-outbound asymmetry. We therefore do not attempt to isolate an intrinsic color residual, and leave a systematic study of nucleus color evolution to denser thermal-infrared time series from future missions.

\subsection{General Trends in Nucleus Sizes \label{sec:general_size}}

The diameters derived in this study (Table~\ref{tab:size_result} and Figure~\ref{Fig03}) form a statistical base large enough to begin asking how nucleus size relates to orbital properties. In this section, we examine the ensemble behavior of the estimated sizes as a function of the orbital elements.

\begin{figure}[htb]
\centering
\includegraphics[width=\textwidth]{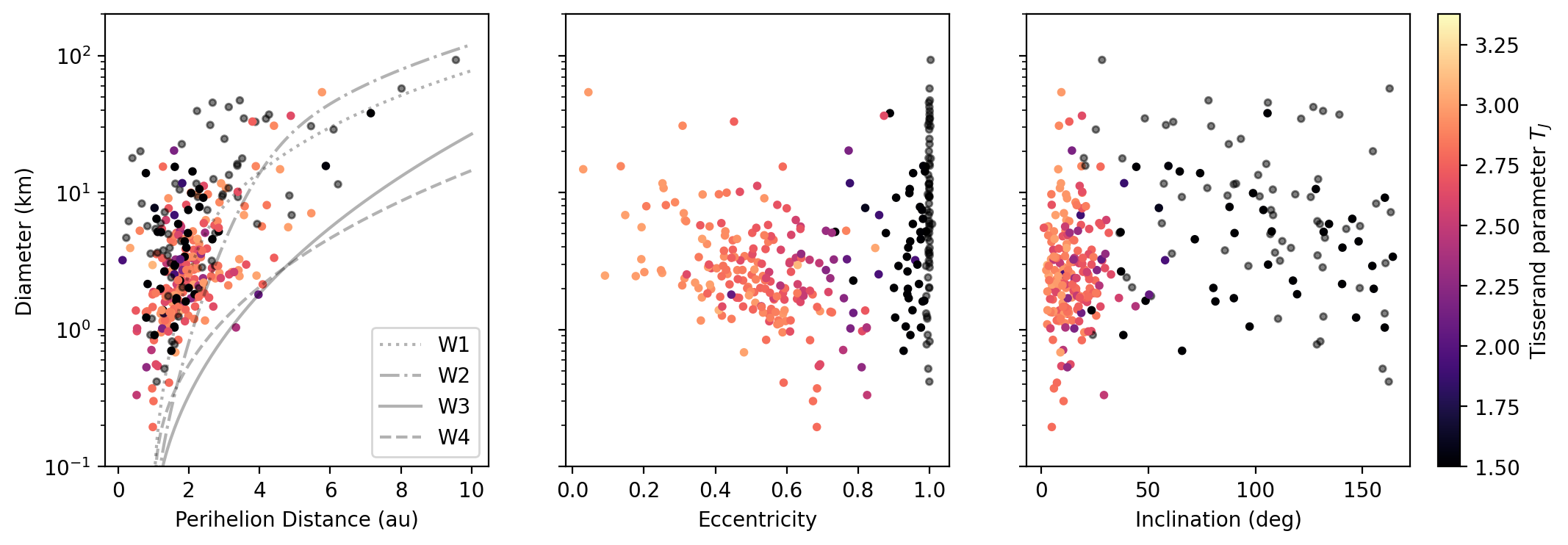}
\caption{Nucleus diameters of the 301 comets (107 LPCs and 194 SPCs) with valid Model~2 estimates as a function of perihelion distance ($q$), eccentricity ($e$), and inclination ($i$), color-coded by the Tisserand parameter with respect to Jupiter ($T_{\rm J}$), taken from the JPL Small-Body Database (\url{https://ssd.jpl.nasa.gov/tools/sbdb\_lookup.html}). The 222 comets (128 LPCs and 94 SPCs) with only Model~1 upper limits are not shown. The four curves in the leftmost panel indicate the approximate smallest diameter detectable in each WISE band as a function of heliocentric distance, computed with NEATM under the same priors as the main analysis (Section~\ref{sec:phot_thermal}) and a representative observing geometry (solar elongation of 90$^{\circ}$). The curves are anchored to the limiting magnitudes of the WISE Explanatory Supplement (\url{https://irsa.ipac.caltech.edu/data/WISE/docs/release/All-Sky/expsup/sec6\_3a.html}), shifted by $\sim$2.5~mag to account for the added depth of our coadded images; the relative offsets among the bands are preserved. These curves are illustrative only, as the actual sensitivity depends on the observing geometry, coadd depth, activity level, and the nucleus extraction. 
\label{Fig05}}
\end{figure}

Figure~\ref{Fig05} shows the estimated diameters as a function of perihelion distance ($q$), eccentricity ($e$), and inclination ($i$), color-coded by the Tisserand parameter with respect to Jupiter ($T_{\rm J}$). We use the 301 comets (107 LPCs and 194 SPCs) with valid Model~2 estimates (Table~\ref{tab:size_result}); comets with only Model~1 upper limits are excluded. In the $q$--diameter panel, we overlay approximate detectability curves for each WISE band \citep{Cutri2012}, computed with NEATM under the same assumptions as the main analysis, to guide the eye on where our survey runs out of sensitivity.

We note at the outset that the sub-kilometer regime is populated by only a handful of objects in all three panels, so we refrain from interpreting that region throughout.

In the $q$--diameter plane, the sample forms a cloud rising diagonally to the upper right. LPCs ($T_{\rm J} < 2$) cluster at $q \lesssim 2$~au, while the SPCs ($T_{\rm J} \gtrsim 2$) cluster at $q \lesssim 3$~au, both reflecting the higher observational completeness of our survey in the inner Solar System (Figure~\ref{Fig01}). The detectability curves make the origin of the diagonal envelope explicit: the W3 and W4 curves broadly trace the lower boundary of the observed distribution, confirming both the central role of the thermal bands in our diameter estimation and the brightness-limited nature of the overall pattern. A direct consequence is that the apparent absence of small nuclei at large $q$ is a selection effect rather than an intrinsic feature: the few comets we detect beyond $q \sim 4$~au are, necessarily, large ones. A similar brightness-limited pattern was reported in the WISE study of Centaurs \citep{Bauer2013}.

In the $e$--diameter plane, there is a mild paucity of kilometer-sized comets at $e \sim 0.15$ to 0.3. Because Figure~\ref{Fig01} shows that we undersample this orbital region relative to the known population, we do not read intrinsic meaning into this gap. Among the SPCs, comets with $T_{\rm J}$ approaching 3, that is those closest to dynamical detachment from Jupiter, tend toward lower eccentricities and somewhat larger diameters. The statistics in this corner of the diagram are small, however, and we thus flag the trend as suggestive rather than established. The LPCs show no discernible trend, with diameters spread nearly uniformly at $e \approx 1$, as their dynamical definition dictates.

In the $i$--diameter plane, our diameter sample traces the known comet population well (Figure~\ref{Fig01}), a fair sampling owed largely to the all-sky survey character of WISE, which carries much less bias against high-inclination objects than ground-based observations \citep{Bauer2013,Bauer2024}. The SPCs concentrate near the ecliptic with a median inclination of $\sim$15$^{\circ}$, consistent with their presumed origin in the trans-Neptunian region \citep{Nesvorny2017}, whereas the LPCs are distributed nearly isotropically, as expected for an Oort-cloud source population \citep{Vokrouhlicky2019}.

Figure~\ref{Fig05} offers a compact summary of where the measured comet diameters currently sit in orbital-element space, but it also makes plain that most of the visible patterns are sculpted by observational biases. Recovering the intrinsic population thus requires a full debiasing, in which each detected object is weighted by the inverse of its detection probability given its size, orbit, and the survey geometry. That debiasing, and the reconstruction of the underlying size distributions (Section~\ref{sec:size_dist}), is the subject of our future work.

\subsection{Size Distribution of Comet Nuclei \label{sec:size_dist}}

The size-frequency distribution (SFD) of comets carries memory of both the formation and the subsequent collisional and dynamical histories of the small-body populations of the Solar System. Its shape, interpreted against numerical models and laboratory results, diagnoses the collisional state and internal strength of the parent population \citep{Dohnanyi1969,OBrien2003}. The cumulative SFD is commonly quantified by fitting simple functional forms: single or broken power laws, or exponentially tapered power laws of the kind that emerge naturally from streaming-instability planetesimal formation \citep{Kavelaars2021,Fraser2024,Simon2024}. The fitted slopes and break locations are then compared with the predictions of formation and evolution models.

Our study adds roughly 200 new nucleus size measurements ($\sim$400 including upper limits), cross-validated against the previous major surveys for the overlapping comets (Section~\ref{sec:verify}). On this basis, we examine the cumulative SFD of each dynamical group.

\subsubsection{Short-Period Comets (SPCs) \label{sec:spcs}}

\begin{figure}[htb]
\centering
\includegraphics[width=\textwidth]{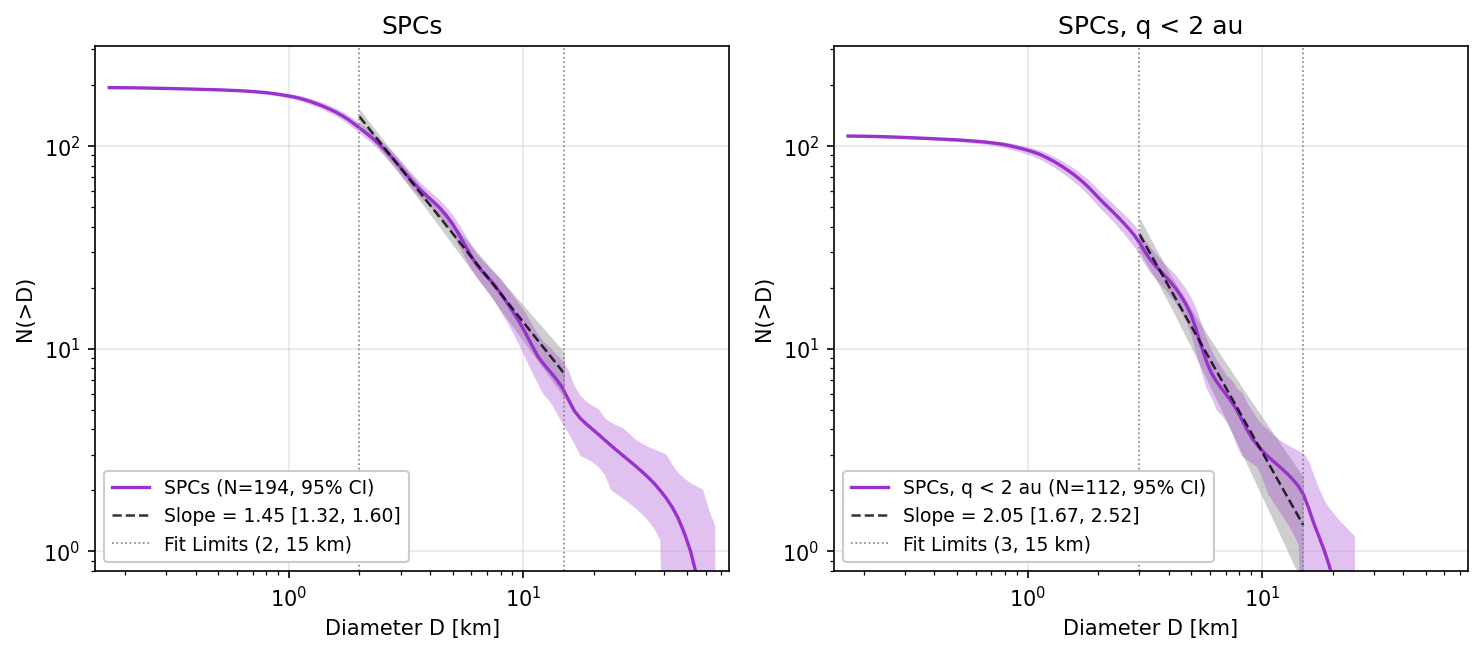}
\caption{Cumulative size-frequency distribution of the SPCs with Model~2 diameters (Table~\ref{tab:size_result}), for the full sample (left, $N = 194$) and for comets with perihelion distance $q < 2$~au (right, $N = 112$). The purple curve is the median of the Monte Carlo realizations drawn from the diameter posteriors and the purple band is their 95\% interval. The dashed line and gray region show the best-fit power law and its uncertainty, fit between the dotted vertical lines (2 to 15~km for the full sample, 3 to 15~km for the $q < 2$~au subset). The legend lists the median slope with its 16th and 84th percentiles.}
\label{Fig06}
\end{figure}

Figure~\ref{Fig06} shows the cumulative SFD $N(>D)$ of the 194 SPCs with Model~2 diameters (Table~\ref{tab:size_result}). Comets with only Model~1 upper limits are excluded. The solid curve traces the median of the distribution and the shaded region its 95\% confidence interval (CI), constructed as follows. Each comet carries 10,000 posterior draws of $D$ from the NEATM MCMC (Section~\ref{sec:phot_thermal}). Drawing one diameter per comet yields one realization of the SFD. Repeating this over all draws yields 10,000 realizations, from which we take the median and the 2.5th and 97.5th percentiles at each $D$\footnote{This interval reflects only the uncertainty of our diameter fits, not that of the population. An alternate can be made by assuming that successful diameter measurements obey a Poisson-process. It treats the observed counts as a Poisson realization of an underlying population SFD, which adds the sampling uncertainty of the population itself. The resulting interval exceeds the fit-only interval by a factor of 2 to 4 at all diameters. Figure~\ref{Fig06} presents the uncertainties from our MCMC fitting alone.}. A caveat of this scheme is that the per-comet posterior widths are dictated by the heterogeneous data coverage of each comet, so the realizations inherit that heterogeneity. Using point estimates (i.e., face value of Model~2) alone would discard the uncertainty information entirely, so we regard the posterior-based construction as the more defensible choice. Homogeneous thermal coverage from future surveys
will reduce this limitation.

The observed distribution shows no clear inflection within the uncertainties before it levels off at the small end, so we fit a single power law, $N(>D) \propto D^{-\gamma}$, over $2 < D < 15$~km. The upper bound excludes the sparsely populated large-diameter tail, where too few comets remain to constrain the slope. For each SFD realization, we perform a linear least-squares fit in log-log space, and we report the median slope with the 16th and 84th percentiles as the 1$\sigma$-equivalent uncertainty. We fit the cumulative distribution by least squares for direct comparability with the literature, noting that the counts are correlated by construction, so the formal uncertainties should be interpreted with care. Roughly half of the sample lies at perihelion distance $q \lesssim 2$~au, where our completeness is highest (Figure~\ref{Fig01}), and we therefore fit two cases: the full SPC sample and the $q < 2$~au subset.

For the full SPC sample we obtain $\gamma = 1.45_{-0.13}^{+0.15}$ (1$\sigma$) with a 95\% CI of [1.20, 1.78] over $2 < D < 15$~km (117 comets in this size bin), and for the $q < 2$~au subset $\gamma = 2.05_{-0.38}^{+0.47}$ with a 95\% CI of [1.35, 3.01] over $3 < D < 15$~km (31 comets in the bin). The small-diameter bins carry far smaller uncertainties than the sparse large-diameter tail, so the fitted slope is sensitive to the adopted fitting range. Below $D \sim 2$~km, the distribution turns shallow and levels off near 1~km, as seen in all previous comet surveys. Two explanations have long been discussed. The first is observational: brightness-limited surveys progressively miss small, faint nuclei, and separating a weak point source from an active coma becomes increasingly challenging at these sizes \citep{Lamy2004}. The second is physical: small nuclei are preferentially removed by fading, whether through mantling into dormancy, disruption, or ejection \citep{Meech2004,Jewitt2004,Fernandez2013}. Our sample is likely incomplete in this regime, so we cannot separate the two effects here. Only a full debiasing can establish how much of the turnover is intrinsic.

Compared with previous work, the full-sample slope is shallower than most published values but agrees well with the $1.45 \pm 0.05$ measured over $\sim$2 to 20~km with HST and Keck \citep{Meech2004}, whose results are also consistent with ours in the median sizes and distributions of both dynamical groups despite entirely different analysis methods and target selection. The full-sample slope also broadly agrees with the $1.6 \pm 0.2$ found for the SPC population when extinct and dormant candidates are included \citep{Lamy2004}. We do not see the ``knee'' at $\sim$4 to 5~km reported in some studies \citep{Meech2004,Fernandez2013}, although a weak feature near that diameter may be present in the $q < 2$~au panel at the 1$\sigma$ level. The $q < 2$~au slope is nominally steeper than the full-sample slope, and its value sits well among the literature results in similar size ranges ($1.91 \pm 0.06$ over $\sim$4 to 10~km, \citealt{Meech2004}; $1.9 \pm 0.3$ down to $\sim$4~km, \citealt{Lamy2004}; $1.92 \pm 0.20$ above $\sim$3~km, \citealt{Snodgrass2011}; $1.92 \pm 0.23$ for $D > 5$~km, \citealt{Fernandez2013}; $1.93 \pm 0.06$, \citealt{Bauer2017}).

The full sample, although less complete at large $q$, probes a broader cross-section of the population than the $q < 2$~au subset, and its shallow slope invites comparison with the source region. The destabilized trans-Neptunian population, widely considered the reservoir of the SPCs \citep{Nesvorny2017,Bottke2023}, is modeled with a cumulative slope of $\gamma \sim 1.2$ in the relevant size range \citep{Bottke2023}, compatible with the kilometer-scale projectile population inferred from the crater records of Charon and Arrokoth ($1.0 \lesssim \gamma \lesssim 1.2$; \citealt{Morbidelli2021}). Both values are consistent with the 95\% CI of our full-sample SPC slope. Considering that large nuclei take longer than small ones to exhaust their near-surface volatile inventories, and hence to fade from observability in the inner Solar System \citep{Jewitt2004,Bottke2023}, one may speculate that the multikilometer SPC SFD has not been strongly reshaped since implantation, in which case the observed distribution retains a memory of the SFD the bodies carried when they were dynamically delivered from the source region. The present uncertainties and the absence of a debiasing correction prevent us from elevating this beyond a hypothesis. Our future work on debiasing, together with independent constraints on nucleus maturity from observations of the other cometary components, is intended to address it directly.

\subsubsection{Long-Period Comets (LPCs) \label{sec:lpcs}}

\begin{figure}[htb]
\centering
\includegraphics[width=\textwidth]{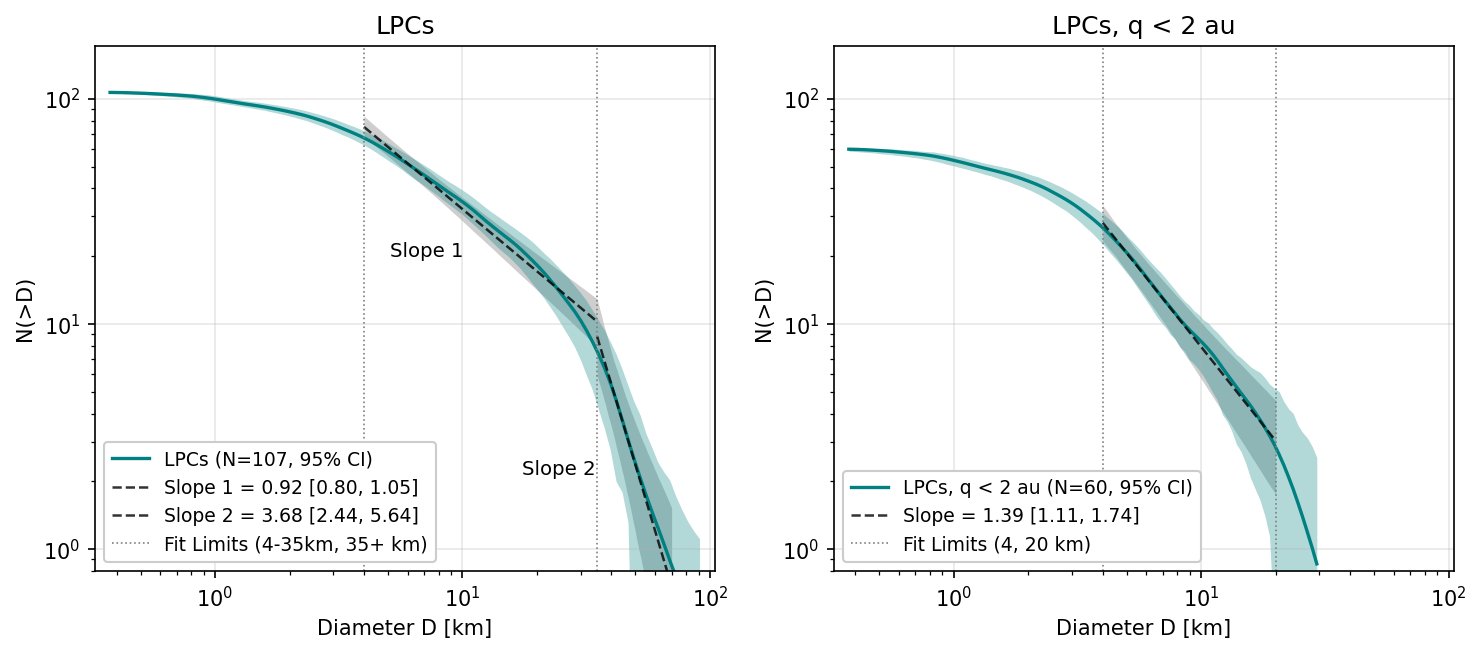}
\caption{Cumulative size-frequency distribution of the 107 LPCs with Model~2 diameters and uncertainties (Table~\ref{tab:size_result}), for the full sample (left) and for comets with perihelion distance $q < 2$~au (right). All conventions follow Figure~\ref{Fig06}.
\label{Fig07}}
\end{figure}

Figure~\ref{Fig07} shows the corresponding cumulative SFDs for the LPCs: the full sample of 107 comets with Model~2 constraints, and $\approx$60\% of the sample with $q < 2$~au. All conventions follow Figure~\ref{Fig06}. Here, however, the purpose of the $q < 2$~au cut is not observational completeness but dynamical purity. LPCs divide into dynamically old and dynamically new comets according to the number of past perihelion passages, associated respectively with the inner and outer parts of the Oort cloud, whose boundary lies near 2,000~au \citep{Brasser2013,Vokrouhlicky2019}. Comets arriving from the outer cloud, beyond the Oort spike at $\sim 10^{4}$~au, are predominantly dynamically new\footnote{Recent studies suggest that this fiducial boundary should be placed farther out than the classical value \citep{Dybczynski2015}.}. Their binding energies are small enough that a single stellar encounter or the Galactic tide can deliver them directly into the inner Solar System, hence called ``jumpers.'' Comets from the inner cloud are predominantly dynamically old ``creepers,'' whose perihelia are reduced gradually in multiple stages by planetary perturbations. Because Oort cloud properties are most cleanly inferred from jumpers, whose surfaces have undergone little or no processing on prior perihelion passages, the $q < 2$~au subset offers the better approximation to a pristine population \citep{Vokrouhlicky2019}.

The full-sample LPC SFD is relatively smoothly concave down, declining gradually toward small sizes, so a (broken) power law may not be a suitable functional form for it. We nevertheless apply a broken power-law fit as an approximation, following the common practice of retaining this parameterization for comparison with published slopes even where the underlying distribution is continuous \citep[e.g.,][]{Vokrouhlicky2019, Heinze2021, Kavelaars2021}. 
For the full LPC sample with the break placed at $D \approx 35$~km, we obtain $\gamma = 0.92_{-0.12}^{+0.13}$ (1$\sigma$) with a 95\% CI of [0.70, 1.20] over $4 < D < 35$~km (59 comets in this size bin), and $\gamma = 3.68_{-1.24}^{+1.96}$ (1$\sigma$) with a 95\% CI of [1.46, 9.78] over $D > 35$~km (6 comets in this size bin). For the $q < 2$~au subset $\gamma = 1.39_{-0.28}^{+0.35}$ with a 95\% CI of [0.89, 2.19] over $4 < D < 20$~km (24 comets in the bin). The break diameter is a descriptive parameter of the fit and should not be read as a discrete transition in the population. The shallow small-size slope agrees with the debiased LPC value of $1.0 \pm 0.1$ reported by \citet{Bauer2017}. The steep large-size slope is constrained by the few largest objects in the sample and therefore carries a correspondingly wide interval. It is similar with the debiased Pan-STARRS1 result of \citet{Boe2019} and steeper than the WISE-based determination of \citet{Bauer2017} over a comparable size range.

For the $q < 2$~au LPCs, a cut that favors dynamically new arrivals over dynamically old ones, we expect the SFD to sit closer to the primordial one, as these nuclei have suffered far less cumulative mass loss and insolation-driven surface processing than SPCs subjected to repeated perihelion passages. The distribution remains smooth but is adequately described by a single power law, with $\gamma = 1.39_{-0.28}^{+0.35}$ (95\% CI [0.89, 2.19]) over $4 < D < 20$~km. Despite the large uncertainties, the fit is stable, and the slope is close to that of the full SPC sample over over $2 < D < 15$~km ($\gamma \simeq 1.45$, 95\% CI [1.20, 1.78]). Though the agreement may be fortuitous, their similarity could reflect a shared parentage and thus deserves further elaboration.

Current models hold that SPCs and LPCs condensed from a common population of primordial planetesimals at roughly 22 to 30~au, and subsequent giant-planet instability, driven principally by Neptune's migration, scattered the planetesimals into the two present reservoirs: the Oort cloud and the scattered disk and Kuiper belt \citep{Brasser2013,Dones2015,Nesvorny2017,Nesvorny2018,Morbidelli2020,Bottke2023}. Under this picture, the two subsets compared above, the multikilometer SPCs (less affected by fading than their smaller counterparts) and the dynamically new small-$q$ LPCs, are the two windows with the best chance of preserving early SFD information. Their mutual consistency in SFD slope, and their broad agreement with the modeled destabilized-population SFD \citep{Morbidelli2021,Bottke2023}, may therefore support common formative circumstances. Indeed, results from a 4.5-Gyr numerical simulation of \citet{Bottke2023} show that the collisional evolution of the destabilized population can reproduce the shapes of both the Jupiter-family and long-period comet SFDs within a single framework, and our measurements are consistent with that account. Given the present uncertainties and observational bias, however, we characterize this agreement as consistency rather than corroboration.

As emphasized throughout this paper, reconstructing the intrinsic populations and their SFDs requires a full debiasing treatment that accounts for the dynamical and physical mechanisms simultaneously, and until then part of the discussion above remains suggestive. What this study establishes is the scientific value of the long temporal baseline of a single observing facility analyzed with a consistent analysis method. This homogeneity makes our sample of comets a suitable testbed for the debiasing and population modeling to come.

\section{Summary and Conclusions \label{sec:summary}}

This second paper of the COSINE project (Cometary Object Study Investigating their Nature and Evolution) presents a separation of the nucleus and extended signals in the WISE/NEOWISE coadded images of Paper~I \citep{Kwon2025} and, combined with Spitzer measurements \citep{Fernandez2013}, derives nucleus sizes for 523 comets (235 LPCs and 288 SPCs). Our key findings are summarized below.

\begin{itemize}
    \item 
    We separate the nucleus from the coma in two stages, refining the standard nucleus--coma separation technique: an azimuthal wedge analysis that identifies tail-bearing wedges and builds a tail-free radial profile from the remaining quiet wedges, followed by a peak-constrained least-squares fit of PSF and $\rho^{-s}$ coma templates to that profile (Section~\ref{sec:models} and Appendix~\ref{sec:app1}). Model~1 (PSF only) always yields a conservative upper limit on the nucleus flux, while Model~2 returns a nucleus-coma decomposition wherever the two templates are non-degenerate; flux-injection tests set the effective uncertainty of each extraction (Section~\ref{sec:model2-unc} and Appendix~\ref{sec:fluxinj}).

    \item 
    Nucleus fluxes are converted to diameters with a NEATM plus reflected-light model fit jointly over all band-epochs of each comet by MCMC, with Gaussian priors on the $V$-band albedo $p_V$, the beaming parameter $\eta$, and the slope parameter $G$. Diameters are reported as the maximum a posteriori (MAP) value bounded with the 16th and 84th uncertainties (Sections~\ref{sec:phot_thermal} and \ref{sec:size_result}).

    \item 
    Model~1 upper limits are available for all 523 comets, and Model~2 diameters with uncertainties for 301 (107 LPCs and 194 SPCs), with median diameters of 5.3 and 2.5~km, respectively. Our estimates agree well with in situ determinations, for instance, 67P and 19P, and the median diameter ratios against five previous infrared and visible surveys are consistent with unity to within 10\%, despite the largely independent systematics of those studies (Section~\ref{sec:verify}).

    \item 
   Nucleus colors $W1-W2$, $W2-W3$, and $W3-W4$ are derived from the Model~2 fluxes for epochs with accepted extractions in both bands. All three colors are positive, with flux rising toward longer wavelengths, and their dependence on heliocentric distance follows the transition from reflected to thermal emission predicted by NEATM and FRM, with no measurable difference between LPCs and SPCs.  The ensemble phase-angle dependence of the nucleus fluxes reflects the coupling of phase angle to heliocentric distance at the fixed 90~degree solar elongation of WISE, combined with the faint-end detection limit, rather than intrinsic surface scattering.

    \item 
    The cumulative SFD of the SPCs gives $\gamma = 1.45_{-0.13}^{+0.15}$ (95\% CI [1.20, 1.78]) over $2 < D < 15$~km for the full sample and $2.05_{-0.38}^{+0.47}$ (95\% CI [1.35, 3.01]) over $3 < D < 15$~km for $q < 2$~au. The LPC distribution is smoothly concave-down without a distinctive inflection point. A break near 35~km separate slopes of $0.92_{-0.12}^{+0.13}$ (4--35~km, 95\% CI [0.70, 1.20]) and $3.68_{-1.24}^{+1.96}$ ($D > 35$~km, 95\% CI [1.46, 9.78]). The dynamically new $q < 2$~au LPC subset gives $\gamma = 1.39_{-0.28}^{+0.35}$ (95\% CI [0.89, 2.19]) over $4 < D < 20$~km. All distributions flatten below the fitted ranges and level off near 1~km, where the sample is demonstrably incomplete (Section~\ref{sec:size_dist}).

    \item 
    The multikilometer SPC slope and the slope of the dynamically new LPCs are mutually consistent and agree with the modeled destabilized trans-Neptunian population and the projectile distributions inferred from the Charon and Arrokoth crater records \citep{Morbidelli2021,Bottke2023}, as expected if the two groups descend from a common planetesimal population. Because the sample remains brightness limited, we characterize this agreement as consistency, not corroboration. A full debiasing is required to convert it into a firm constraint (Sections~\ref{sec:spcs} and \ref{sec:lpcs}).
\end{itemize}

In future COSINE papers, we will address the remaining cometary components, including the dust coma, tails and trails, and the gas coma, together with a new debiasing analysis based on this new database.

\begin{acknowledgments}
This publication makes use of data products from the Wide-Field Infrared Survey Explorer, which is a joint project of the University of California, Los Angeles, and the Jet Propulsion Laboratory/California Institute of Technology, funded by the National Aeronautics and Space Administration. This publication also makes use of data products from NEOWISE, which is a project of the Jet Propulsion Laboratory/California Institute of Technology, funded by the Planetary Science Division of the National Aeronautics and Space Administration. This research has made use of the NASA/IPAC Infrared Science Archive, which is funded by the National Aeronautics and Space Administration and operated by the California Institute of Technology. This publication also makes use of software and data products from the NEO Surveyor, which is a joint project of the University of California, Los Angeles and the Jet Propulsion Laboratory/California Institute of Technology, funded by the National Aeronautics and Space Administration.
\end{acknowledgments}

\facilities{WISE, NEOWISE, Spitzer}

\software{
NumPy \citep{Harris2020}, Kete \citep{Dahlen2025}, SciPy \citep{Virtanen2020}, Matplotlib \citep{Hunter2007}
}

\appendix

\section{Detailed Description of Profile Fitting Modeling Techniques \label{sec:app1}}
\counterwithin{figure}{section}
\counterwithin{table}{section}

This appendix details the radial-profile fitting used to separate the point-source and extended components of each cutout image. The procedure converts a two-dimensional cutout into a tail-free radial profile and then performs the peak-constrained fit of Section~\ref{sec:models}. The steps below define the quantities used there; their inputs and outputs are summarized in Table~\ref{tab:m1m2_io}.

\subsection{Step 1: Preprocessing}

Three quantities are measured directly from the cutout before the profile is built: the centroid, the background level $b$, and the peak data value $d_{\rm peak}$. The centroid is initialized at the brightest pixel within a disk of radius $r_{\rm search} = \mathrm{FWHM}$ (about $2.2$~pix in the W1, W2, and W3 bands and $4.4$~pix in W4; \citealt{Wright2010}) centered on the geometric center, then refined to sub-pixel precision by a two-dimensional quadratic fit over the surrounding $5 \times 5$ pixels; the refined offset is clamped to $r_{\rm search}$ to prevent a noise spike near the search boundary from displacing it. The resulting offset shifts the PSF and coma templates and defines the origin of the wedge grid. The background $b$ and sky noise $\sigma_{\rm b}$ are taken from the photometry of Paper~I, where they were measured from off-source regions of the frames contributing to each coadded image; if either value is missing or invalid we substitute the cutout median for $b$ and the standard deviation of the edge pixels for $\sigma_{\rm b}$. The background is held fixed throughout the fit. The per-pixel variance combines the sky term with Poisson shot noise from the source,
\begin{equation}
  \sigma_{\rm pix}^2(\rho) \;=\; \sigma_{\rm b}^2 \;+\; \frac{\max[\,0,\,d(\rho) - b\,]}{g_{\rm eff}}~,
  \label{eq:sigpix}
\end{equation}
\noindent where $d(\rho)$ is the measured intensity at radial distance $\rho$ and $g_{\rm eff}$ is the effective gain, namely the WISE per-frame gain\footnote{Explanatory Supplement to the WISE
All-Sky Data Release Products: \url{https://irsa.ipac.caltech.edu/data/WISE/docs/release/All-Sky/expsup/sec4\_4a.html\#gainrn}} scaled by the number of stacked frames. The peak value $d_{\rm peak}$ is the brightest pixel within a $5$~pix disk of the geometric center, read from the image itself.

\subsection{Step 2: Per-wedge Radial Profile}

Within each wedge $k$ and annulus centered at radius $\rho$, the mean intensity is the area-weighted average over the enclosed pixels,
\begin{equation}
  \mu_k(\rho) \;=\; \frac{\sum_{i \in (k,\rho)} a_i\,I_i}{\sum_{i \in (k,\rho)} a_i}~,
  \label{eq:muk}
\end{equation}
\noindent where $I_i$ is the image value of pixel $i$ and $a_i$ is its effective area within the cell, which accounts for sub-pixel overlap, fractional coverage at cell edges, and masked pixels. This yields an $N \times N_\rho$ matrix of per-wedge mean profiles.

A symmetric source contributes equally to all wedges, so the symmetric component is removed by subtracting, at each radius, the median of $\mu_k(\rho)$ across wedges,
\begin{equation}
  m(\rho) \;=\; \operatorname*{median}_k \mu_k(\rho)~, \qquad
  R_k(\rho) \;=\; \mu_k(\rho) - m(\rho)~.
  \label{eq:resid}
\end{equation}
By construction at most half the wedges lie above the median and at most half below, so the median traces the symmetric component and the residual $R_k(\rho)$ retains only the asymmetric part. Each wedge is then assigned a scalar asymmetry score by averaging its residual over the outer part of the profile,
\begin{equation}
  \zeta_k \;=\; \big\langle R_k(\rho) \big\rangle_{\rho \,\ge\, 3\,\rho_{\rm ap}/2}~, 
  \label{eq:score}
\end{equation}
\noindent where $\rho_{\rm ap}$ is the photometric aperture radius, adopted from Paper~I (6, 9, 11.5, 23 arcseconds for W1--W4). Restricting the average to large radii avoids the high-amplitude PSF core, which would otherwise dilute the asymmetry signal, and targets the region where a tail dominates and the symmetric coma has faded. A wedge with $\zeta_k > 0$ lies on the bright (tail) side. We identify the tail as the longest contiguous circular arc of positive-score wedges; if this arc spans at least four wedges ($\ge 60\arcdeg$) it is declared the tail, and otherwise the source is treated as symmetric. A profile whose scores share a single sign is also treated as symmetric, since the median subtraction has then failed to isolate a one-sided excess. Selecting the longest positive arc, rather than flagging individual high-score wedges, keeps the detection robust even when the tail spans many wedges.
 
The quiet wedges $\mathcal{Q}$ are the complement of the tail. The radial profile fed to the fit is their area-weighted mean, and the corresponding summed area sets the per-bin noise,
\begin{equation}
  d(\rho) \;=\; \frac{\sum_{k \in \mathcal{Q}} a_{k,\rho}\,\mu_k(\rho)}{\sum_{k \in \mathcal{Q}} a_{k,\rho}}~, \qquad
  A_{\rm bin}(\rho) \;=\; \sum_{k \in \mathcal{Q}} a_{k,\rho}~, \qquad
  \sigma(\rho) \;=\; \frac{\sigma_{\rm pix}(\rho)}{\sqrt{A_{\rm bin}(\rho)}}~.
  \label{eq:quietprof}
\end{equation}
Bins with no quiet-wedge area ($A_{\rm bin} = 0$) are assigned infinite variance and excluded from the fit.

\subsection{Step 3: Peak-constrained Model}

The fit consumes the quiet-wedge profile $d(\rho)$, its uncertainty $\sigma(\rho)$, the peak value $d_{\rm peak}$, the background $b$, and the PSF and coma templates evaluated on the same radial bins. The peak equality of Eq.~\ref{eq:peak} eliminates the nuclear amplitude in favor of the coma amplitude,
\begin{equation}
  A(B) \;=\; \frac{(d_{\rm peak} - b) - B\,\max[\,C(\cdot;s)\,]}{\max P}~,
  \label{eq:AofB}
\end{equation}
\noindent so that, at fixed slope $s$, the model $M(\rho;B) = A(B)\,P(\rho) + B\,C(\rho;s) + b$ is linear in the single parameter $B$.

\subsection{Step 4: Chi-square Minimization}

\begin{figure}[!thb]
\centering
\includegraphics[width=0.7\textwidth]{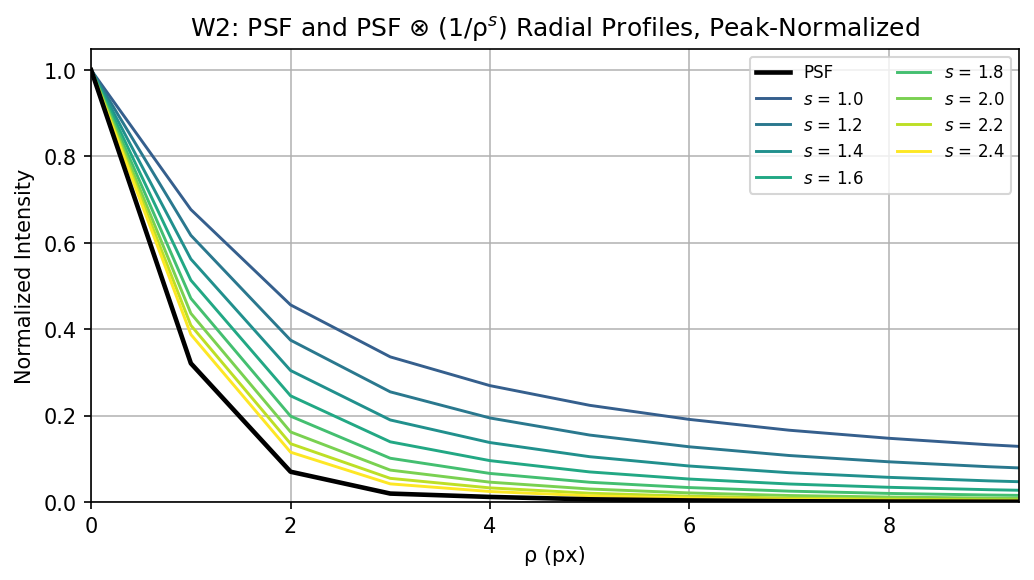}
\caption{Profile degeneracy between the PSF and the convolved coma template as a function of coma slope $s$, shown against projected radius $\rho$. As $s$ increases toward $\simeq 2$, the convolved coma profile approaches the PSF, so the decomposition of Eq.~\ref{eq:phot_obs_signal} becomes ill-conditioned; this motivates the peak constraint of Eq.~\ref{eq:peak} and the Model~1 fallback.
\label{Figap_01}}
\end{figure}

The weighted residual sum of squares,
\begin{equation}
  \chi^2(B) \;=\; \sum_\rho w_\rho\,[\,d(\rho) - M(\rho;B)\,]^2~, \qquad
  w_\rho = \sigma(\rho)^{-2}~,
  \label{eq:chi2B}
\end{equation}
\noindent is quadratic in $B$ and has a unique minimum. Setting $\mathrm{d}\chi^2/\mathrm{d}B = 0$ gives the closed-form weighted-least-squares estimate
\begin{equation}
  B_{\rm unc} \;=\; \frac{\sum_\rho w_\rho\,g(\rho;s)\,[\,d(\rho) - M_1(\rho)\,]}{\sum_\rho w_\rho\,g(\rho;s)^2}~, \qquad \text{with} \quad
  g(\rho;s) \;\equiv\; C(\rho;s) - \frac{\max[\,C(\cdot;s)\,]}{\max P}\,P(\rho)~,
  \label{eq:Bunc}
\end{equation}
\noindent that is, the inverse-variance-weighted projection of the residual about the Model~1 profile, $d(\rho) - M_1(\rho)$, onto the coma direction $g(\rho;s)$, the coma template with its peak-locked PSF component removed. Nonnegativity of both amplitudes ($A \ge 0$ and $B \ge 0$) restricts $B$ to
\begin{equation}
  0 \;\le\; B \;\le\; B_{\rm max}(s) \;=\; \frac{d_{\rm peak} - b}{\max[\,C(\cdot;s)\,]}~,
  \label{eq:Bbox}
\end{equation}
\noindent and we clip the estimate accordingly, $B = \min[\,\max(B_{\rm unc}, 0),\, B_{\rm max}\,]$, recovering $A$ from Eq.~\ref{eq:AofB}. The lower bound $B = 0$ recovers Model~1, with the entire peak assigned to the nucleus; the upper bound, where $A = 0$, corresponds to a pure-coma solution. Figure~\ref{Figap_01} compares the modeled radial profiles at several coma slopes $s$ with the PSF profile. Near $s \simeq 2$ the convolved coma template becomes nearly indistinguishable from the PSF, so $B$ is poorly constrained. The same situation arises observationally when a comet is inactive or when its coma is too compact to resolve, in which case the profile is likewise PSF-like. In the exact degeneracy, $g \rightarrow 0$, the denominator of Eq.~\ref{eq:Bunc} vanishes, $B$ is unidentifiable, and the fit reduces to the Model~1 limit.

\subsection{Step 5: Slope Selection}

For each slope on the grid, we build the templates, solve Eq.~\ref{eq:Bunc}, apply the clip, recover $A_{\rm M2}$, and evaluate the residual
\begin{equation}
  \chi^2(s) \;=\; \sum_\rho w_\rho\,[\,d(\rho) - b - A_{\rm M2}(s)\,P(\rho) - B(s)\,C(\rho;s)\,]^2~.
  \label{eq:chi2s}
\end{equation}
The adopted solution is the $(A_{\rm M2}, B, s)$ with the smallest $\chi^2(s)$. Because only $B$ is free under the peak constraint, the fit has $N_{\rm bin} - 1$ degrees of freedom, where $N_{\rm bin}$ is the number of fitted radial bins, and the one-parameter standard errors are
\begin{equation}
  \sigma_B \;=\; \Big(\sum_\rho w_\rho\,g(\rho;s)^2\Big)^{-1/2}~, \qquad
  \sigma_A \;=\; \frac{\max[\,C(\cdot;s)\,]}{\max P}\,\sigma_B~,
  \label{eq:errs}
\end{equation}
\noindent with $A_{\rm M2}$ and $B$ perfectly anti-correlated (correlation coefficient $-1$) along the peak-equality line.

\begin{deluxetable}{lll}
\tablecaption{Inputs and outputs of the profile-fitting procedure.\label{tab:m1m2_io}}
\tablewidth{0pt}
\tablehead{\colhead{Step} & \colhead{Inputs} & \colhead{Outputs}}
\startdata
Preprocessing      & raw cutout; $b$, $\sigma_{\rm b}$, gain (Paper~I)                 & centroid; $b$; $\sigma_{\rm b}$; $d_{\rm peak}$ \\
Per-wedge profile & centered cutout; $N = 24$                                         & $\mu_k(\rho)$; cell areas $a_{k,\rho}$ \\
Tail detection     & $\mu_k(\rho)$                                                     & quiet-wedge set $\mathcal{Q}$; tail position angle \\
Quiet profile      & $\mu_k(\rho)$ for $k \in \mathcal{Q}$                             & $d(\rho)$; $\sigma(\rho)$ \\
Per-slope fit      & $d(\rho)$, $\sigma(\rho)$, $d_{\rm peak}$, $b$, $P$, $C(\cdot;s)$  & $A_{\rm M2}(s)$, $B(s)$, $\chi^2(s)$ \\
Slope selection    & $\chi^2(s)$ on the grid                                           & $A_{\rm M2}$, $B$, $s$; $\sigma_A$, $\sigma_B$; $\chi^2_{\rm red}$\\
\enddata
\tablecomments{$\rho$ is the projected radial distance from the centroid and $\mathcal{Q}$ the quiet (non-tail) wedges. Model~1 corresponds to the $B = 0$ limit of the per-slope fit. Each variable and procedure is described in Appendix~\ref{sec:app1}.}
\end{deluxetable}

\section{Flux Injection \label{sec:fluxinj}}

The formal nucleus uncertainty $\sigma_A$ from the peak-constrained fit (Appendix~\ref{sec:app1}) captures the statistical degeneracy between nucleus and coma but assumes the $1/\rho^s$ coma model is exact. Real comae can exhibit features that do not match the radially symmetric ideal model. To quantify the resulting systematic error, we performed an empirical injection-recovery analysis independently for each coadded image.

The procedure adds a synthetic point source of known amplitude at the nucleus location, refits Model~2, and measures the change in recovered amplitude. If the decomposition is robust, the recovered nucleus amplitude should increase by exactly the injected flux. We defined a function $\text{Fit}_{M2}$ that accepts an image and returns the best-fit nucleus amplitude $A_{\rm M2}$. We first drew 100 independent noise realizations $n_k \sim \mathcal{N}(0, \sigma_{\rm b})$, where $\sigma_{\rm b}$ is the background flux uncertainty, and fit the baseline image under each to obtain $A_{\rm base,k} = \text{Fit}_{M2}(\text{image} + n_k)$.

For each of four injected amplitudes,
\begin{equation}
  A_{\rm true} \;=\; (0.1,\, 0.25,\, 0.5,\, 1.0) \times A_{\rm M1}~,
\end{equation}
\noindent where $A_{\rm M1}$ is the peak amplitude from Model~1, we added a scaled PSF template to the coadd, refit under the \emph{same} 100 noise fields, and computed the matched-noise recovery increment
\begin{equation}
  \Delta A_k \;=\; \text{Fit}_{M2}\big(\text{image} + A_{\rm true}\,\text{PSF} + n_k\big) \;-\; \text{Fit}_{M2}(\text{image} + n_k)~.
  \label{eq:deltaA}
\end{equation}
Sharing the noise field $n_k$ between the injected and baseline fits cancels the noise common to both, so $\Delta A_k$ isolates the recovery of the injected source alone. In the ideal case, $\Delta A_k = A_{\rm true}$ for all realizations. Deviations from this equality reveal where the model struggles to correctly separate the nucleus signal from the coma.

At each injected amplitude, we computed the $\texttt{bias} = \text{median}_k(\Delta A_k) - A_{\rm true}$ and $\texttt{uncertainty} = \text{std}_k(\Delta A_k)$ of the reconstructed injected fluxes. The fraction of the injected nucleus that is successfully recovered by the fit,
\begin{equation}
  f \;=\; \frac{A_{\rm true} + \texttt{bias}}{A_{\rm true}}~,
\end{equation}
\noindent was evaluated at the brightest injected amplitude and clamped to $[0.05, 1]$ to prevent excessive inflation. The $\texttt{uncertainty}$ is interpolated along the four-point curve to the observed nucleus amplitude $A_{\rm M2}$, and the injection uncertainty is their ratio,
\begin{equation}
  \sigma_{\rm inj} \;=\; \frac{\texttt{uncertainty}(A_{\rm M2})}{f}~.
\end{equation}
This per-stack uncertainty is independent of the formal $\sigma_A$ and typically exceeds it when the real coma deviates significantly from the $1/\rho^s$ template.

The final effective uncertainty used in the thermal modeling is
\begin{equation}
  \sigma_{\rm eff} \;=\; \max(\sigma_A,\, \sigma_{\rm inj})~,
\end{equation}
\noindent taking whichever is larger. A band-epoch enters the NEATM fit as a nucleus measurement, $F_b = A_{\rm M2} \pm \sigma_{\rm eff}$, only when both recoverability criteria are satisfied: $A_{\rm M2} \ge \sigma_{\rm eff}$ (nucleus detected above noise) and $A_{\rm M2} < 0.95\,A_{\rm M1}$ (decomposition differs from pure PSF). Band-epochs failing either test revert to the conservative Model~1 upper limit.

\section{Thermal Origin of the Heliocentric Color Trends \label{sec:app2}}

\begin{figure}[htb]
\centering
\includegraphics[width=0.57\textwidth]{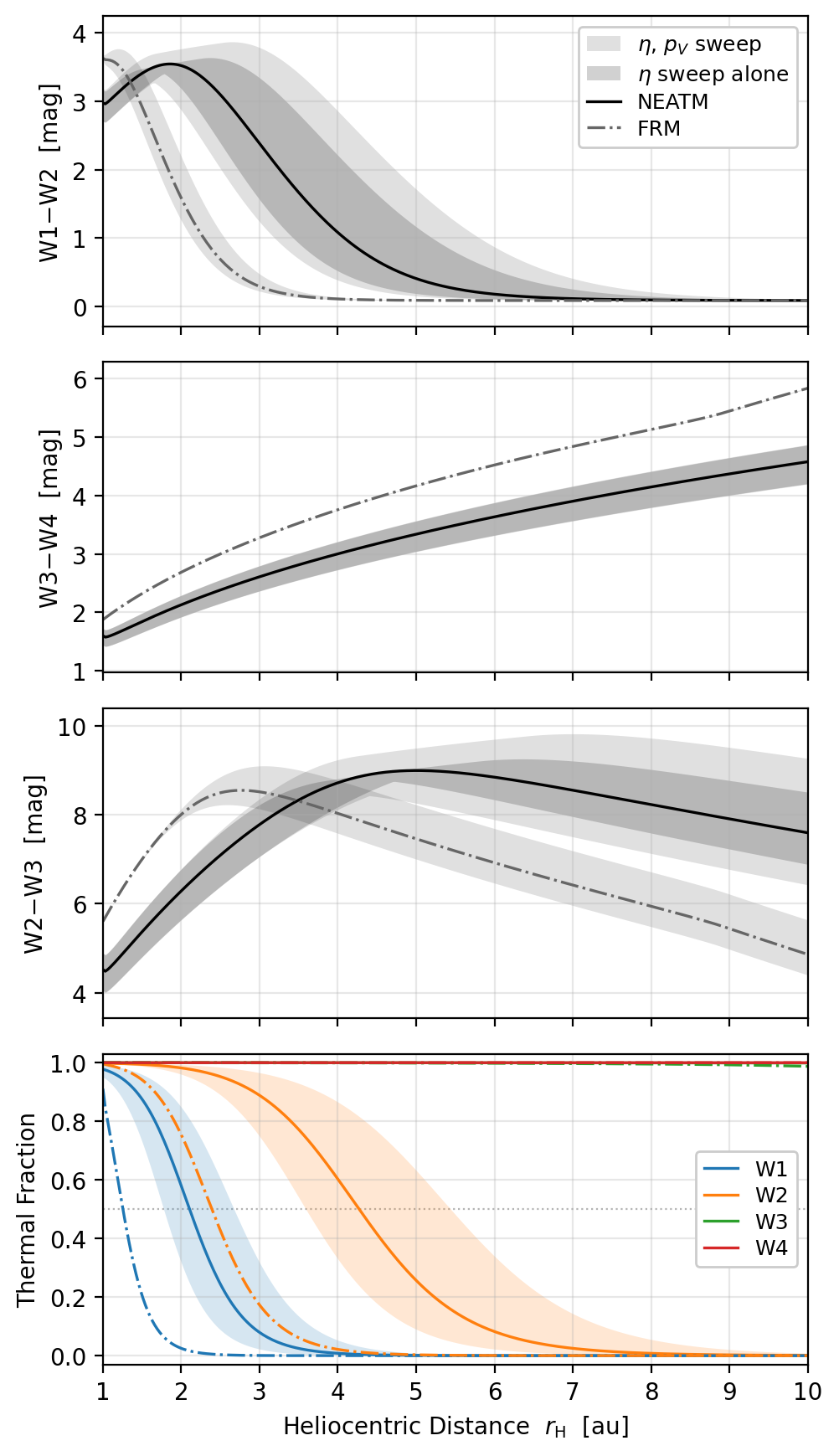}
\caption{Model predictions for the heliocentric dependence of nucleus colors, computed for a 5~km nucleus at 90~degree solar elongation. Top three panels: $W1-W2$, $W3-W4$, and $W2-W3$ from NEATM (solid) and the FRM (dash-dot), for $p_V = 0.04$ and $\eta = 1.03$. Two shading spans 1) $\eta = 0.70$ to 1.36 at fixed albedo and 2) $p_V$ from 0.02 to 0.06. The FRM band reflects the albedo sweep only. Bands are absent where the swept parameter has negligible effect, as for the albedo in $W3-W4$. Bottom panel: thermal fraction of each WISE band, with line styles as above. The dotted line at 0.5 marks equal thermal and reflected contributions. The diameter enters as a model input but does not affect any quantity shown.
\label{Figap_02}}
\end{figure}

This appendix details the model predictions used in Section~\ref{sec:general_phot} to interpret the heliocentric dependence of the nucleus colors. Figure~\ref{Figap_02} shows the $W1-W2$, $W3-W4$, and $W2-W3$ colors predicted by NEATM \citep{Harris1998} and the FRM \citep{Lebofsky1989} as a function of $r_{\rm H}$, together with the thermal fraction of each band, defined as the ratio of thermal to total (thermal plus reflected) flux. The models are evaluated for a 5~km nucleus at the 90~degree solar elongation of the WISE survey. The diameter is required as a model input but cancels in every color and fraction. The reference case uses $p_V = 0.04$ and, for NEATM, $\eta = 1.03$; the shaded bands sweep $p_V$ from 0.02 to 0.06 \citep{Lamy2004} and $\eta$ from 0.70 to 1.36 (3$\sigma$ of the beaming priors in Table~\ref{tab:neatm_priors}). The FRM has no beaming parameter, so its band reflects the albedo sweep alone. The two models bracket the plausible surface temperature distribution, from emission concentrated near the subsolar point (NEATM) to isothermal latitude bands (FRM). The cooler FRM surface shifts every transition inward and raises the colors between thermal bands.

The bottom panel establishes which component dominates each band. W3 and W4 are thermal throughout the range plotted. W1 is thermal only inside about 2~au ($\sim$1.3~au for the FRM), and W2 crosses from thermal to reflected near 4~au ($\sim$2.4~au for the FRM). W2 is therefore the only band in which both components contribute comparably over the distances sampled here, with a thermal fraction that depends on $\eta$ as well as on $r_{\rm H}$.

These crossings set the shapes of the color curves. $W1-W2$ is near zero beyond 4~au, where both bands are reflected, rises to a maximum near 2~au, where W2 has become thermal while W1 remains reflected, and declines inside 1.5~au as thermal emission enters W1. $W3-W4$ increases monotonically with distance because both bands remain thermal throughout. Both band pairs lie on the short-wavelength side of the thermal peak and therefore the longer-to-shorter flux ratio rises as the surface cools. $W2-W3$ rises for the same reason while W2 is thermal, but peaks near 5~au (2.7~au for the FRM) and declines beyond it once W2 has become reflected and its flux no longer follows the steep thermal dependence on $r_{\rm H}$. Our $W2-W3$ measurements all lie inside 3~au, on the rising branch, which is why the trend appears monotonic in Figure~\ref{Fig_nuchelio}. In summary, the ensemble color distribution of multiple comets seen here is largely shaped by basic principles of thermal physics, rather than albedo variations in individual comets.

\bibliography{sample701}{}
\bibliographystyle{aasjournalv7}

\end{document}